\documentclass[onecolumn,epjc3]{svjour3} 
 
\usepackage[T1]{fontenc}
\usepackage{epsfig,amsmath,amssymb}
\usepackage{graphicx}
\usepackage{xspace}
\usepackage{listings}
\usepackage{ulem}
\usepackage{slashed}
\usepackage{cite}
\usepackage[colorlinks,citecolor=blue,urlcolor=blue,linkcolor=blue]{hyperref}
\journalname{Eur. Phys. J. C}

\def\hc{\text{h.c.}}
\def\BR{\text{BR}}
\def\IM{\, \text{Im}}
\def\RE{\, \text{Re}}

\newcommand\SARAH{{\tt SARAH}\xspace}
\newcommand\NewPackage{{\tt PreSARAH}\xspace}

\newcommand\FeynArts{{\tt FeynArts}\xspace}
\newcommand\FormCalc{{\tt FormCalc}\xspace}

\newcommand\SPheno{{\tt SPheno}\xspace}

\newcommand\FlavorKit{{\tt FlavorKit}\xspace}

\newcommand\Mathematica{{\tt Mathematica}\xspace}

\newcommand{\Fortran}{\texttt{Fortran}\xspace}

\begin{document}

\title{A Flavor Kit for BSM models}
\author{
   Werner Porod \thanksref{a1} \and
   Florian Staub \thanksref{a2} \and
   Avelino Vicente \thanksref{a3}}
 
\institute{
Institut f\"ur Theoretische Physik und Astronomie, 
Universit\"at W\"urzburg, 97074 W\"urzburg, Germany\label{a1} 
\and
BCTP \& Physikalisches Institut der 
Universit\"at Bonn, Nu{\ss}allee 12, 53115 Bonn, Germany\label{a2} 
\and
IFPA, Dep. AGO, Universit\'e de Li\`ege, Bat B5, Sart-Tilman B-4000
Li\`ege 1, Belgium\label{a3}   
}

\date{BONN-TH-2014-07}

\maketitle

\lstset{frame=shadowbox}
\lstset{prebreak=\raisebox{0ex}[0ex][0ex]
        {\ensuremath{\hookrightarrow}}}
\lstset{postbreak=\raisebox{0ex}[0ex][0ex]
        {\ensuremath{\hookleftarrow\space}}}
\lstset{breaklines=true, breakatwhitespace=true}
\lstset{numbers=left, numberstyle=\scriptsize}
 
\begin{abstract}
We present a new kit for the study of flavor observables in models
beyond the standard model. The setup is based on the public codes
\SARAH and \SPheno and allows for an easy implementation of new
observables. The Wilson coefficients of the corresponding operators in
the effective lagrangian are computed by \SPheno modules written by
\SARAH. New operators can also be added by the user in a modular
way. For this purpose a handy \Mathematica package called \NewPackage
has been developed. This uses \FeynArts and \FormCalc to derive the
generic form factors at tree- and 1-loop levels and to generate the
necessary input files for \SARAH.  This framework has been used to
implement BR($\ell_\alpha¸\to \ell_\beta \gamma$), BR($\ell_\alpha \to
3\, \ell_\beta$), CR($\mu-e,A$), BR($\tau \to P \, \ell$), BR($h\to
\ell_\alpha \ell_\beta$), BR($Z\to \ell_\alpha \ell_\beta$),
BR($B_{s,d}^0 \to \ell \bar{\ell}$), BR($\bar B \to X_s\gamma$),
BR($\bar B \to X_s \ell \bar{\ell}$), BR($\bar B \to X_{d,s} \nu \bar
\nu$), BR($K^+ \to \pi^+ \nu \bar \nu$), BR($K_L \to \pi^0 \nu \bar
\nu$), $\Delta M_{B_s,B_d}$, $\Delta M_K$, $\varepsilon_K$, BR($B \to
K \mu \bar{\mu}$), BR($B\to \ell \nu$), BR($D_s \to \ell \nu$) and
BR($K \to \ell \nu$) in \SARAH. Predictions for these observables can
now be obtained in a wide range of SUSY and non-SUSY models. Finally,
the user can use the same approach to easily compute additional
observables.
\end{abstract}

\tableofcontents

\section{Introduction}
With the exploration of the terascale, particle physics has entered a
new era. On the one hand, the discovery of a Higgs boson at the LHC
\cite{Aad:2012tfa,Chatrchyan:2012ufa} seemingly completed the Standard
Model (SM) of particle physics, even though there is still quite some
room for deviations from the SM predictions.  The observed mass of
about 125 GeV in combination with a top quark mass of 173.34~GeV
\cite{ATLAS:2014wva} implies within the SM that we potentially live in a
meta-stable vacuum \cite{Buttazzo:2013uya}. This, together with other
observations, like the dark matter relic density or the  
unification of gauge forces, indicates that there is physics
beyond the SM (BSM). Although no sign of new physics has been found so
far at the LHC, colliders are not the only places where one can search
for new physics. Low energy experiments focused on flavor observables
can also play a major role in this regard, since new particles leave
their traces via quantum effects in flavor violating processes such
as $b\to s \gamma$, $B_s \to \mu^+ \mu^-$ or $\mu\to e \gamma$. In the
last few years there has been a tremendous progress in this field,
both on the experimental as well as on the theoretical side. In
particular, observables from the Kaon- and B-meson sectors, rare
lepton decays and electric dipole moments have put stringent bounds on
new flavor mixing parameters and/or additional phases in models
beyond the SM.

There are several public tools on the market which predict 
the rates of several flavor observables: 
{\tt superiso} \cite{Mahmoudi:2007vz,Mahmoudi:2008tp,Mahmoudi:2009zz}, {\tt
SUSY\_Flavor} \cite{Rosiek:2010ug,Crivellin:2012jv},
{\tt NMSSM-Tools} \cite{Ellwanger:2006rn},
{\tt MicrOmegas} \cite{Belanger:2001fz,Belanger:2004yn,Belanger:2006is,Belanger:2008sj,Belanger:2013oya}, 
{\tt SuperBSG} \cite{Degrassi:2007kj},
{\tt SupeLFV} \cite{Murakami:2013rca}, {\tt SuseFlav} \cite{Chowdhury:2011zr},
{\tt IsaJet} with {\tt IsaTools} \cite{Paige:2003mg,Baer:2003xc,Baer:1999sp,Paige:1998xm,Paige:1998ux,Baer:1993ae} 
or {\tt SPheno} \cite{Porod:2003um,Porod:2011nf}. 
However, all of these codes have in common that they are only valid in
the Two-Higgs-doublet model or in the MSSM or simple extensions of it
(NMSSM, bilinear R-parity violation). In addition, none of these tools
can be easily extended by the user to calculate additional
observables. This has made flavor studies beyond the SM a cumbersome
task. The situation has changed with the development of \SARAH
\cite{Staub:2008uz,Staub:2009bi,Staub:2010jh,Staub:2012pb,Staub:2013tta}. This
\Mathematica package can be used to generate modules for \SPheno,
which then can calculate flavor observables at the 1-loop level in a
wide range of supersymmetric and non-supersymmetric models
\cite{Dreiner:2012mx,Dreiner:2012dh,Dreiner:2013jta}. However, so far
all the information about the underlying Wilson
coefficients\footnote{Sometimes the \emph{Wilson coefficients} are
  also referred to as \emph{form factors}. We will nevertheless stick
  to the name \emph{Wilson coefficients} in the following, also 
  for lepton flavor violating processes.} for the
operators triggering the flavor violation as well as the calculation
of the flavor observables had been hardcoded in \SARAH. Therefore, it
was also very difficult for the user to extend the list of calculated
observables. The implementation of new operators was even more difficult.

We present a new kit for the study of flavor observables beyond the
standard model. In contrast to previous flavor codes, \FlavorKit is
not restricted to a single model, but can be used to obtain
predictions for flavor observables in a wide range of models (SUSY and
non-SUSY).  \FlavorKit can be used in two different ways. The basic
usage of \FlavorKit allows for the computation of a large number of
lepton and quark flavor observables, using generic analytical
expressions for the Wilson coefficients of the relevant operators. The
setup is based on the public codes \SARAH and \SPheno, and thus allows
for the analytical and numerical computation of the observables in the
model defined by the user. If necessary, the user can also go beyond
the basic usage and define his own operators and/or observables. For
this purpose, a \Mathematica package called \NewPackage has been
developed. This tool uses \FeynArts/\FormCalc
\cite{Hahn:1998yk,Hahn:2000kx,Hahn:2000jm,Hahn:2004rf,Hahn:2005vh,Nejad:2013ina}
to compute generic expressions for the required Wilson coefficients at
the tree- and 1-loop levels. Similarly, the user can easily implement
new observables. With all these tools properly combined, the user can
obtain analytical and numerical results for the observables of his
interest in the model of his choice. To calculate new flavor
observables with \SPheno for a given model the user only needs the
definition of the operators and the corresponding expressions for the
observables as well as the model file for \SARAH. All necessary
calculations are done automatically.  We have used this setup to
implement BR($\ell_\alpha¸\to \ell_\beta \gamma$), BR($\ell_\alpha \to
3\, \ell_\beta$), CR($\mu-e,A$), BR($\tau \to P \, \ell$), BR($h\to
\ell_\alpha \ell_\beta$), BR($Z\to \ell_\alpha \ell_\beta$),
BR($B_{s,d}^0 \to \ell \bar{\ell}$), BR($\bar B \to X_s\gamma$),
BR($\bar B \to X_s \ell \bar{\ell}$), BR($\bar B \to X_{d,s} \nu \bar
\nu$), BR($K^+ \to \pi^+ \nu \bar \nu$), BR($K_L \to \pi^0 \nu \bar
\nu$), $\Delta M_{B_s,B_d}$, $\Delta M_K$, $\varepsilon_K$, BR($B \to
K \mu \bar{\mu}$), BR($B\to \ell \nu$), BR($D_s \to \ell \nu$) and
BR($K \to \ell \nu$) in \SARAH.

This manual is structured as follows: in the next section we give a
brief introduction into the calculation of flavor observables focusing
on the main steps that one has to follow. Then we present \FlavorKit,
our setup to combine \FeynArts/\FormCalc, \SPheno and \SARAH in
Section \ref{sec:setup}. In Section \ref{sec:observables} we explain how new
observables can be added and in Section \ref{sec:operators} how the
list of operators can be extended by the user. A  comparison between 
\FlavorKit and the other public codes is presented in Section
\ref{sec:validation} taking the MSSM as an example 
before we conclude in Section
\ref{sec:conclusion}.  The appendix contains information about the
existing operators and how they have been combined to compute the
different flavor observables.

\section{General strategy: calculation of flavor observables in a nutshell}
\label{sec:strategy}

Once we have chosen a BSM model~\footnote{The current version of
  \FlavorKit can only handle renormalizable operators at this stage of
  the computation.}, our general strategy for the computation of
a flavor observable follows these steps:

\begin{itemize}

\item {\bf Step 1:} We first consider an effective Lagrangian
  that includes the operators relevant for the flavor observable of our
  interest,
 \begin{equation}
  {\cal L}_{eff} = \sum_i C_i {\cal O}_i \, .
 \end{equation}
 This Lagrangian consists of a list of (usually) higher-dimensional
 operators ${\cal O}_i$. The Wilson coefficients $C_i$ can be induced
 either at tree or at higher loop levels and include both the SM and
 the BSM contributions ($C_i = C_i^\text{SM} + C_i^\text{BSM}$). They
 encode the physics of our model.

\item {\bf Step 2:} The Wilson coefficients are computed
  diagrammatically, taking into account all possible tree-level and
  1-loop topologies leading to the ${\cal O}_i$ operators~\footnote{In
    principle, one can go beyond the 1-loop level, although in our
    case we will restrict our computation to the addition of a few NLO
    corrections.}.

\item {\bf Step 3:} The results for the Wilson coefficients are
  plugged in a general expression for the observable and a final
  result is obtained.

\end{itemize}

The user has to make a choice in step 1. The list of operators in
the effective Lagrangian can be restricted to the most relevant ones
or include additional operators beyond the leading contribution,
depending on the required level of precision. Usually, the complete
set of renormalizable operators contributing to the observable of
interest is considered, although in some well motivated cases one may
decide to concentrate on a smaller subset of operators.  This freedom
is not present in step 2. Once the list of operators has been
arranged, the computation of the corresponding $C_i$ coefficients
follows from the consideration of all topologies (penguin diagrams,
box diagrams, \dots) leading to the ${\cal O}_i$ operators. This is
the most complicated and model dependent step, since it demands a full
knowledge of all masses and vertices in the model under
study. Furthermore, it may be necessary to compute the coefficients at
an energy scale and then obtain, by means of their renormalization
group running, their values at a different scale. Finally, step 3 is
usually quite straightforward since, like step 1, is model
independent. In fact, the literature contains general expressions for
most flavor observables, thus facilitating the final step. However,
one should be aware that the formulas given in the literature assume
that certain operators contribute only sub-dominantly and, thus, omit
the corresponding contributions. This is in general justified for the
SM but not in a general BSM model. In particular, this is the case for
processes involving external neutrinos, which are often assumed to be
purely left-handed, making the operators associated to their
right-handed components to be neglected.

We will exemplify our strategy using a simple example: BR($\mu \to e
\gamma$) in the Standard Model extended by right-handed neutrinos and
Dirac neutrino masses. The starting point is, as explained above, to
choose the relevant operators. In this case, it is well known that
only dipole interactions can contribute to to the radiative decay
$\ell_\alpha \to \ell_\beta \gamma$ at leading order~\footnote{At next
  to leading order, one would also have to consider operators like
  $\bar{\mu} \gamma_\nu e \, \bar{q} \gamma^\nu q$, to be combined
  with a $q - q - \gamma$ dipole interaction.}. Therefore, the
relevant operators are contained in the $\ell - \ell - \gamma$ dipole
interaction Lagrangian. This is in general given by
\begin{equation} 
{\cal L}_{\ell \ell \gamma}^{\text{dipole}} = i e \, m_{\ell_\alpha} \, \bar \ell_\beta \sigma^{\mu \nu} q_\nu \left(K_2^L P_L + K_2^R P_R \right) \ell_\alpha A_\mu + \hc
\end{equation}
Here $e$ is the electric charge, $q$ the photon momentum, $P_{L,R} =
\frac{1}{2} (1 \mp \gamma_5)$ are the usual chirality projectors and
$\ell_{\alpha,\beta}$ denote the lepton flavors. This concludes step 1.

The information about the underlying model is encoded in the
coefficients $K_2^{L,R}$.  In the next step, these coefficients have
to be calculated by summing up all Feynman diagrams contributing at a
given loop level. Expressions for these coefficients for many
different models are available in the literature. In the SM only
neutrino loops contribute and one finds \cite{Ma:1980gm}
\begin{align}
K_2^L = & \frac{G_F}{2 \sqrt{2} \pi^2} m_\mu \sum_i \lambda_{i\mu} \lambda^*_{i e}  
 (F_1 + F_2)\\
K_2^R = & \frac{G_F}{2 \sqrt{2} \pi^2} m_e \sum_i \lambda_{i\mu} \lambda^*_{i e}  
 (F_1 - F_2)
\end{align}
Here, $\lambda_{ij}$ denote the entries of the
Pontecorvo-Maki-Nakagawa-Sakata matrix and $F_1$ and $F_2$ are
loop functions. One finds approximately $F_1 \simeq
-\frac{1}{4}\left(\frac{m_\nu}{m_W}\right)^2$ and $F_2 \simeq 0$.
Finally, we just need to proceed to the last step, the computation of
the observable. After computing the Wilson coefficients $K_2^{L,R}$ it is
easy to relate them to BR($\mu \to e \gamma$) by
using~\cite{Hisano:1995cp}
\begin{equation}
\Gamma \left( \ell_\alpha \to \ell_\beta \gamma \right) =
\frac{\alpha m_{\ell_\alpha}^5}{4} \left( |K_2^L|^2 + |K_2^R|^2 \right) \, ,
\end{equation}                 
This expression holds for all models. With this final step, the
computation concludes.

As we have seen, the main task to get a prediction for BR($\mu \to e
\gamma$) in a new model is to calculate $K_2^{L,R}$. However, this
demands the knowledge of all masses and vertices involved. Moreover,
in most cases a numerical evaluation of the resulting loop integrals
is also welcome. Therefore, even for a \emph{simple} process like $\mu
\to e \gamma$, a computation from scratch in a new model can be a hard
work. In order to solve this practical problem, we are going to
present here a fully automatized way to calculate a wide range of
flavor observables for several classes of models.

\section{Setup}
\label{sec:setup}

\subsection{\FlavorKit: usage and goals}


As we have seen, the calculation of flavor observables in a specific
model is a very demanding task. A detailed knowledge about the model
is required, including
\begin{enumerate}
 \item expressions for all involved masses and vertices
 \item optionally, renormalization group equations to get the running parameters at the considered scale
 \item expressions to calculate the operators
 \item formulae to obtain the observables from the operators
\end{enumerate}
Nearly all codes devoted to flavor physics have those pieces
hardcoded, and they are only valid for a few specific
models~\footnote{Recently, {\tt Peng4BSM@LO} \cite{Bednyakov:2013tca}
  was made public. This code derives analytical expressions for vector
  penguins for a model defined in the corresponding \FeynArts model
  file.}. The only exception is \SPheno, thanks to its extendability
with new modules for additional models. These modules are generated by
the \Mathematica package \SARAH and provide all necessary information
about the calculation of the (loop corrected) mass spectrum, the
vertices and the 2-loop RGEs. These expressions, derived from
fundamental principles for any (renormalizable) model, contain all the
information required for the computation of flavor observables.  In
fact, \SARAH also provides \Fortran code for a set of flavor
observables. For this output, generic expressions of the necessary
Wilson coefficients have been included. These are matched to the model
chosen by the user and related to the observables by the standard
formulae available in the literature. However, it was hardly possible
for the user to extend the list of observables or operators included
in \SARAH without a profound knowledge of either the corresponding
\Mathematica or \Fortran code.

We present a new setup to fill this gap in \SARAH: \FlavorKit. As
discussed in Sec. \ref{sec:strategy}, the critical step in the
computation of a flavor observable is the derivation of analytical
expressions for the Wilson coefficients of the relevant
operators. This step, being model dependent, requires information
about the model spectrum and interactions. However, generic
expressions can be derived, later to be matched to the specific
spectrum and interaction Lagrangian of a given model. For this
purpose, we have created a new \Mathematica package called
\NewPackage. This package uses the power of \FeynArts and \FormCalc to
calculate generic 1-loop amplitudes, to extract the coefficients of
the demanded operators, to translate them into the syntax needed for
\SARAH and to write the necessary wrapper code. \NewPackage works for
any 4-fermion or 2-fermion-1-boson operators and will be extended in
the future to include other kinds of operators. The current version
already contains a long list of fully implemented operators (see
Appendix \ref{app:operators}). The results for the Wilson coefficients
obtained with \NewPackage are then interpreted by \SARAH, which adapts
the generic expressions to the specific details of the model chosen by
the user and uses snippets of \Fortran code to calculate flavor
observables from the resulting Wilson coefficients. As for the
operators, there is a long list of observables already implemented
(see Appendices \ref{sec:LFV} and \ref{sec:QFV}). Finally, \SARAH can
be used to obtain analytical output in \LaTeX\ format
or to create \Fortran modules for
\SPheno, thus making possible numerical studies.

\begin{table}
\centering
\begin{tabular}{c c}
\hline
{\bf Lepton flavor} & {\bf Quark flavor} \\
\hline
$\ell_\alpha \to \ell_\beta \gamma$ & $B_{s,d}^0 \to \ell^+ \ell^-$ \\
$\ell_\alpha \to 3 \, \ell_\beta$ & $\bar B \to X_s \gamma$ \\
$\mu-e$ conversion in nuclei & $\bar B \to X_s \ell^+ \ell^-$ \\
$\tau \to P \, \ell$ & $\bar B \to X_{d,s} \nu \bar \nu$ \\
$h \to \ell_\alpha \ell_\beta$ & $B \to K \ell^+ \ell^-$ \\
$Z \to \ell_\alpha \ell_\beta$ & $K \to \pi \nu \bar \nu$ \\
 & $\Delta M_{B_{s,d}}$ \\
 & $\Delta M_{K}$ and $\varepsilon_K$ \\
 & $P \to \ell \nu$ \\
\hline
\end{tabular}
\caption{List of flavor violating processes and observables which have
  been already implemented in \FlavorKit. To the left, observables
  related to lepton flavor, whereas to the right observables
  associated to quark flavor. See appendices \ref{sec:LFV} and
  \ref{sec:QFV} for the definition of the observables and the relevant
  references for their calculation.}
\label{tab:observables}
\end{table}

\FlavorKit can be used in two ways:

\begin{itemize}

\item {\bf Basic usage:} This is the approach to be followed by the
  user who does not need any operator nor observable beyond what is
  already implemented in \FlavorKit. In this case, \FlavorKit reduces
  to the standard \SARAH package. The user can use \SARAH to obtain
  analytical results for the flavor observables and, if he wants
  to make numerical studies, to produce \Fortran modules for
  \SPheno. For the list of implemented operators we refer to Appendix
  \ref{app:operators}, whereas the list of implemented observables is
  given in Table \ref{tab:observables}.

\item {\bf Advanced usage:} This is the approach to be followed by the
  user who needs an operator or an observable not included in
  \FlavorKit. In case the user is interested in an operator that is
  not implemented in \FlavorKit, he can define his own operators and
  get analytical results for their coefficients using
  \NewPackage. Then the output can be passed to \SARAH in order to
  continue with the basic usage. In case the user is interested in an
  observable that is not implemented in \FlavorKit, this can be easily
  implemented by the addition of a \Fortran file, with a few lines of
  code relating the observable to the operators in \FlavorKit
  (implemented by default or added by the user).  The \Fortran files
  just have to be put together with a short steering file into a
  specific directory located in the main \SARAH directory. Then one
  can continue with the basic usage.

\end{itemize}

\begin{figure}[hbt]
\begin{center}
\includegraphics[width=0.66\linewidth]{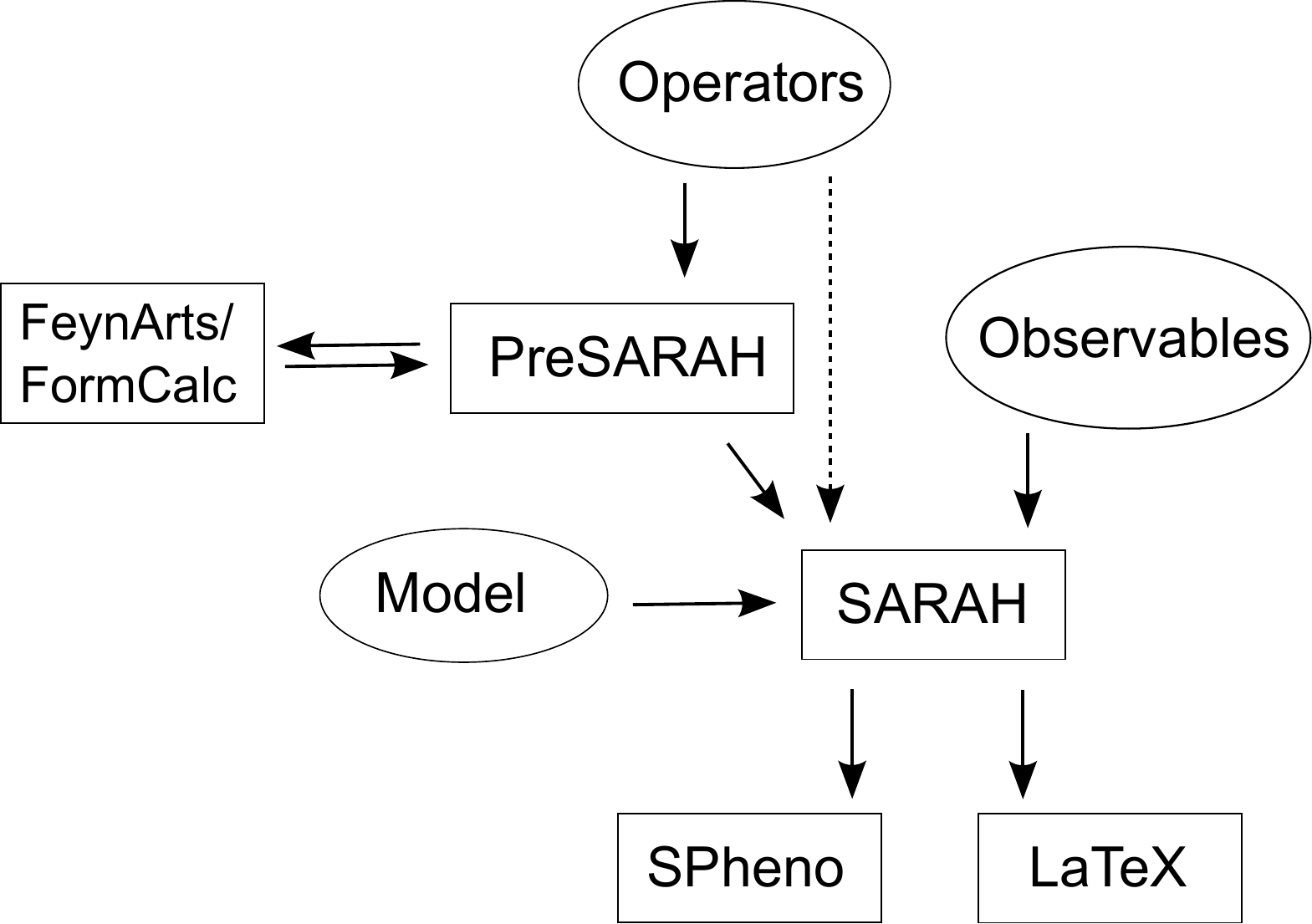}
\end{center}
\caption{Schematic way to use \FlavorKit: the user can define new
  operators in \NewPackage, which then calculates the coefficients in
  a generic form using \FeynArts and \FormCalc and creates the
  necessary input files for \SARAH. In addition, \Fortran code can be
  provided to relate the Wilson coefficients to specific flavor
  observables. This information is used by \SARAH to generate \SPheno
  code for the numerical calculation of the observables.}
\label{fig:setup}
\end{figure}
The combination of \NewPackage together with \SARAH and \SPheno allows
for a modular and precise calculation of flavor observables
in a wide range of particles physics models. We have summarized the
setup in Fig.~\ref{fig:setup}: the user provides as input \SARAH model
files for his favorite models or takes one of the models which are
already implemented in \SARAH (see Appendix \ref{app:models} for a list of
models available in \SARAH). New observables are implemented by
providing the necessary \Fortran code to \SARAH while new operators
can be either implemented by hand or by using \NewPackage which then
calls \FeynArts and \FormCalc for the calculation of the necessary
diagrams. However, most users will not require to implement new
operators or observables. In this case, the user can simply use \SARAH
in the standard way and (1) derive analytical results for the Wilson
coefficients and observables, and (2) generate \Fortran modules for
\SPheno in order to run numerical analysis.

\subsection{Download and installation}

\FlavorKit involves several public codes. We proceed to describe how
to download and install them.

\begin{enumerate}

 \item \FeynArts/\FormCalc\\
 \FeynArts and \FormCalc can be downloaded from
 \begin{center}
 {\tt www.feynarts.de/} 
 \end{center}
 It is also possible to use the script {\tt FeynInstall}, to be found
 on the same site, for an automatic installation.
 
 \item \SARAH  and \NewPackage \\
 \SARAH can be downloaded from
 \begin{center}
  {\tt sarah.hepforge.org/}
 \end{center}
 No installation or compilation is necessary. Both packages just need to be extracted by using {\tt tar}. \\
  {\tt 
  > tar -xf SARAH-4.2.0 \\
  > tar -xf PreSARAH-1.0.0 \\
 }
 \NewPackage needs the paths to load \FeynArts and \FormCalc. These have to be provided by the user in the file {\tt PreSARAH.ini}
 \begin{lstlisting}
FeynArtsPackage = "FeynArts/FeynArts.m";
FormCalcPackage = "FormCalc/FormCalc.m";  
 \end{lstlisting}
 This would work if \FeynArts and \FormCalc have been installed in
 the {\tt Application} directory of the local \Mathematica
 installation. Otherwise, absolute paths should be used, e.g.
 \begin{lstlisting}
FeynArtsPackage = "/home/$user/$path/FeynArts-3.7/FeynArts.m";
FormCalcPackage = "/home/$user/$path/FormCalc-8.1/FormCalc.m";  
 \end{lstlisting}

 \item \SPheno\\
 \SPheno can be downloaded from
 \begin{center}
  {\tt spheno.hepforge.org/}
 \end{center}
 After extracting the package, {\tt make} is used for the compilation.  \\
 {\tt 
  > tar -xf SPheno-3.3.0.tar.gz \\
  > cd SPheno-3.3.0 \\
  > make \\
 }

\end{enumerate}

\subsection{Basic usage}

As explained above, \FlavorKit can be used in several ways, depending
on the user's needs and interests. The advanced usage, which involves
the introduction of new observables and/or the computation of new
operators, is explained in detail in Secs. \ref{sec:observables} and
\ref{sec:operators}. Here we focus on the basic usage, which just
requires the codes \SARAH and \SPheno.

\SARAH can handle the analytical derivation of all the relevant Wilson
coefficients in the model defined by the user. The resulting
expressions can be then extracted in \LaTeX\ form or used to generate
a \SPheno module for numerical evaluation. These are the steps to
follow in order to use \SARAH:

\begin{enumerate}
 \item {\bf Loading \SARAH}: after starting \Mathematica, \SARAH is loaded via \\
 {\tt 
 <<SARAH-4.2.0/SARAH.m \\
 }
 or via \\
 {\tt 
 <<[\$path]/SARAH-4.2.0/SARAH.m \\
 }
 The first choice works if \SARAH has been installed in the {\tt Application} directory of \Mathematica. Otherwise,
 the absolute path ({\tt [\$path]}) to the local \SARAH installation must been used. 
 \item {\bf Initialize a model:} as example for the initialization of a model in \SARAH we consider the NMSSM: \\
 {\tt 
 Start[\textquotedblleft NMSSM\textquotedblright]; \\
 }
 \item {\bf Obtaining the \LaTeX\ output}: the user can get \LaTeX\ output with all the information about the model 
 (including the coefficients for the flavor operators) via \\
 {\tt 
 ModelOutput[EWSB]; \\
 MakeTeX[]; \\
 }
 \item {\bf Obtaining the \SPheno code}: to create the \SPheno output the user should run \\
 {\tt 
 MakeSPheno[]; \\
 }
\end{enumerate}

Thanks to \FlavorKit, \SARAH can also write \LaTeX\ files with the
analytical expressions for the Wilson coefficients. These are given
individually for each Feynman diagram contributing to the
coefficients, and saved in the folder
\begin{center}
{\tt  [\$SARAH]/Output/[\$MODEL]/EWSB/TeX/FlavorKit/ }
\end{center}
For the 4-fermion operators the results are divided into separated
files for tree-level contributions, penguins contributions and box
contributions. The corresponding Feynman diagrams are drawn by using
{\tt FeynMF} \cite{Ohl:1995kr}. To compile all Feynman diagrams at
once and to generate the pdf file, a shell script called 
{\tt MakePDF\_[\$OPERATOR].sh} is written as well by \SARAH.

In case the user is interested in the numerical evaluation of the
flavor observables, a \SPheno module must be created as explained
above. Once this is done, the resulting \Fortran code can be used for
the numerical analysis of the model. This can be achieved in the
following way:
\begin{enumerate}
 \item {\bf building \SPheno}: as soon as the \SPheno output is finished, 
 open a terminal and enter the root directory of the \SPheno installation, and create a new subdirectory,
 copy the \SARAH output to that directory and compile it \\
 {\tt 
 > cd [\$SPheno] \\
 > mkdir NMSSM \\
 > cp [\$SARAH]/Output/NMSSM/EWSB/SPheno/* NMSSM/ \\
 > make Model=NMSSM
 }
 \item {\bf Running \SPheno:} After the compilation, a new binary {\tt SPhenoNMSSM} is created.
 This file can be executed providing a standard Les Houches input file (\SARAH provides an example file, see the \SARAH output folder). 
 Finally, \SPheno is executed via  \\
 {\tt 
 > ./bin/SPhenoNMSSM NMSSM/LesHouches.in.NMSSM
 }

 This generates the output file {\tt SPheno.spc.NMSSM}, which contains
 the blocks {\tt QFVobservables} and {\tt LFVobservables}. In those
 two blocks, the results for quark and lepton flavor violating
 observables are given.
\end{enumerate}

Finally, an even easier way to implement new models in \SARAH is the 
{\tt butler} script provided with the {\tt SUSY Toolbox} \cite{Staub:2011dp}
\begin{center}
  {\tt sarah.hepforge.org/Toolbox/} 
\end{center}

\subsection{Limitations}

\FlavorKit is a tool intended to be as general as possible. For this
reason, there are some limitations compared to codes which perform
specific calculations in a specific model. Here we list the main
limitations of \FlavorKit:

\begin{itemize}

\item Chiral resummation is not included because of its large model
  dependence, see e.g.\ \cite{Crivellin:2011jt} and references
  therein.

\item Even though we have included some of the higher order
  corrections for the SM part of some observables in a parametric way,
  2- or higher loop corrections, calculated in the context of the SM
  or the MSSM for specific observables, are not considered, see for
  instance
  \cite{Buras:1990fn,Buchalla:1993wq,Ciuchini:1996sr,Buras:1999st,Misiak:2006ab,Buras:2002tp,Boughezal:2007ny,Buras:2012fs}.

\end{itemize}

\section{Advanced usage I: Implementation of new observables using existing operators}
\label{sec:observables}

In order to introduce new observables to the \SPheno output of \SARAH,
the user can add new definitions to the directories
\begin{center}
 {\tt [\$SARAH]/FlavorKit/[\$Type]/Processes/}
\end{center}
{\tt [\$Type]} is either {\tt LFV} for lepton flavor violating or {\tt
  QFV} for quark flavor violating observables. The definition of the
new observables consists of two files
\begin{enumerate}
 \item A steering file with the extension {\tt .m}
 \item A \Fortran body with the extension {\tt .f90}
\end{enumerate}
The steering file contains the following information:
\begin{itemize}
 \item {\tt NameProcess}: a string as name for the set of observables. 
 \item {\tt NameObservables}: names for the individual observables and
   numbers which are used to identify them later in the \SPheno
   output.  The value is a three dimensional list. The first part of
   each entry has to be a symbol, the second one an integer 
   and the third one a comment to be
   printed in the \SPheno output file ({\tt
     \{\{name1,number1,comment1\},\dots\}}).
 \item {\tt NeededOperators}: The operators which are needed to
   calculate the observables.  A list with all operators already
   implemented in \FlavorKit is given in Appendix~
   \ref{app:operators}. In case the user needs additional operators,
   this is explained in Sec. \ref{sec:operators}.
\item {\tt Body}: The name (as string) of the file which contains the
  \Fortran code to calculate the observables from the operators.
\end{itemize}
For instance, the corresponding file to calculate $\ell_\alpha \to \ell_\beta \gamma$ reads
\begin{lstlisting}
NameProcess = "LLpGamma";
NameObservables = {{muEgamma, 701, "BR(mu->e gamma)"}, 
                   {tauEgamma, 702, "BR(tau->e gamma)"}, 
                   {tauMuGamma, 703, "BR(tau->mu gamma)"}};
NeededOperators = {K2L, K2R};
Body = "LLpGamma.f90"; 
\end{lstlisting}
The observables will be saved in the variables {\tt muEgamma}, {\tt
  tauEgamma}, {\tt tauMuGamma} and will show up in the spectrum file
written by \SPheno in the block {\tt FlavorKitLFV} as numbers 701 to
703.

The file which contains the body to calculate the observables should
be standard \Fortran {\tt 90} code. For our example it reads
\begin{lstlisting}
Real(dp) :: width
Integer :: i1, gt1, gt2

Do i1=1,3 

If (i1.eq.1) Then         ! mu -> e gamma
 gt1 = 2
 gt2 = 1
Elseif (i1.eq.2) Then     !tau -> e gamma
 gt1 = 3
 gt2 = 1
Else                      ! tau -> mu gamma
 gt1 = 3
 gt2 = 2
End if

width=0.25_dp*mf_l(gt1)**5*(Abs(K2L(gt1,gt2))**2 &
           & +Abs(K2R(gt1,gt2))**2)*Alpha

If (i1.eq.1) Then
 muEgamma = width/(width+GammaMu)
Elseif (i1.eq.2) Then 
 tauEgamma = width/(width+GammaTau)
Else
 tauMuGamma = width/(width+GammaTau)
End if

End do
\end{lstlisting}

{\tt Real(dp)} is the \SPheno internal definition of double precision
variables.  Similarly one would have to use {\tt Complex(dp)} for
complex double precision variables when necessary.

Besides the operators, the SM parameters given in Table~\ref{tab:SM}
and the hadronic parameters given in Tables~\ref{tab:hadronic} and
\ref{tab:decay_constants} can be used in the calculations. For
instance, we used {\tt Alpha} for $\alpha(0)$ and {\tt mf\_l} which
contains the poles masses of the leptons as well as {\tt GammaMu} and
{\tt GammaTau} for the total widths of $\mu$ and $\tau$ leptons.
\begin{table}[hbt]
\centering
\begin{tabular}{|cc||cc||cc|} 
\hline
\multicolumn{6}{|c|}{Real Variables} \\
\hline
{\tt AlphaS\_MZ} & $\alpha_S(M_Z)$ & {\tt AlphaS\_160} & $\alpha_S(Q)$  &  & \\
{\tt sinW2\_MZ} & $\sin(\Theta_W)^2$ at $M_Z$ & {\tt sinW2\_160} & $\sin(\Theta_W)^2$ at $Q$  & {\tt sinW2}  & $\sin(\Theta_W)^2$  \\
{\tt Alpha\_MZ} & $\alpha(M_Z)$ & {\tt Alpha\_160} & $\alpha(Q)$  & {\tt Alpha} & $\alpha(0)$ \\
{\tt MW\_MZ} & $M_W(M_Z)$ & {\tt MW\_160} & $M_W(Q)$  & {\tt MW} & $M_W$ \\
{\tt GammaMu} & Width $\Gamma_\mu$ of $\mu$ & {\tt GammaTau} & Width $\Gamma_\tau$ of $\tau$ \\
\hline 
\hline
\multicolumn{6}{|c|}{Real Vectors of length 3} \\
\hline
{\tt mf\_d\_160} & $m_d(Q)$ &  {\tt mf\_d\_MZ} & $m_d(M_Z)$ & {\tt mf\_d} & $m_d$ \\
{\tt mf\_u\_160} & $m_u(Q)$ & {\tt mf\_u\_MZ} & $m_u(M_Z)$  & {\tt mf\_u} & $m_u$ \\
{\tt mf\_l\_160} & $m_l(Q)$ & {\tt mf\_l\_MZ} & $m_l(M_Z)$  & {\tt mf\_l} & $m_l$  \\
\hline 
\hline
\multicolumn{6}{|c|}{Complex Arrays of dimension $3 \times 3$} \\
\hline
{\tt CKM\_MZ} & CKM at $(M_Z)$ & {\tt CKM\_160} & CKM at $Q$  & {\tt CKM} & input \\
\hline
\end{tabular}
\caption{List of SM parameters available in \FlavorKit. All hadronic
  observables are calculated at $Q=160$~GeV.}
\label{tab:SM}
\end{table}

\begin{table}
\centering
\begin{tabular}{|c |c c | c c| c|}
\hline 
Particle       & Life time          & default [s] & Mass                 & default [GeV]  & PDG number \\
\hline
$\pi^0$        & {\tt tau\_pi0}           & $8.52\cdot10^{-17}$  & {\tt mass\_pi0}      &   0.13498      & 111 \\
$\pi^+$        & {\tt tau\_pip}           & $2.60\cdot10^{-8}$   & {\tt mass\_pip}      &   0.13957      & 211 \\
$\rho(770)^0$  & {\tt tau\_rho0}          & $4.41\cdot10^{-24}$  & {\tt mass\_rho0}     &   0.77549      & 113 \\
$D^0$          & {\tt tau\_D0}            & $4.10\cdot10^{-13}$  & {\tt mass\_D0}       &   1.86486      & 421 \\
$D^+$          & {\tt tau\_Dp}            & $1.04\cdot10^{-12}$  & {\tt mass\_Dp}       &   1.86926      & 411 \\
$D_s^+$        & {\tt tau\_DSp}           & $5.00\cdot10^{-13}$  & {\tt mass\_DSp}      &   1.96849      & 431 \\
$D_s^{*+}$     & {\tt tau\_DSsp}          &        -             & {\tt mass\_DSsp}     &   2.1123       & 433 \\
$\eta$         & {\tt tau\_eta}           & $5.06\cdot10^{-19}$  & {\tt mass\_eta}      &   0.54785      & 221 \\
$\eta^\prime(958)$  & {\tt tau\_etap}     & $3.31\cdot10^{-21}$  & {\tt mass\_etap}     &   0.95778      & 331 \\
$\omega(782)$  & {\tt tau\_omega}         & $7.75\cdot10^{-23}$  & {\tt mass\_omega}    &   0.78265      & 223 \\
$\phi(1020)$   & {\tt tau\_phi}           & $1.54\cdot10^{-22}$  & {\tt mass\_phi}      &   1.01946      & 333 \\
$K_L^0$        & {\tt tau\_KL0}           & $5.12\cdot10^{-8}$   & {\tt mass\_KL0}      &     -          & 130 \\
$K_S^0$        & {\tt tau\_KS0}           & $0.90\cdot10^{-10}$  & {\tt mass\_KS0}      &     -          & 310 \\
$K^0$          & {\tt tau\_K0}            &          -           & {\tt mass\_K0}       &   0.49761      & 311 \\
$K^+$          & {\tt tau\_Kp}            & $1.24\cdot 10^{-8}$  & {\tt mass\_Kp}       &   0.49368      & 321 \\
$B^0_d$        & {\tt tau\_B0d}           & $1.52\cdot 10^{-12}$ & {\tt mass\_B0d}      &   5.27958      & 511 \\
$B^0_s$        & {\tt tau\_B0s}           & $1.50\cdot 10^{-12}$ & {\tt mass\_B0s}      &   5.36677      & 531 \\
$B^+$          & {\tt tau\_Bp}            & $1.64\cdot 10^{-12}$ & {\tt mass\_Bp}       &   5.27925      & 521 \\
$B^{*0}$       & {\tt tau\_B0c}           & $1.43\cdot 10^{-23}$ & {\tt mass\_B0c}      &   5.3252       & 513 \\
$B^{*+}$       & {\tt tau\_Bpc}           & $1.43\cdot 10^{-23}$ & {\tt mass\_Bpc}      &   5.3252       & 523 \\ 
$B^+_c$        & {\tt tau\_Bcp}           & $4.54\cdot 10^{-13}$ & {\tt mass\_Bcp}      &   6.277        & 541 \\
$K^{*0}(892)$  & {\tt tau\_K0c}           & $1.42\cdot 10^{-23}$ & {\tt mass\_K0c}      &   0.8959       & 313 \\
$K^{*+}(892)$  & {\tt tau\_Kpc}           & $1.30\cdot 10^{-23}$ & {\tt mass\_Kpc}      &   0.8917       & 323 \\
$\eta_c(1S)$   & {\tt tau\_etac}          & $2.22\cdot 10^{-23}$ & {\tt mass\_etac}     &   2.9810       & 441 \\
$J/\Psi(1S)$   & {\tt tau\_JPsi}          & $7.08\cdot 10^{-24}$ & {\tt mass\_JPSi}     &   3096.92      & 443 \\
$\Upsilon(1S)$ & {\tt tau\_Ups}           & $1.21\cdot 10^{-23}$ & {\tt mass\_Ups}      &   9.4603       & 553 \\
\hline
\end{tabular}
\caption{Hadronic parameters used in \FlavorKit. These can be changed
  via {\tt FMASS} and {\tt and FLIFE} in the Les Houches input file.}
\label{tab:hadronic}
\end{table}

\begin{table}[hbt]
\centering
\begin{tabular}{|c|cc|c|}
\hline 
Decay constant & Variable        & default [MeV]  & FLHA \\
\hline 
$f_K$          & {\tt f\_k\_CONST}          &  176     & {\tt FCONST[321,1]}  \\
$f_{K^+}$      & {\tt f\_Kp\_CONST}         &  156     & {\tt FCONST[323,1]}  \\
$f_\pi$        & {\tt f\_pi\_CONST}         &  118     & {\tt FCONST[111,1]} \\
$f_{B^0_d}$    & {\tt f\_B0d\_CONST}        &  194     & {\tt FCONST[511,1]} \\
$f_{B^0_s}$    & {\tt f\_B0s\_CONST}        &  234     & {\tt FCONST[531,1]} \\
$f_{B^+}$      & {\tt f\_Bp\_CONST}         &  234     & {\tt FCONST[521,1]} \\               
$f_{\eta'}$    & {\tt f\_etap\_CONST}       &  172     & {\tt FCONST[231,1]} \\
$f_{\rho}$     & {\tt f\_rho\_CONST}        &  220     & {\tt FCONST[213,1]} \\
$f_{D^+}$      & {\tt f\_Dp\_CONST}         &  256     & {\tt FCONST[411,1]} \\
$f_{D_s}$      & {\tt f\_Ds\_CONST}         &  248     & {\tt FCONST[431,1]} \\   
\hline 
\end{tabular}
\caption{Decay constants available in the \SPheno output of
  \SARAH. The values can be changed according to the FLHA conventions
  using the block {\tt FCONST} in the Les Houches input file.}
\label{tab:decay_constants}
\end{table}
By extending or changing the file {\tt hadronic\_parameters.m} in the
{\tt FlavorKit} directory, it is possible to add new variables for the
mass or life time of mesons. These variables are available globally in
the resulting \SPheno code. The numerical values for the hadronic 
parameters can be changed in the Les Houches input file by using the 
blocks {\tt FCONST} and {\tt FMASS} defined in the Flavor Les Houches
Accord (FLHA) \cite{Mahmoudi:2010iz}.

It may happen that the calculation of a specific observable has to be
adjusted for each model. This is for instance the case when (1) the
calculation requires the knowledge of the number of generations of
fields, (2) the mass or decay width of a particle, calculated by
\SPheno, is needed as input, or (3) a rotation matrix of a specific
field enters the analytical expressions for the observable. For these
situations, a special syntax has been created. It is possible to
start a line with {\tt @} in the \Fortran file. This line will then be
parsed by \SARAH, and \Mathematica commands, as well as \SARAH
specific commands, can be used. We made use of this functionality in
the implementation of $h \to \ell_\alpha \ell_\beta$. The lines in
{\tt hLLp.f90} read
\begin{lstlisting}
! Check for SM like Higgs
@ If[getGen[HiggsBoson]>1, "hLoc = MaxLoc(Abs(" <> ToString[HiggsMixingMatrix]<>"(2,:)),1)", "hLoc = 1"]

! Get Higgs mass
@ "mh ="<>ToString[SPhenoMass[HiggsBoson]] <> If[getGen[HiggsBoson]>1,"(hLoc)", ""]

! Get Higgs width
@ "gamh ="<>ToString[SPhenoWidth[HiggsBoson]] <> If[getGen[HiggsBoson]>1,"(hLoc)", ""] 
\end{lstlisting}
In this implementation we define an integer {\tt hLoc} that gives the
generation index of the SM-like Higgs, to be found among all CP even
scalars. In the first line it is checked if more than one scalar Higgs
is present.  If this is the case, the {\tt hLoc} is set to the
component which has the largest amount of the up-type Higgs, if not,
it is just put to {\tt 1}. Of course, this assumes that the
electroweak basis in the Higgs sector is always defined as
$(\phi_d,\phi_u,\dots)$ as is the case for all models delivered with
\SARAH.  In the second and third lines, the variables {\tt mh} and {\tt
  gamh} are set to the mass and total width of the SM-like Higgs,
respectively.  For this purpose, the \SARAH commands {\tt
  SPhenoMass[x]} and {\tt SPhenoWidth[x]} are used. They return the
name of the variable for the mass and width in \SPheno and it is
checked if these variables are arrays or not \footnote{ 
The user can define in the {\tt parameters.m} and {\tt particles.m} file 
for a given model in {\tt SARAH} the particles which should be taken to be the CP-even 
or CP-odd Higgs and the parameter that corresponds to their rotation matrices. This is done by using the
{\tt Description} statements {\tt Higgs} or {\tt Pseudo-Scalar Higgs} as
well as {\tt Scalar-Mixing-Matrix} or {\tt Pseudo-Scalar-Mixing-Matrix}.
If the particle or parameter needed to calculate an observable is not present 
or has not been defined, the observable is skipped in the {\tt SPheno} output.}. 
For the MSSM, the above
lines lead to the following code in the \SPheno output:
\begin{lstlisting}
! Check for SM like Higgs
hLoc = MaxLoc(Abs(ZH(2,:)),1)

! Get Higgs mass
mh =Mhh(hLoc)

! Get Higgs width
gamh =gThh(hLoc)
\end{lstlisting}
We give in Table \ref{tab:SARAHcommands} the most important \SARAH
commands which might be useful in this context.
\begin{table}
\begin{tabular}{|l|l|}
\hline 
{\tt getGen[x]} & returns the number of generations of a particle {\tt x} \\
{\tt getDim[x]} & returns the dimension of a variable {\tt x} \\
{\tt SPhenoMass[x]} & returns the name used for the mass of a particle {\tt x} in the \SPheno output \\
{\tt SPhenoMassSq[x]} & returns the name used for the mass squared of a particle {\tt x} in the \SPheno output \\
{\tt SPhenoWidth[x]} & returns the name used for the width of a particle {\tt x} in the \SPheno output \\
\hline 
{\tt HiggsMixingMatrix} & name of the mixing matrix for the CP even Higgs states in a given model \\
{\tt PseudoScalarMixingMatrix} & name of the mixing matrix for the CP odd Higgs states in a given model \\
\hline 
\end{tabular}
\caption{\SARAH commands which can be used in the input file for the
  calculation of an observable.}
\label{tab:SARAHcommands} 
\end{table}

Many more examples are given in Appendix~\ref{sec:LFV}, where we have
added all input files for the calculations of flavor observables
delivered with \SARAH.

\section{Advanced usage II: Implementation of new operators}
\label{sec:operators}

The user can also implement new operators and obtain analytical
expressions for their Wilson coefficients. In this case, he will need
to use \NewPackage which, with the help of {\tt FeynArts} and
{\tt FormCalc}, provides generic expressions for the coefficients,
later to be adapted to specific models with \SARAH.

\subsection{Introduction}
New operators can be implemented by extending the content of the
folder
\begin{center}
 {\tt [\$SARAH]/FlavorKit/[\$Type]/Operators/}
\end{center}
In the current version of \FlavorKit, 3- and 4-point operators are
supported.  Each operator is defined by a {\tt .m}-file.  These files
contain information about the external particles, the kind of
considered diagrams (tree-level, self-energies, penguins, boxes) as
well as generic expressions for the coefficients. These expressions,
derived from the generic Feynman diagrams contributing to the
coefficients, are written in the form of a \Mathematica code, which
can be used to generate \Fortran code.

\begin{figure}[hbt]
\centering
\begin{tabular}{ccc}
& \includegraphics[width=0.2\linewidth]{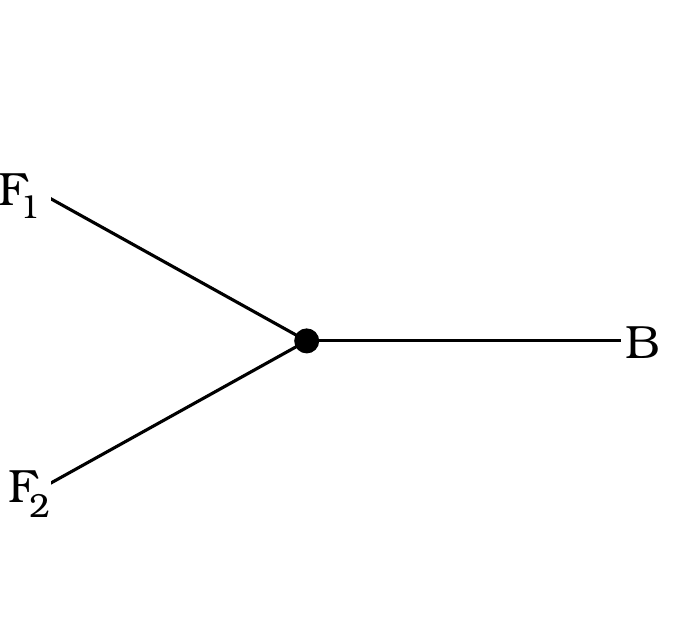}  & \\
\includegraphics[width=0.2\linewidth]{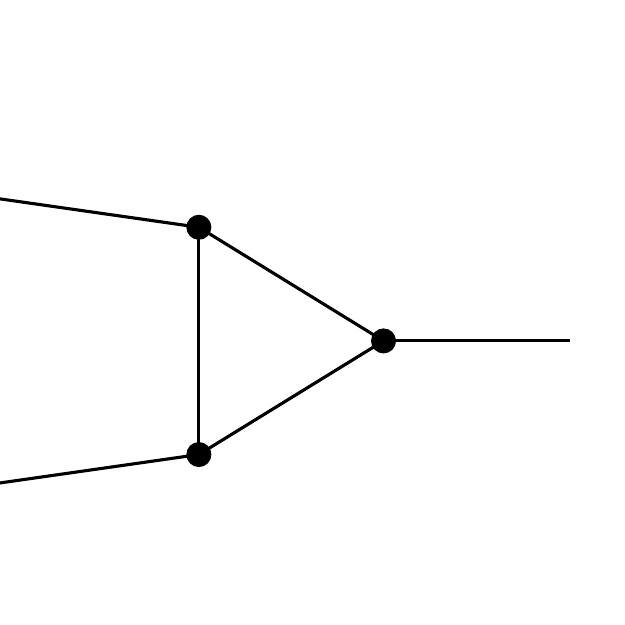} &
\includegraphics[width=0.2\linewidth]{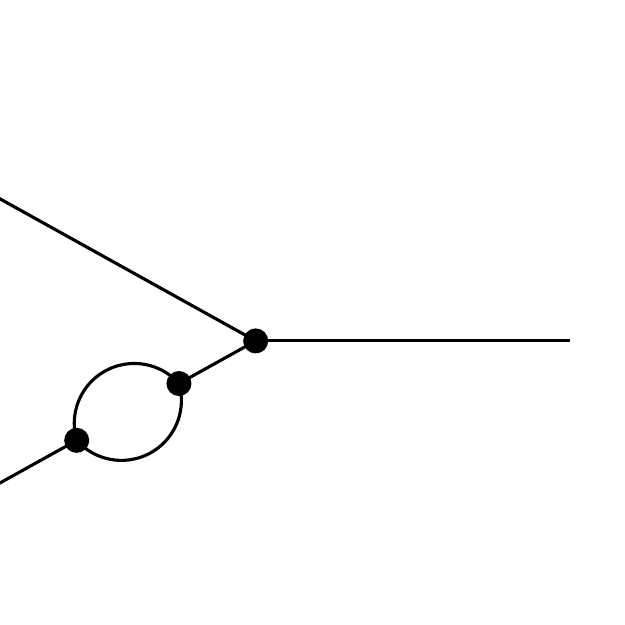} &
\includegraphics[width=0.2\linewidth]{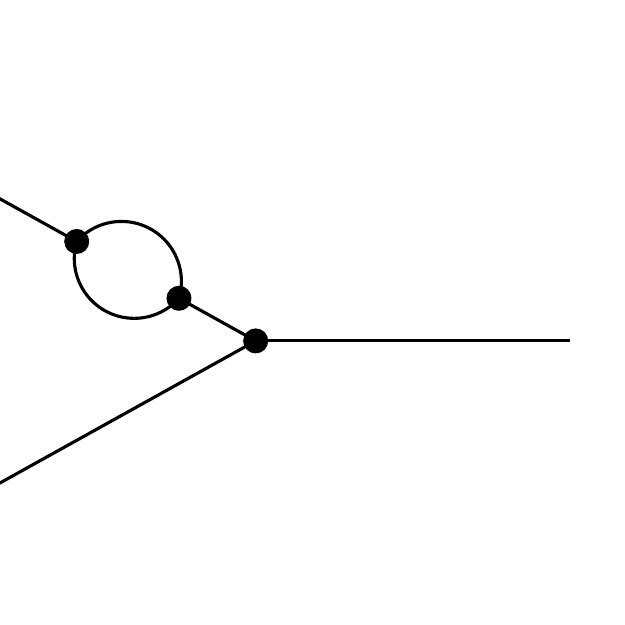} \\
\end{tabular}
\caption{All topologies considered by \NewPackage to calculate the
  Wilson coefficients of 2-fermion-1-boson operators. All possible generic
  combinations of the internal fields are taken into account.}
\label{fig:topologies2F}
\end{figure}

\begin{figure}[hbt]
\centering
\begin{tabular}{ccc}
\includegraphics[width=0.2\linewidth]{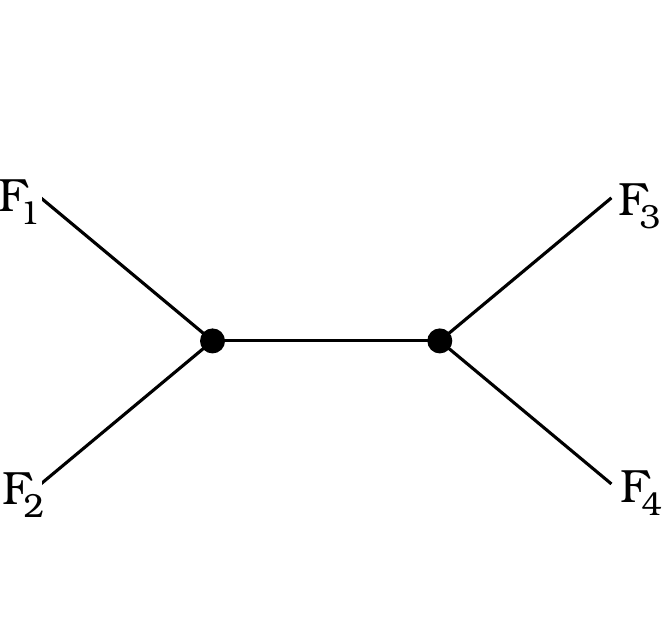} &
\includegraphics[width=0.2\linewidth]{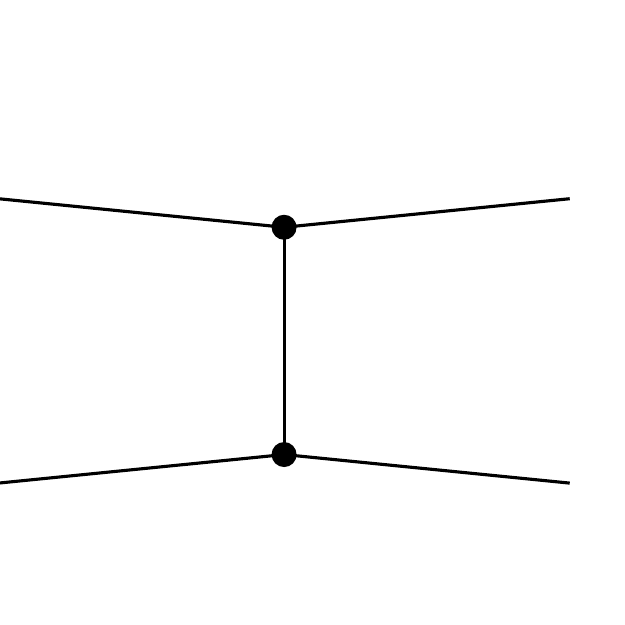} &
\includegraphics[width=0.2\linewidth]{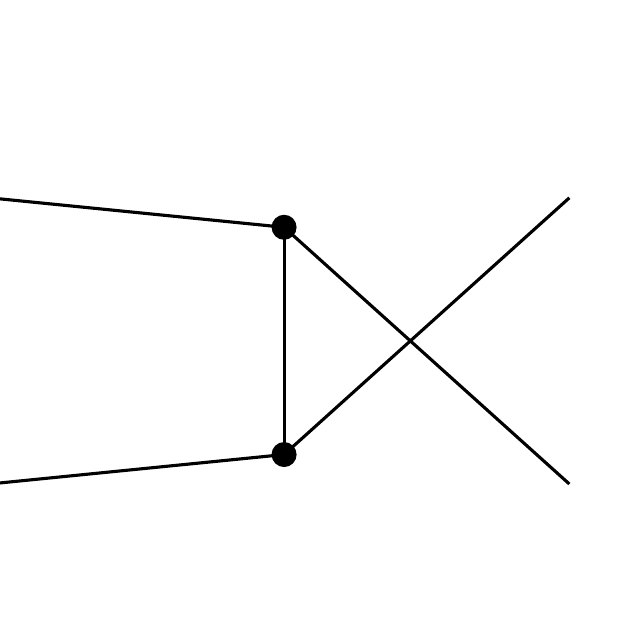} 
\end{tabular}
\caption{All tree topologies considered by \NewPackage to calculate
  the Wilson coefficients of 4-fermion operators. All possible generic
  combinations of the internal fields are taken into account.}
\label{fig:topologies4Ftree}
\end{figure}

\begin{figure}[hbt]
\centering
\begin{tabular}{ccc}
\includegraphics[width=0.2\linewidth]{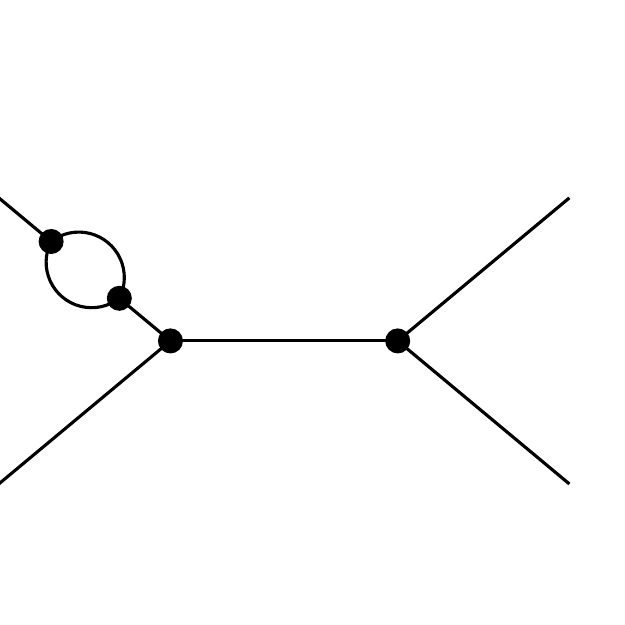} &
\includegraphics[width=0.2\linewidth]{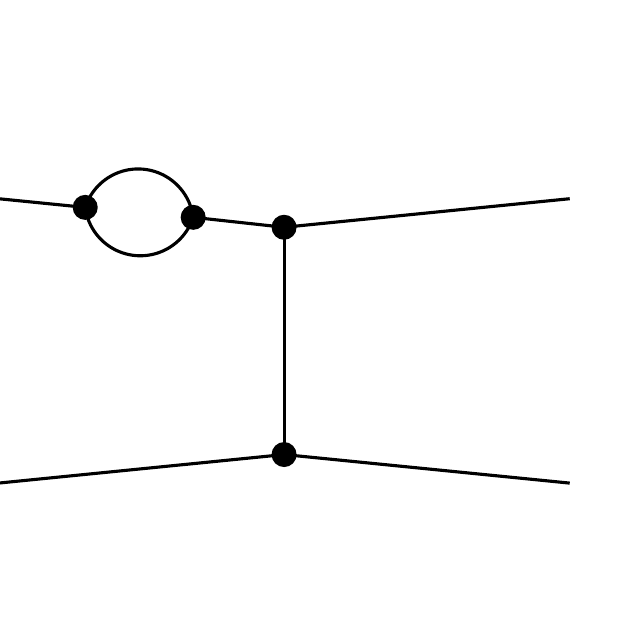} &
\includegraphics[width=0.2\linewidth]{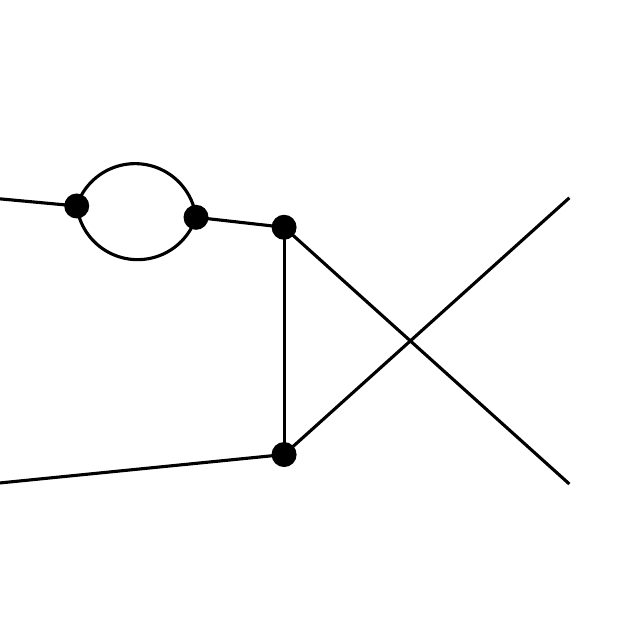}  \\
\includegraphics[width=0.2\linewidth]{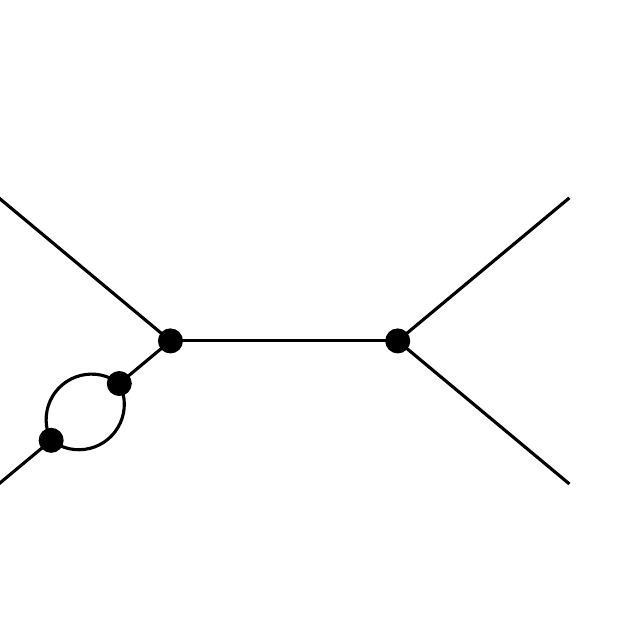} &
\includegraphics[width=0.2\linewidth]{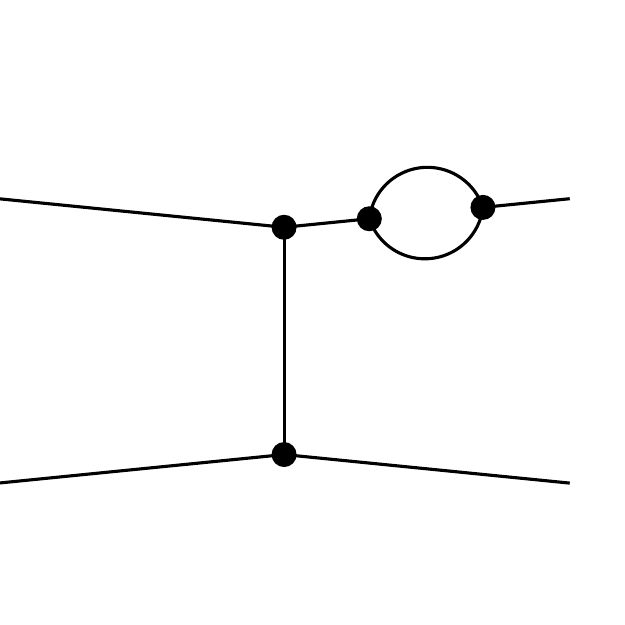} &
\includegraphics[width=0.2\linewidth]{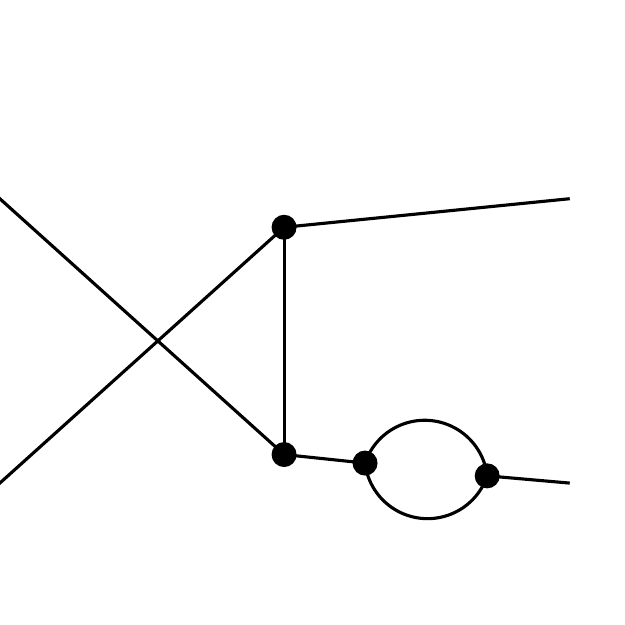}  \\
\includegraphics[width=0.2\linewidth]{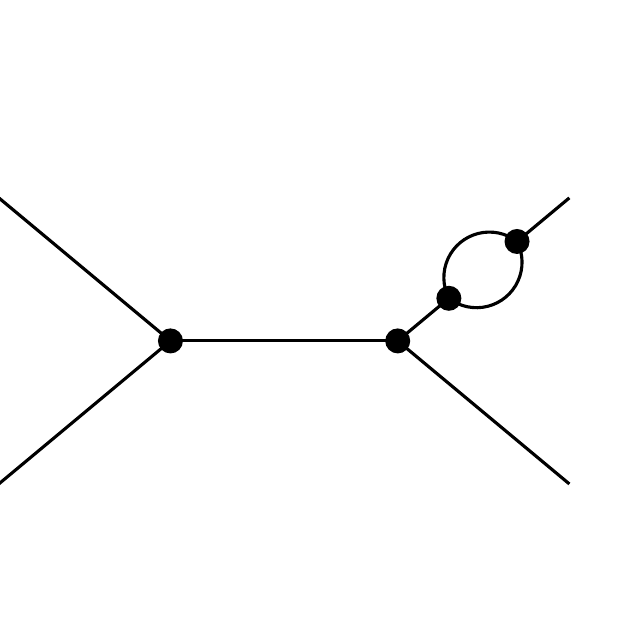} &
\includegraphics[width=0.2\linewidth]{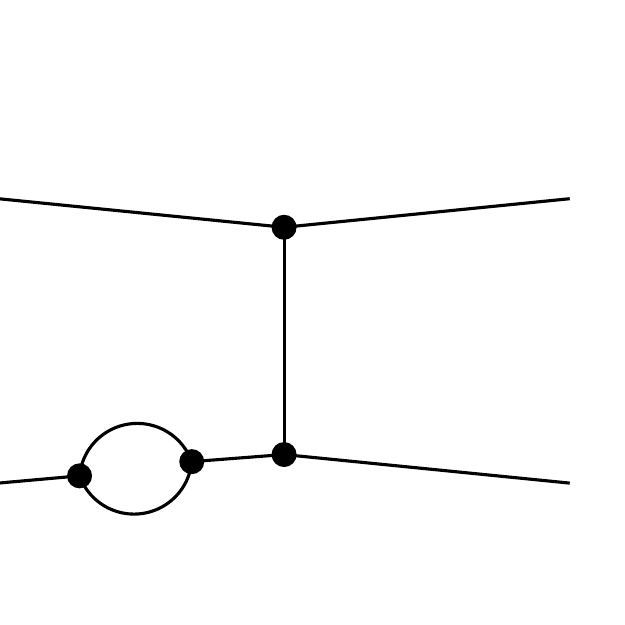} &
\includegraphics[width=0.2\linewidth]{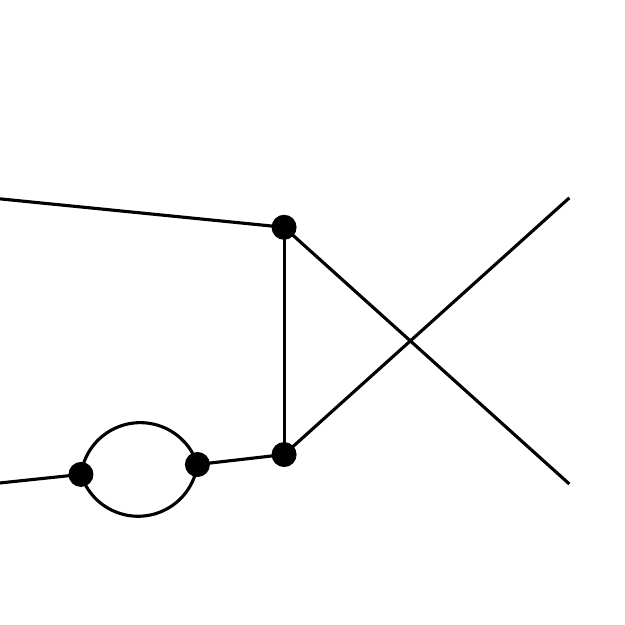}  \\
\includegraphics[width=0.2\linewidth]{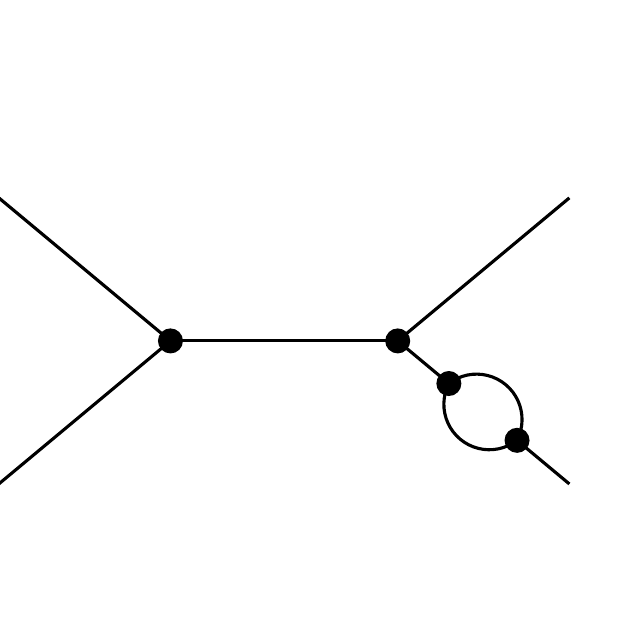} &
\includegraphics[width=0.2\linewidth]{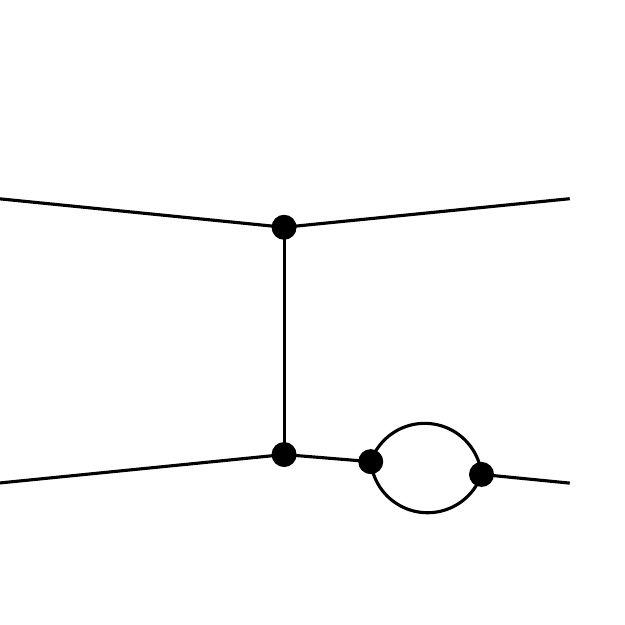} &
\includegraphics[width=0.2\linewidth]{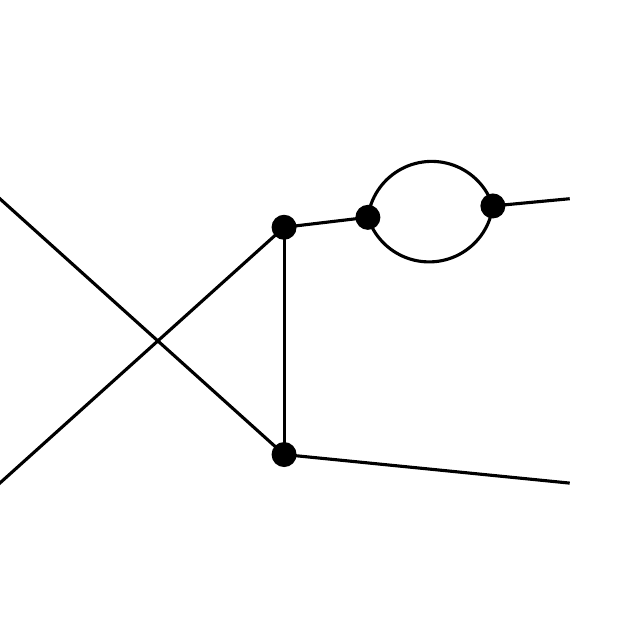}  
\end{tabular}
\caption{All self-energy topologies considered by \NewPackage to
  calculate the Wilson coefficients of 4-fermion operators. All possible
  generic combinations of the internal fields are taken into account.
}
\label{fig:topologies4Fwave}
\end{figure}

\begin{figure}[hbt]
\centering
\begin{tabular}{ccc}
\includegraphics[width=0.2\linewidth]{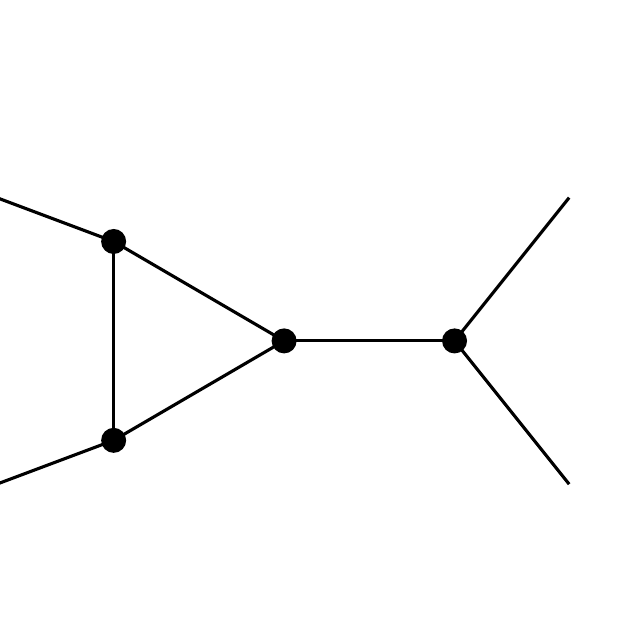} &
\includegraphics[width=0.2\linewidth]{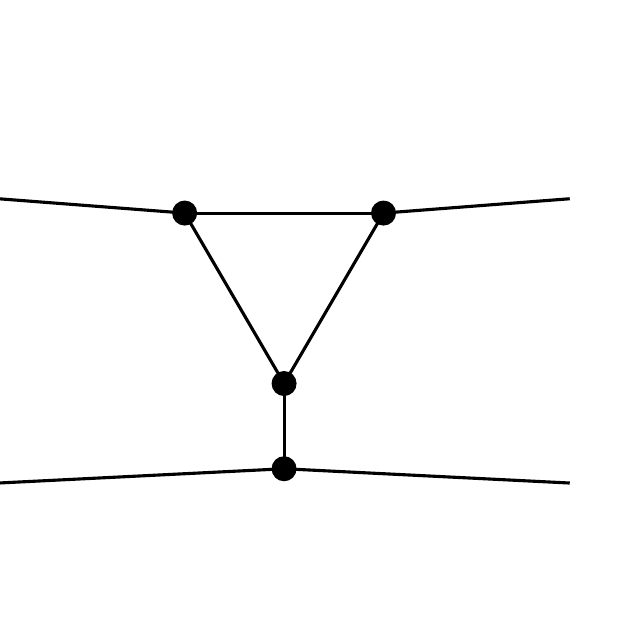} &
\includegraphics[width=0.2\linewidth]{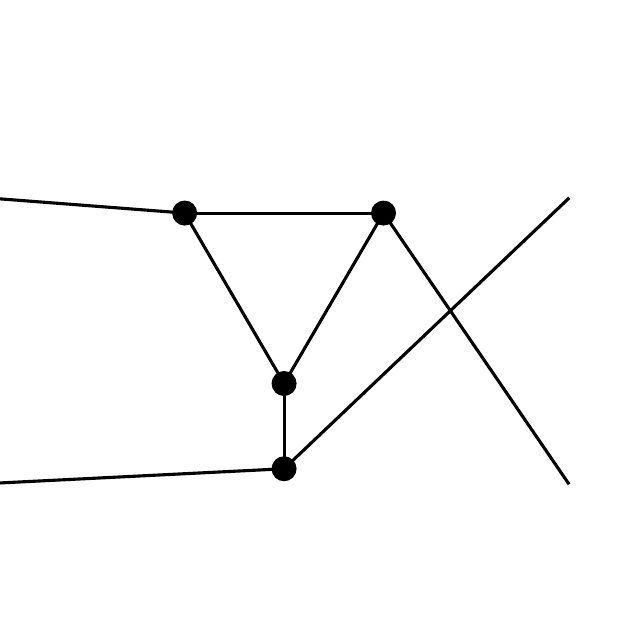} \\
\includegraphics[width=0.2\linewidth]{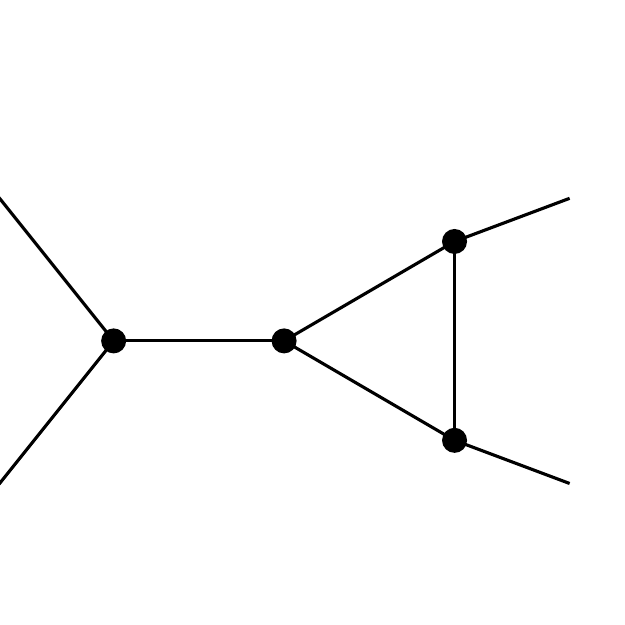} &
\includegraphics[width=0.2\linewidth]{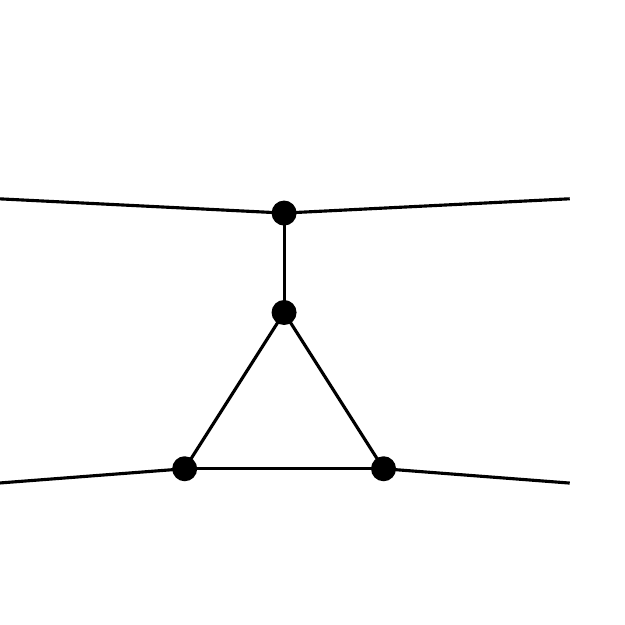} &
\includegraphics[width=0.2\linewidth]{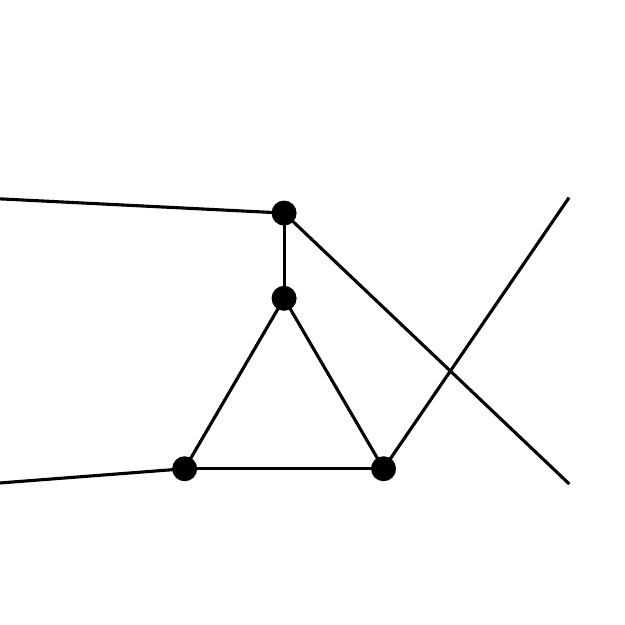} 
\end{tabular}
\caption{All penguin topologies considered by \NewPackage to calculate
  the Wilson coefficients of 4-fermion operators. All possible generic
  combinations of the internal fields are taken into account.  }
\label{fig:topologies4Fpeng}
\end{figure}

\begin{figure}[hbt]
\centering
\begin{tabular}{ccc}
\includegraphics[width=0.2\linewidth]{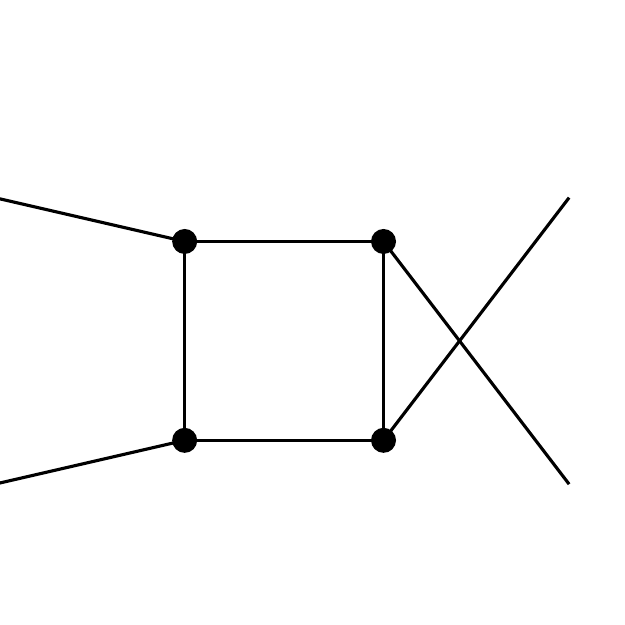} &
\includegraphics[width=0.2\linewidth]{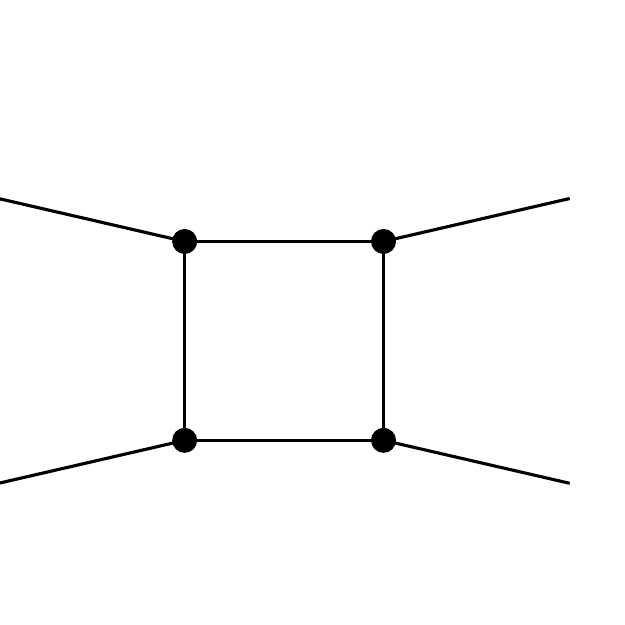} &
\includegraphics[width=0.2\linewidth]{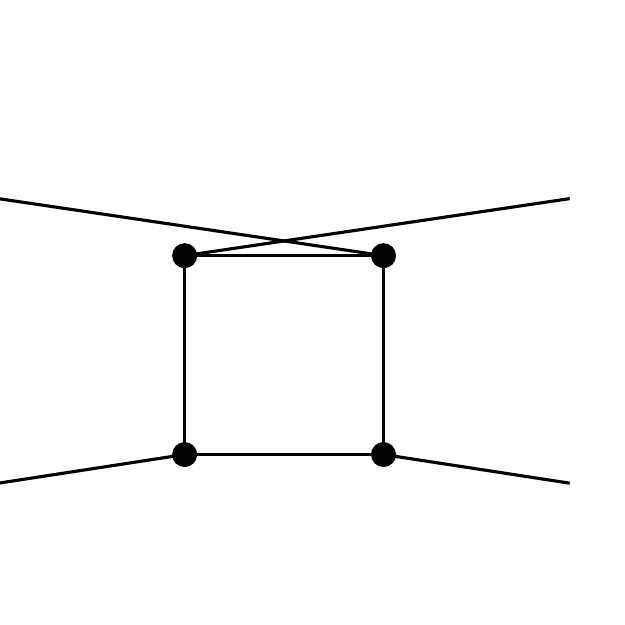}
\end{tabular}
\caption{All box topologies considered by \NewPackage to calculate the
  Wilson coefficients of 4-fermion operators. All possible generic
  combinations of the internal fields are taken into account.}
\label{fig:topologies4Fbox}
\end{figure}

For the automatization of the underlying calculations we have created
an additional \Mathematica package called \NewPackage, which can be
used to create the files for all 4-fermion as well as
2-fermion-1-boson operators. This package creates not only the
infrastructure to include the operators in the \SPheno output of
\SARAH but makes also use of {\tt FeynArts} and {\tt FormCalc} to
calculate the amplitudes and to extract the coefficient of the
demanded operators. It takes into account all topologies depicted in
Figs.~\ref{fig:topologies2F} to \ref{fig:topologies4Fbox}.

\subsection{Input for \NewPackage}
In order to derive the results for the Wilson coefficients,
\NewPackage needs an input file with the following information:
\begin{itemize}
 \item {\tt ConsideredProcess}: A string which defines the generic type for the process
   \begin{itemize}
    \item {\tt \textquotedblleft4Fermion\textquotedblright}
    \item {\tt \textquotedblleft2Fermion1Scalar\textquotedblright}
    \item {\tt \textquotedblleft2Fermion1Vector\textquotedblright}
   \end{itemize}
 \item {\tt NameProcess}: A string to uniquely define the process
 \item {\tt ExternalFields}: The external fields. Possible names are
   {\tt ChargedLepton}, {\tt Neutrino}, {\tt DownQuark}, {\tt
     UpQuark}, {\tt ScalarHiggs}, {\tt PseudoScalar}, {\tt Zboson},
   {\tt Wboson} \footnote{ 
   The {\tt particles.m} file is used to define for each model which particle corresponds to SM states using the {\tt Description}
statement together with {\tt Leptons}, {\tt Neutrinos}, {\tt Down-Quarks}, {\tt Up-Quarks}, {\tt Higgs}, {\tt Pseudo-Scalar Higgs}, {\tt Z-Boson}, {\tt W-Boson}.
If there is a mixture between the SM particles and other states (like in $R$-parity violating SUSY or in models with
additional vector quarks/leptons) the combined state has to be labeled according to the description for the SM state. Notice that in the SM {\tt Pseudo-Scalar Higgs} is just the neutral Goldstone boson. If an external state is not present in a given model or has not been defined as
such in the {\tt particles.m} file the corresponding Wilson coefficients are not calculated by \SPheno.
   }
 \item {\tt FermionOrderExternal}: the fermion order to apply the
   Fierz transformation (see the \FormCalc manual for more details)
 \item {\tt NeglectMasses}: which external masses can be neglected (a
   list of integers counting the external fields)
 \item {\tt ColorFlow}: defines the color flow in the case of four
   quark operators. To contract the colors of external fields, {\tt
     ColorDelta} is used, i.e {\tt ColorFlow =
     ColorDelta[1,2]*ColorDelta[3,4]} assigns $(\bar{q}^\alpha \Gamma
   q_\alpha)(\bar{q}^\beta \Gamma' q_\beta)$.
 \item {\tt AllOperators}: a list with the definition of the
   operators. This is a two dimensional list, where the first entry
   defines the name of the operator and the second one the Lorentz
   structure. The operators are expressed in the chiral basis and the
   syntax for Dirac chains in \FormCalc is used:
 \begin{itemize}
  \item {\tt 6} for $P_L = \frac{1}{2}(1-\gamma_5)$, {\tt 7} for $P_R = \frac{1}{2}(1-\gamma_5)$
  \item {\tt Lor[1]}, {\tt Lor[2]} for $\gamma_\mu$, $\gamma_\nu$
  \item {\tt ec[3]} for the helicity of an external gauge boson. 
  \item {\tt k[N]} for the momentum of the external particle {\tt N} ({\tt N} is an integer). 
  \item {\tt Pair[A,B]} is used to contract Lorentz indices. For instance, {\tt Pair[k[1],ec[3]]} stands for $k^1_\mu \epsilon^{\mu,*}$
  \item A Dirac chain starting with a negative first entry is taken to be anti-symmetrized. 
 \end{itemize}
 
 See the \FormCalc manual for more details.\\ 
To make the definitions more readable, not the full {\tt DiracChain}
object of {\tt FeynArts}/{\tt FormCalc} has to be defined: \NewPackage
puts everything with the head {\tt Op} into a Dirac chain using the
defined fermion order. For 4-fermion operators the combination of both
operators is written as dot product. For instance {\tt Op[6].Op[6]} is
internally translated into
\begin{verbatim}
 DiracChain[Spinor[k[1],MassEx1,-1],6,Spinor[k[2],MassEx2,1]]*
             DiracChain[Spinor[k[3],MassEx3,-1],6,Spinor[k[4],MassEx4,1]]
\end{verbatim}
while {\tt Op[6] Pair[ec[3],k[1]} becomes 
\begin{verbatim}
 DiracChain[Spinor[k[1],MassEx1,-1],6,Spinor[k[2],MassEx2,1]] Pair[ec[3],k[1]]
\end{verbatim}
 \item {\tt CombinationGenerations}: the combination of external generations for which the operators are calculated by \SPheno
 \item {\tt Filters}: a list of filters to drop specific
   diagrams. Possible entries are {\tt NoBoxes}, {\tt NoPenguins},
   {\tt NoTree}, {\tt NoCrossedDiagrams}.
 \begin{itemize}
  \item {\tt Filters = \{NoBoxes,  NoPenguins\}} can be used for processes which are already triggered at tree-level
  \item {\tt Filters = \{NoPenguins\}} might be useful for processes which at the 1-loop level are only induced by  box diagrams
  \item {\tt Filters = \{NoCrossedDiagrams\}} is used to drop diagrams which only differ by a permutation of the external fields.
\end{itemize}
\end{itemize}
For instance, the \NewPackage input to calculate the coefficient of
the $(\bar{\ell}\Gamma \ell)(\bar{d} \Gamma' d)$ operator reads
\begin{lstlisting}
NameProcess="2L2d";
ConsideredProcess = "4Fermion";
ExternalFields={{ChargedLepton,bar[ChargedLepton],
                   DownQuark,bar[DownQuark]}};
                   
FermionOrderExternal={2,1,4,3};
NeglectMasses={1,2,3,4}; 
                   

AllOperators={
   (* scalar operators*)
   {OllddSLL,Op[7].Op[7]},
   {OllddSRR,Op[6].Op[6]},
   {OllddSRL,Op[6].Op[7]},
   {OllddSLR,Op[7].Op[6]},

   (* vector operators*)
   {OllddVRR,Op[7,Lor[1]].Op[7,Lor[1]]},
   {OllddVLL,Op[6,Lor[1]].Op[6,Lor[1]]},
   {OllddVRL,Op[7,Lor[1]].Op[6,Lor[1]]},
   {OllddVLR,Op[6,Lor[1]].Op[7,Lor[1]]},

   (* tensor operators*)
   {OllddTLL,Op[-7,Lor[1],Lor[2]].Op[-7,Lor[1],Lor[2]]},
   {OllddTLR,Op[-7,Lor[1],Lor[2]].Op[-6,Lor[1],Lor[2]]},
   {OllddTRL,Op[-6,Lor[1],Lor[2]].Op[-7,Lor[1],Lor[2]]},
   {OllddTRR,Op[-6,Lor[1],Lor[2]].Op[-6,Lor[1],Lor[2]]}
};

CombinationGenerations = {{2,1,1,1}, {3,1,1,1}, {3,2,1,1},
                          {2,1,2,2}, {3,1,2,2}, {3,2,2,2}};

Filters = {};                          
\end{lstlisting}
Here, we neglect all external masses in the operators ({\tt
  NeglectMasses=\{1,2,3,4\}}), and the different coefficients of the
scalar operators $(\bar{\ell}P_X \ell)(\bar{d} P_Y d)$ are called {\tt
  OllddSXY}, the ones for the vector operators $(\bar{\ell} P_X
\gamma_\mu \ell)(\bar{d} P_Y \gamma^\mu d)$ are called {\tt OllddVYX}
and the ones for the tensor operators $(\bar{\ell} P_X \sigma_{\mu
  \nu} \ell)(\bar{d} \sigma^{\mu \nu} P_Y d)$ {\tt OllddTYX}, with
X,Y=L,R.  Notice that \FormCalc returns the results in form of $P_X
\gamma_\mu$ while in the literature the order $\gamma_\mu P_X$ is
often used.  Finally, \SPheno will not calculate all possible
combinations of external states, but only some specific cases:
$\mu e d d$, $\tau e d d$, $\tau \mu d d$, $\mu e s s$, $\tau e s s$,
$\tau \mu s s$ \footnote{Here we used $d$ for the first generation of
  down-type quarks while in the rest of this manual it is used to
  summarize all three families.}.

The input file to calculate the coefficients of the $\ell-\ell-Z$
operators $(\bar{\ell} \gamma_\mu P_{L,R} \ell) Z^\mu$ and
$(\bar{\ell} p_\mu P_{L,R} \gamma_\mu \ell) Z^\mu$ is
\begin{lstlisting}
NameProcess="Z2l";

ConsideredProcess = "2Fermion1Vector";
FermionOrderExternal={1,2};
NeglectMasses={1,2};  


ExternalFields= {ChargedLepton,bar[ChargedLepton],Zboson};
CombinationGenerations = {{1,2},{1,3},{2,3}};


AllOperators={
   {OZ2lSL,Op[7]}, {OZ2lSR,Op[6]},
   {OZ2lVL,Op[7,ec[3]]}, {OZ2lVR,Op[6,ec[3]]}
};

OutputFile = "Z2l.m";

Filters = {}; 
\end{lstlisting}
Note that {\tt ExternalFields} must contain first the involved
fermions and the boson at the end.  Furthermore, in the case of
processes involving scalars one can define
\begin{lstlisting}
ExternalFields= {ChargedLepton,bar[ChargedLepton],ScalarHiggs};
CombinationGenerations = {{1,2,ALL}, {1,3,ALL}, {2,3,ALL}};
\end{lstlisting}
In this case the operators for all Higgs states present in the
considered model will be computed.

\subsection{Operators with massless gauge bosons}
We have to add a few more remarks concerning 2-fermion-1-boson
operators with massless gauge bosons since those are treated in a
special way.  It is common for these operators to include terms in the
amplitude which are proportional to the external masses. Therefore, if
one proceeds in the usual way and neglects the external momenta, some
inconsistencies would be obtained. For this reason, a special
treatment is in order. In \NewPackage, when one uses
\begin{lstlisting}
ConsideredProcess = "2Fermion1Vector";
FermionOrderExternal={1,2};
NeglectMasses={3};   
\end{lstlisting}
the dependence on the two fermion masses is neglected in the resulting
Passarino-Veltman integrals but terms proportional to $m_{f_1}$ and
$m_{f_2}$ are kept. This solves the aforementioned potential
inconsistencies.

Furthermore, for the dipole operators, defined by
\begin{lstlisting}
   {DipoleL,Op[6] Pair[ec[3],k[1]]}, 
   {DipoleR,Op[7] Pair[ec[3],k[1]]}, 
\end{lstlisting}
we are using the results obtained by \FeynArts and \FormCalc and have
implemented all special cases for the involved loop integrals ($C_0,
C_{00}, C_1, C_2, C_{11}, C_{12}, C_{22}$) with identical or vanishing
internal masses in \SPheno. This guarantees the numerical stability of
the results~\footnote{We note that the coefficients for the operators
  defined above ($\bar{f} \gamma_\mu f \, V^\mu$) are by a factor of 2
  (4) larger than the coefficients of the standard definition for the
  dipole operators $\bar{f} \sigma_{\mu\nu} P_L f q^\nu V^\mu$
  ($\bar{f} \sigma_{\mu\nu}P_L f F^{\mu\nu}$).}.

The monopole operators of the form $q^2 (\bar{f} \gamma_\mu f) V^\mu$
are only non-zero for off-shell external gauge bosons, while
\NewPackage always treats all fields as on-shell. Because of this, and
to stabilize the numerical evaluation later on, these operators are
treated differently to all other operators: the coefficients are not
calculated by \FeynArts and \FormCalc but instead we have included the
generic expressions in \NewPackage using a special set of loop
functions in \SPheno. In these loop functions the resulting
Passarino-Veltman integrals are already combined, leading to
well-known expressions in the literature, see
\cite{Hisano:1995cp,Ilakovac:1994kj}. They have been cross-checked
with the package {\tt Peng4BSM@LO} \cite{Bednyakov:2013tca}. To get
the coefficients for the monopole operators, these have to be given
always in the form
\begin{lstlisting}
   {MonopoleL,Op[6,ec[3]] Pair[k[3],k[3]]},
   {MonopoleR,Op[7,ec[3]] Pair[k[3],k[3]]} 
\end{lstlisting}
in the input of \NewPackage. 

\subsection{Combination and normalization of operators}
The user can define new operators as combination of existing
operators. For this purpose wrapper files containing the definition of
the operators can be included in the FlavorKit directories. These
files have to begin with {\tt ProcessWrapper = True;}. This function
is also used by \NewPackage in the case of 4-fermion operators: for
these operators the contributions stemming from tree-level, box- and
penguin- diagrams are saved separately and summed up at the
end. Thus, the wrapper code for the 4-lepton operators written by
\NewPackage reads
\begin{lstlisting}
ProcessWrapper = True; 
NameProcess = "4L" 
ExternalFields = {ChargedLepton, bar[ChargedLepton], ChargedLepton, bar[ChargedLepton]}; 
SumContributionsOperators["4L"] = { 
{O4lSLL, BO4lSLL + PSO4lSLL + PVO4lSLL + TSO4lSLL + TVO4lSLL}, 
{O4lSRR, BO4lSRR + PSO4lSRR + PVO4lSRR + TSO4lSRR + TVO4lSRR}, 
...
};
\end{lstlisting}
It is also possible to use these wrapper files to change the
normalization of the operators. We have made use of this functionality
for the operators with external photons and gluons to match the
standard definition used in literature: it is common to write these
operators as $e \, m_f (\bar{f} \sigma_{\mu\nu} f) F^{\mu\nu}$,
i.e. with the electric coupling (or strong coupling for gluon
operators) and fermion mass factored out. This is done with the files
{\tt Photon\_wrapper.m} and {\tt Gluon\_wrapper.m}, included in the
FlavorKit directory of \SARAH:
\begin{lstlisting}
ProcessWrapper = True; 
NameProcess = "Gamma2Q" 
ExternalFields = {bar[BottomQuark], BottomQuark, Photon}; 

SumContributionsOperators["Gamma2Q"] = { 
{CC7, OA2qSL}, 
{CC7p, OA2qSR}
}; 

NormalizationOperators["Gamma2Q"] ={
"CC7(3,:) = 0.25_dp*CC7(3,:)/sqrt(Alpha_160*4*Pi)/mf_d_160(3)",
"CC7p(3,:) = 0.25_dp*CC7p(3,:)/sqrt(Alpha_160*4*Pi)/mf_d_160(3)",

"CC7SM(3,:) = 0.25_dp*CC7SM(3,:)/sqrt(Alpha_160*4*Pi)/mf_d_160(3)",
"CC7pSM(3,:) = 0.25_dp*CC7pSM(3,:)/sqrt(Alpha_160*4*Pi)/mf_d_160(3)"
};
\end{lstlisting}
First, the coefficients {\tt OA2qSL} and {\tt OA2qSR} derived with
\NewPackage are matched to the new coefficients {\tt CC7} and {\tt
  CC7p}. The same matching is automatically applied also to the SM
coefficients {\tt OA2qSLSM} and {\tt OA2qSRSM}. In a second step,
these operators are re-normalized to the standard definition of the
Wilson coefficients $C_7$ and $C'_7$.  The initial coefficients {\tt
  OA2qSR}, {\tt OA2qSL} are not discarded, but co-exist besides {\tt
  CC7}, {\tt CC7p}. Thus, the user can choose in the implementation of
the observables which operators are more suitable for his purposes.

\section{Validation}
\label{sec:validation}
\begin{figure}[!h]
\includegraphics[width=0.45\linewidth]{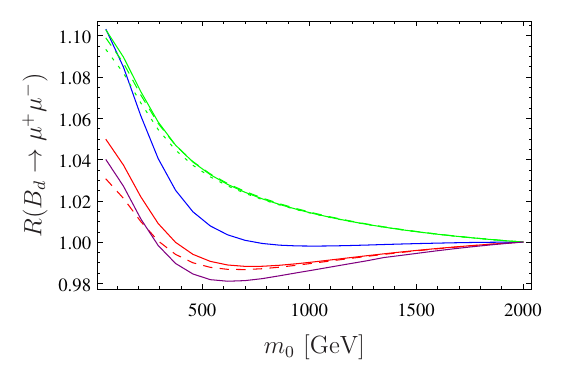} \hfill
\includegraphics[width=0.45\linewidth]{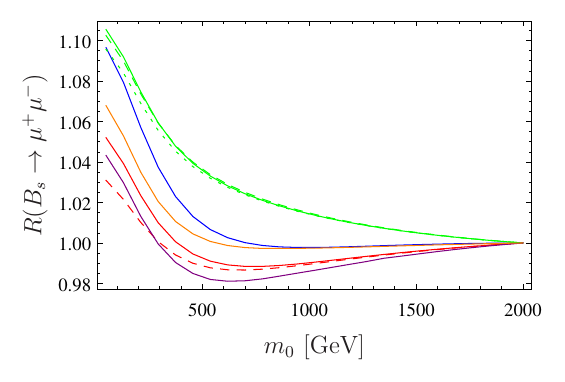} \\
\includegraphics[width=0.45\linewidth]{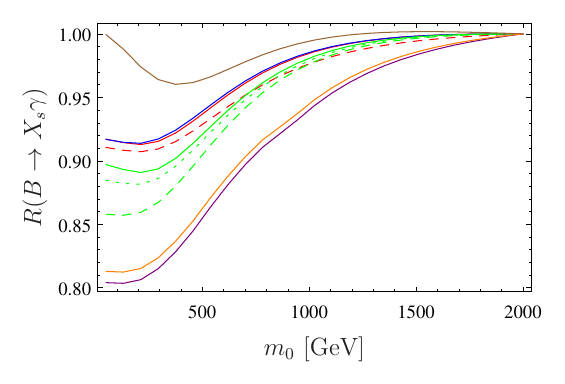} \hfill 
\includegraphics[width=0.45\linewidth]{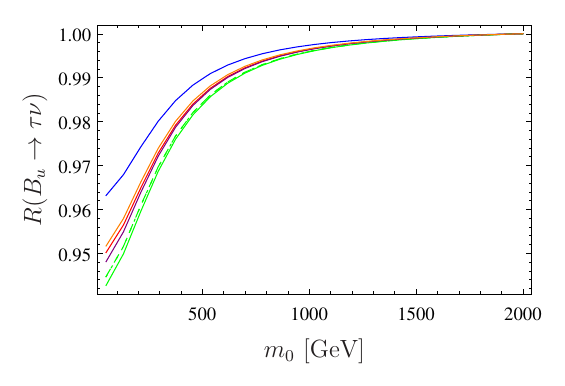} \\
\includegraphics[width=0.45\linewidth]{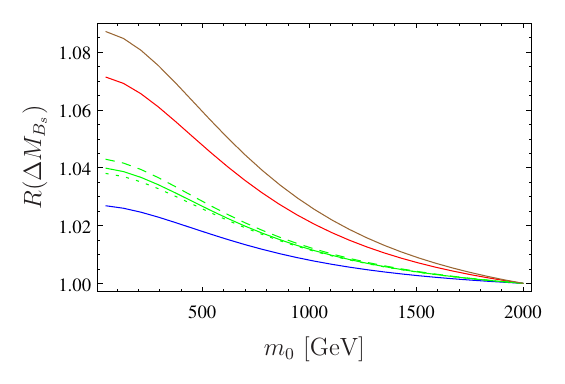} \hfill 
\includegraphics[width=0.45\linewidth]{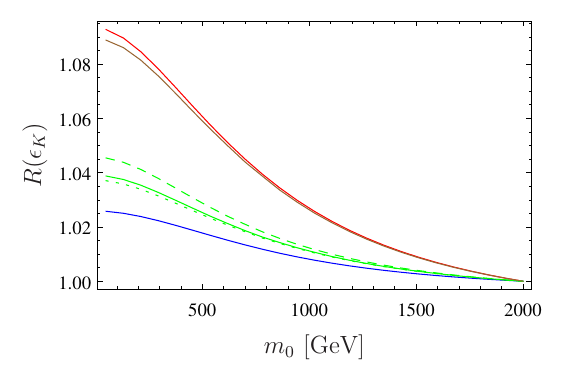} \\
\includegraphics[width=0.45\linewidth]{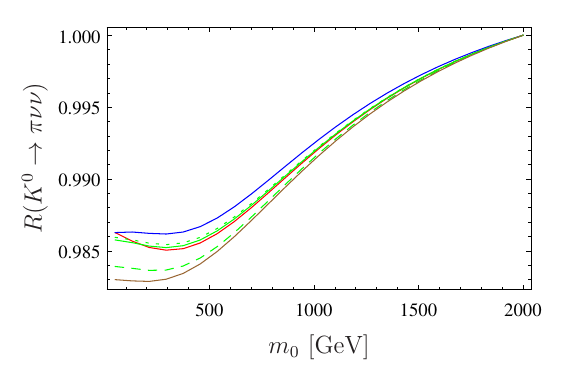} \hfill 
\includegraphics[width=0.45\linewidth]{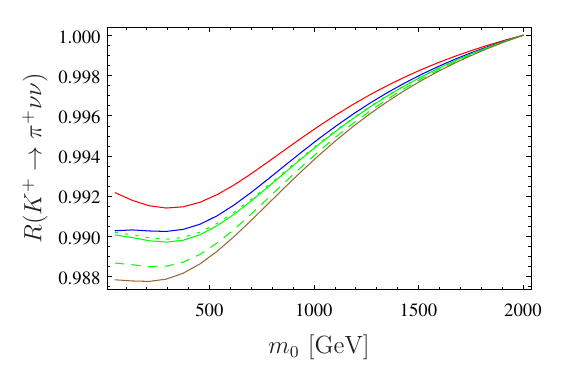} 
\caption{Comparison of the results for BR$(B^0_{s,d} \to \mu \mu)$,
  BR($\bar B \to X_s\gamma$), BR($B\to \tau \nu$), $\Delta M_{B_s}$,
  $\varepsilon_K$, BR($K_L \to \pi^0 \nu \bar \nu$), BR($K^+ \to \pi^+
  \nu \bar \nu$) as a function of $m_0$ using the \FlavorKit (red),
      {\tt superiso} (purple), {\tt SUSY\_Flavor 1} (brown), {\tt
        SUSY\_Flavor 2} (green), \SPheno (blue), {\tt MicrOmegas}
      (orange) and the old implementation in \SARAH (red dashed).  The
      three lines for {\tt SUSY\_Flavor 2} correspond to different
      options of the chiral resummation.  We used $M_{1/2} = 200$~GeV,
      $A_0 = 0$, $\tan\beta=10$, $\mu >0$.}
\label{fig:m0}
\end{figure}

\begin{figure}[!h]
\includegraphics[width=0.45\linewidth]{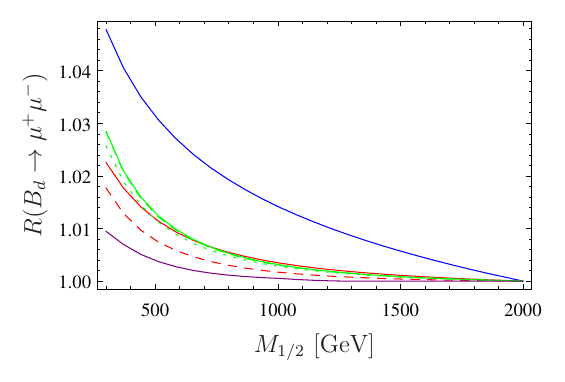} \hfill
\includegraphics[width=0.45\linewidth]{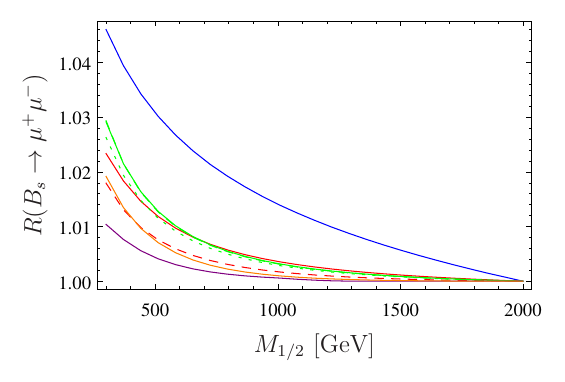} \\
\includegraphics[width=0.45\linewidth]{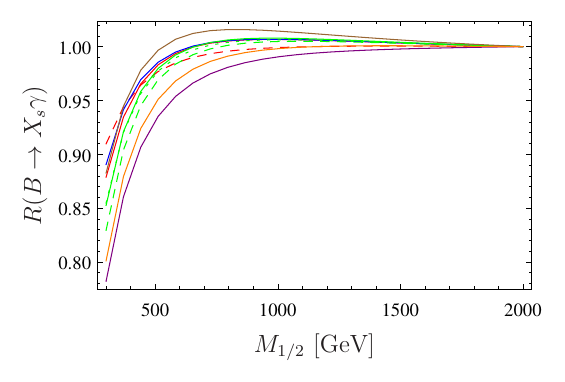} \hfill 
\includegraphics[width=0.45\linewidth]{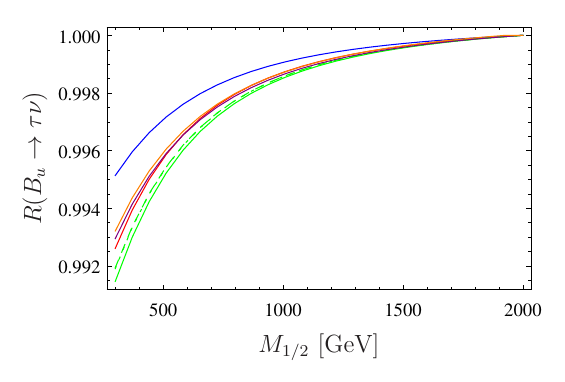} \\
\includegraphics[width=0.45\linewidth]{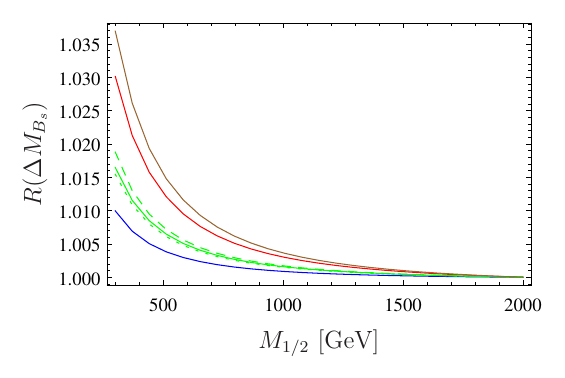} \hfill 
\includegraphics[width=0.45\linewidth]{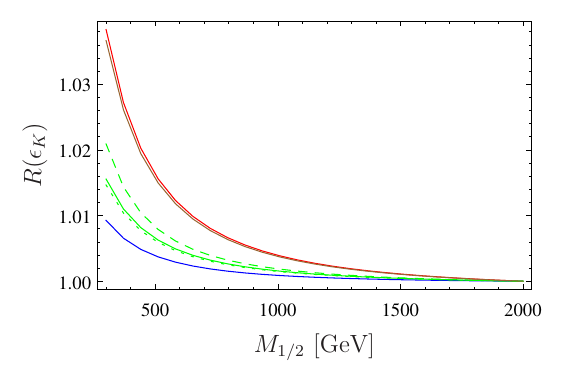} \\
\includegraphics[width=0.45\linewidth]{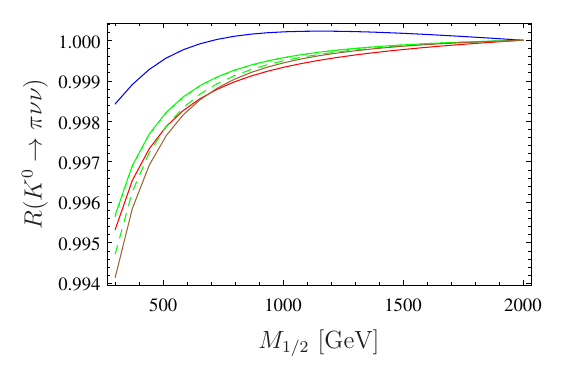} \hfill 
\includegraphics[width=0.45\linewidth]{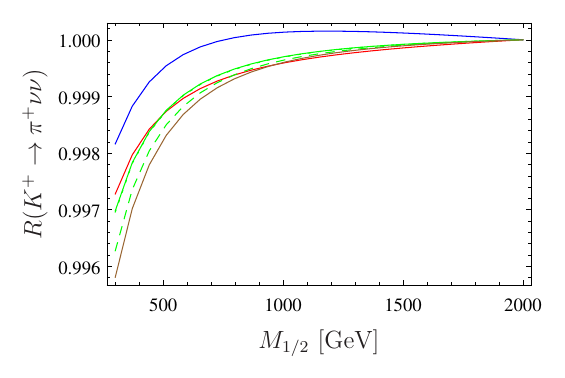} 
\caption{Comparison of the results for different flavor observables as
  function of $M_{1/2}$. The color code is the same as in
  Fig.~\ref{fig:m0}. We used $m_0 = 500$~GeV, $A_0 = -1000$~GeV,
  $\tan\beta=10$, $\mu >0$.}
\label{fig:m12}
\end{figure}

\begin{figure}[!h]
\includegraphics[width=0.45\linewidth]{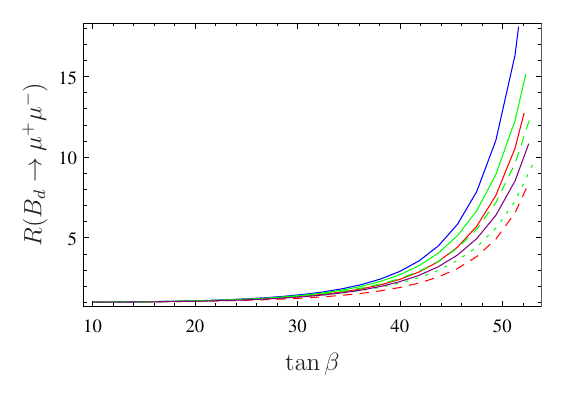} \hfill
\includegraphics[width=0.45\linewidth]{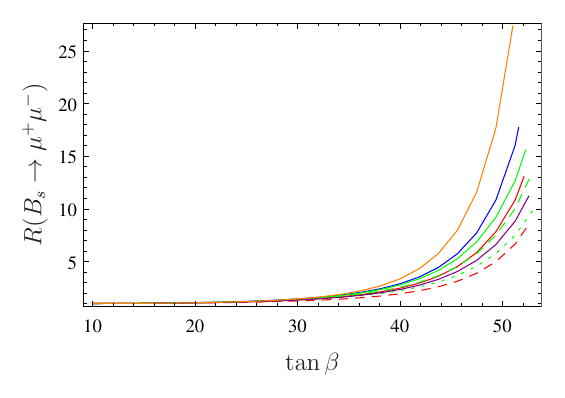} \\
\includegraphics[width=0.45\linewidth]{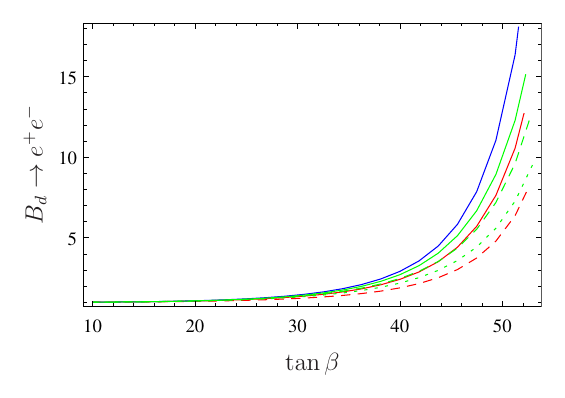} \hfill
\includegraphics[width=0.45\linewidth]{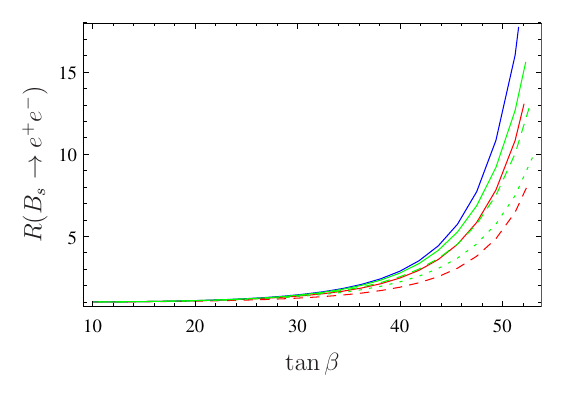} 
\caption{Comparison of BR$(B^0_{s,d} \to \mu \mu)$ (first row) and
  BR$(B^0_{s,d} \to e e)$ (second row) as function of $\tan\beta$. The
  color code is the same as in Fig.~\ref{fig:m0}. We used $m_0 =
  M_{1/2}=500$~GeV, $A_0 = 0$, $\mu >0$. }
\label{fig:tb}
\end{figure}

The validation of the \FlavorKit results happened in three steps:
\begin{enumerate}
 \item {\bf Agreement with SM results}: we checked that the SM prediction for the observables agree with the results given in literature
 \item {\bf Independence of scale in loop function}: the loop integrals for two and three point functions ($B_i, C_i$) depend on the renormalization scale $Q$. However, this dependence has to drop out for a given process at leading order. We checked numerically that the sum of all diagrams is indeed
 independent of the choice of $Q$. 
 \item {\bf Comparison with other tools}: as we have pointed out in the introduction, there are several public tools which calculate flavor observables mostly in the context of the MSSM. We did a detailed comparison with these tools using the \SPheno code produced by \SARAH for the MSSM. Some results are presented in the following. 
\end{enumerate}
We have compared the \FlavorKit results using \SARAH {\tt 4.2.0} and \SPheno {\tt 3.3.0} with 
\begin{itemize}
 \item {\tt superiso 3.3}
 \item {\tt SUSY\_Flavor 1} and {\tt 2.1}
 \item {\tt MicrOmegas 3.6.7}
 \item {\tt SPheno 3.3.0}
 \item a \SPheno version produced by \SARAH\ {\tt 4.1.0} without the \FlavorKit functionality
\end{itemize}
Since these codes often use different values for the hadronic parameters and calculate the flavor observables at different loop levels,
we are not going to compare the absolute numbers obtained by these tools. Instead, we compare the results normalized to the SM prediction of 
each code and thus define, for an observable $X$, the ratio
\begin{equation}
 R(X) = \frac{X^{MSSM}}{X^{SM}} \, . 
\end{equation}
$X^{SM}$ is obtained by taking the value of $X$ calculated by each code in the limit of a very heavy SUSY spectrum. As test case we have used the CMSSM. 
The dependence of a set of flavor observables as function of $m_0$ is shown in Fig.~\ref{fig:m0} and as function of $M_{1/2}$ in Fig.~\ref{fig:m12}. \\
We see that all codes show in general the same dependence. However, it
is also obvious that the lines are not on top of each other but
differences are present. These differences are based on the treatment
of the resummation of the bottom Yukawa couplings, the different order
at which SM and SUSY contributions are implemented, the different
handling of the Weinberg angle, and the different level at which the
RGE running is taken into account by the tools. Even if a detailed
discussion of the differences of all codes might be very interesting
it is, of course, far beyond the scope of this paper and would require
a combined effort. The important point is that the results of
\FlavorKit agree with the codes specialized for the MSSM to the same
level as those codes agree among each other. Since the \FlavorKit
results for all observables are based on the same generic routines it
might be even more trustworthy than human implementations of the
lengthy expressions needed to calculate these observables because it
is less error prone. Of course, known 2-loop corrections for the
MSSM which are implemented in other tools are missing. \\
Finally, it is well known that the process $B_{s,d}^0 \to \ell \bar
\ell$ has a strong dependence on the value of $\tan\beta$.  We show in
Fig.~\ref{fig:tb} that this is reproduced by all codes.

\section{Conclusion}
\label{sec:conclusion}
We have presented \FlavorKit, a new setup for the calculation of
flavor observables for a wide range of BSM models. Generic expressions
for the Wilson coefficients are derived with \NewPackage, a
\Mathematica package that makes use of \FeynArts and \FormCalc. The
output of \NewPackage is then passed to \SARAH, which generates the
\Fortran code that allows to calculate numerically the values of these
Wilson coefficients with \SPheno. The observables are derived by
providing the corresponding pieces of \Fortran code to \SARAH, which
incorporates them into the \SPheno output.  We made use of this code
chain to fully implement a large set of important flavor observables
in \SARAH and \SPheno. In fact, due the simplicity of this kit, the
user can easily extend the list with his own observables and
operators. In conclusion, \FlavorKit allows the user to easily obtain
analytical and numerical results for flavor observables in the BSM
model of his choice.

\section*{Acknowledgments}
We thank Asmaa Abada, Martin Hirsch, Farvah Mahmoudi, Manuel
E. Krauss, Kilian Nickel, Ben O'Leary and C\'edric Weiland for helpful
discussions.  FS is supported by the BMBF PT DESY Verbundprojekt
05H2013-THEORIE 'Vergleich von LHC-Daten mit supersymmetrischen
Modellen'.  WP is supported by the DFG, project No. PO-1337/3-1. AV is
partially supported by the EXPL/FIS-NUC/0460/2013 project financed by
the Portuguese FCT.

\appendix 
\lstset{ basicstyle=\footnotesize}

\section{Lagrangian}
\label{sec:lagrangian}
In this section we present our notation and conventions for the
operators (and their corresponding Wilson coefficients) implemented in
\NewPackage. Although a more complete list of flavor violating operators
can be built, we will concentrate on those implemented in \NewPackage. If
necessary, the user can extend it by adding his/her own operators.

The interaction Lagrangian relevant for flavor violating processes can
be written as
\begin{equation} \label{eq:L-total}
{\cal L}_{\text{FV}} = {\cal L}_{\text{LFV}}  + {\cal L}_{\text{QFV}} \, .
\end{equation}
The first piece contains the operators that can trigger lepton flavor
violation whereas the second piece contains the operators responsible
for quark flavor violation.

The general Lagrangian relevant for lepton flavor violation can be
written as
\begin{equation} \label{eq:L-LFV}
{\cal L}_{\text{LFV}} = {\cal L}_{\ell \ell \gamma} + {\cal L}_{\ell \ell Z} 
+ {\cal L}_{\ell \ell h} + {\cal L}_{4 \ell} + {\cal L}_{2 \ell 2q} \, .
\end{equation}
The first term contains the $\ell - \ell - \gamma$ interaction,
given by
\begin{equation} \label{eq:L-llg}
{\cal L}_{\ell \ell \gamma} = e \, \bar \ell_\beta \left[ \gamma^\mu \left(K_1^L P_L + K_1^R P_R \right) + i m_{\ell_\alpha} \sigma^{\mu \nu} q_\nu \left(K_2^L P_L + K_2^R P_R \right) \right] \ell_\alpha A_\mu + \hc
\end{equation}
Here $e$ is the electric charge, $q$ the photon momentum, $P_{L,R} =
\frac{1}{2} (1 \mp \gamma_5)$ are the usual chirality projectors and
$\ell_{\alpha,\beta}$ denote the lepton flavors. For practical
reasons, we will always consider the photonic contributions
independently, and we will not include them in other vector
operators. On the contrary, the $Z$- and Higgs boson contributions
will be included whenever possible. Therefore, the $\ell - \ell - Z$
and $\ell - \ell - h$ interaction Lagrangians will only be used for
observables involving real $Z$- and Higgs bosons. These two
Lagrangians can be written as
\begin{equation} \label{eq:L-llZ}
{\cal L}_{\ell \ell Z} = \bar \ell_\beta \left[ \gamma^\mu \left(R_1^L P_L + R_1^R P_R \right) + p^\mu \left(R_2^L P_L + R_2^R P_R \right) \right] \ell_\alpha Z_\mu \, ,
\end{equation}
where $p$ is the $\ell_\beta$ 4-momentum, and
\begin{equation} \label{eq:L-llh}
{\cal L}_{\ell \ell h} = \bar \ell_\beta \left(S_L P_L + S_R P_R \right) \ell_\alpha h \, .
\end{equation}

The general $4 \ell$ 4-fermion interaction Lagrangian can be written
as
\begin{equation}
{\cal L}_{4 \ell} = \sum_{\substack{I=S,V,T\\X,Y=L,R}} A_{XY}^I \bar \ell_\beta \Gamma_I P_X \ell_\alpha \bar \ell_\delta \Gamma_I P_Y \ell_\gamma + \hc \, , \label{eq:L-4L}
\end{equation}
where $\ell_{\alpha,\beta,\gamma,\delta}$ denote the lepton flavors
and $\Gamma_S = 1$, $\Gamma_V = \gamma_\mu$ and $\Gamma_T =
\sigma_{\mu \nu}$. We omit flavor indices in the Wilson coefficients
for the sake of clarity. This Lagrangian contains the most general
form compatible with Lorentz invariance. The Wilson coefficients
$A_{LR}^S$ and $A_{RL}^S$ were included in \cite{Ilakovac:2012sh}, but
absent in \cite{Hisano:1995cp,Arganda:2005ji}. As previously stated,
the coefficients in Eq.\eqref{eq:L-4L} do not include photonic
contributions, but they include Z-boson and scalar ones. Finally, the
general $2 \ell 2 q$ four fermion interaction Lagrangian at the quark
level is given by
\begin{equation}
{\cal L}_{2 \ell 2q} = {\cal L}_{2 \ell 2d} + {\cal L}_{2 \ell 2u} \label{eq:L-2L2Q}
\end{equation}
where
\begin{eqnarray}
{\cal L}_{2 \ell 2d} = 
&& \sum_{\substack{I=S,V,T\\X,Y=L,R}} B_{XY}^I \bar \ell_\beta \Gamma_I P_X \ell_\alpha \bar d_\gamma \Gamma_I P_Y d_\gamma + \hc \label{eq:L-2L2D} \\
{\cal L}_{2 \ell 2u} = && \left. {\cal L}_{2 \ell 2d} \right|_{d \to u, \, B \to C} \label{eq:L-2L2U} \, .
\end{eqnarray}
Here $d_{\gamma}$ denotes the d-quark flavor.

Let us now consider the Lagrangian relevant for quark flavor
violation. This can be written as
\begin{equation} \label{eq:L-QFV}
{\cal L}_{\text{QFV}} = {\cal L}_{q q \gamma} + {\cal L}_{q q g} + {\cal L}_{4 d} + 
{\cal L}_{2d2l} + {\cal L}_{2d2\nu} + {\cal L}_{du\ell\nu} + {\cal L}_{d d H} \, .
\end{equation}
The first two terms correspond to operators that couple quark
bilinears to massless gauge bosons. These are
\begin{eqnarray}
{\cal L}_{q q \gamma} = && e \left[ \bar d_\beta \sigma_{\mu \nu} \left( m_{d_\beta} Q_1^L P_L + m_{d_\alpha} Q_1^R P_R \right) d_\alpha \right] F^{\mu \nu} \label{eq:L-qqgamma} \\
{\cal L}_{q q g} = && g_s \left[ \bar d_\beta \sigma_{\mu \nu} \left( m_{d_\beta} Q_2^L P_L + m_{d_\alpha} Q_2^R P_R \right) T^a d_\alpha \right] G_a^{\mu \nu} \label{eq:L-qqgluon} \, .
\end{eqnarray}
Here $T^a$ are $SU(3)$ matrices. The Wilson coefficients
$Q_{1,2}^{L,R}$ can be easily related to the usual
$C_{7,8}^{(\prime)}$ coefficients, sometimes normalized with an
additional $\frac{1}{16 \pi^2}$ factor. The $4 d$ four fermion
interaction Lagrangian can be written as
\begin{equation}
{\cal L}_{4 d} = \sum_{\substack{I=S,V,T\\X,Y=L,R}} D_{XY}^I \bar d_\beta \Gamma_I P_X d_\alpha \bar d_\delta \Gamma_I P_Y d_\gamma + \hc \, , \label{eq:L-4D}
\end{equation}
where $d_{\alpha,\beta,\gamma,\delta}$ denote the lepton
flavors. Again, we omit flavor indices in the Wilson coefficients for
the sake of clarity. The $2 d 2 \ell$ four fermion interaction
Lagrangian is given by
\begin{equation}
{\cal L}_{2d 2 \ell} = \sum_{\substack{I=S,V,T\\X,Y=L,R}} E_{XY}^I \bar d_\beta \Gamma_I P_X d_\alpha \bar \ell_\gamma \, \Gamma_I P_Y \ell_\gamma + \hc \label{eq:L-2D2L} \, .
\end{equation}
Here $\ell_{\gamma}$ denotes the lepton flavor. ${\cal L}_{2d 2 \ell}$
should not be confused with ${\cal L}_{2 \ell 2d}$. In the former case
one has QFV operators, whereas in the latter one has LFV
operators. This distinction has been made for practical reasons. The
$2d 2 \nu$ and $du\ell\nu$ terms of the QFV Lagrangian are
\begin{eqnarray}
{\cal L}_{2d 2 \nu} &=& \sum_{X,Y=L,R} F_{XY}^V \bar d_\beta \gamma_\mu P_X d_\alpha \bar \nu_\gamma \gamma^\mu P_Y \nu_\gamma + \hc \label{eq:L-2D2V} \\
{\cal L}_{du\ell\nu} &=& \sum_{\substack{I=S,V\\X,Y=L,R}} G_{XY}^I \bar d_\beta \Gamma_I P_X u_\alpha \bar \ell_\gamma \, \Gamma_I P_Y \nu_\gamma + \hc \label{eq:L-DULV} \, .
\end{eqnarray}
Note that we have not introduced scalar or tensor $2d2\nu$ operators,
nor tensor $du\ell\nu$ ones, and that lepton flavor (denoted by the
index $\gamma$) is conserved in these operators. Finally, we have also
included a term in the Lagrangian accounting for operators of the type
$(\bar{d} \Gamma d) S$ and $(\bar{d} \Gamma d) P$, where $S$ ($P$) is
a virtual~\footnote{We would like to emphasize that our implementation
  of these operators is only valid for virtual scalars and
  pseudoscalars. They have been introduced in order to provide the
  1-loop vertices necessary for the computation of the double penguin
  contributions to $\Delta M_{B_q}$. Therefore, they are not valid for
  observables in which the scalar or pseudoscalar states are real
  particles.} scalar (pseudoscalar) state. This piece can be written
as
\begin{equation} \label{eq:L-ddSP}
{\cal L}_{d d H} = \bar d_\beta \left(H_L^S P_L + H_R^S P_R \right) d_\alpha S
+ \bar d_\beta \left(H_L^P P_L + H_R^P P_R \right) d_\alpha P \, .
\end{equation}

\section{Operators available by default in the \SPheno output of \SARAH}
\label{app:operators}
The operators presented in Appendix \ref{sec:lagrangian} have been
implemented by using the results of \NewPackage in \SARAH. Those are
exported to \SPheno. We give in the following the list of all internal
names for these operators, which can be used in the calculation of new
flavor observables.

\subsection{2-Fermion-1-Boson operators}
These operators are arrays with either two or three elements. While
operators involving vector bosons have always dimension $3\times 3$,
those with scalars have dimension $3\times 3\times n_g$. $n_g$ is the
number of generations of the considered scalar and for $n_g=1$ the
last index is dropped.

\begin{center}
$(\bar d_\beta  \sigma_{\mu\nu} \Gamma d_\alpha) F^{\mu\nu}$  and $(\bar d_\beta \sigma_{\mu\nu} \Gamma d_\alpha) G^{\mu\nu}$
\end{center}
\begin{center}
\begin{tabular}{|ccc||ccc|}
\hline 
Variable & Operator & Name & Variable & Operator & Name\\
\hline 
CC7 & $e m_{d_\beta} (\bar d_\beta \sigma_{\mu\nu} P_L d_\alpha) F^{\mu\nu}$ & $Q_1^L$ & CC8 & $g_s m_{d_\beta} (\bar d_\beta \sigma_{\mu\nu} P_L d_\alpha) G^{\mu\nu}$ & $Q_2^L$\\
CC7p & $e m_{d_\alpha} (\bar d_\beta \sigma_{\mu\nu} P_R d_\alpha) F^{\mu\nu}$ & $Q_1^R$ & CC8p & $g_s m_{d_\alpha} (\bar d_\beta \sigma_{\mu\nu} P_R d_\alpha) G^{\mu\nu}$ & $Q_2^R$ \\
\hline 
\end{tabular}
\end{center}
These operators are derived by \NewPackage with the following input files
\begin{lstlisting}[caption=PhotonQQp.m]
NameProcess="Gamma2Q";

ConsideredProcess = "2Fermion1Vector";
FermionOrderExternal={1,2};
NeglectMasses={3};  


ExternalFields= {bar[BottomQuark], BottomQuark,Photon};
CombinationGenerations = {{3,2}};


AllOperators={
   {OA2qSL,Op[7] Pair[ec[3],k[1]]}, 
   {OA2qSR,Op[6] Pair[ec[3],k[1]]},
   {OA2qVL,Op[7,ec[3]]},
   {OA2qVR,Op[6,ec[3]]}
};

OutputFile = "Gamma2Q.m";

Filters = {}; 
\end{lstlisting}

\begin{lstlisting}[caption=GluonQQp.m]
NameProcess="Gluon2Q";

ConsideredProcess = "2Fermion1Vector";
FermionOrderExternal={1,2};
NeglectMasses={3};  


ExternalFields= {bar[BottomQuark], BottomQuark,Gluon};
CombinationGenerations = {{3,2}};


AllOperators={
   {OG2qSL,Op[7] Pair[ec[3],k[1]]}, 
   {OG2qSR,Op[6] Pair[ec[3],k[1]]}
};

OutputFile = "Gluon2Q.m";

Filters = {}; 
\end{lstlisting}

The normalization is changed to match the standard definitions by 
\begin{lstlisting}[caption=Photon\_wrapper\_QFV.m]
ProcessWrapper = True; 
NameProcess = "Gamma2Q" 
ExternalFields = {bar[BottomQuark], BottomQuark, Photon}; 

SumContributionsOperators["Gamma2Q"] = { 
{CC7, OA2qSL}, 
{CC7p, OA2qSR}
}; 

NormalizationOperators["Gamma2Q"] ={
"CC7(2,:) = 0.25_dp*CC7(2,:)/sqrt(Alpha_160*4*Pi)/MFd(2)",
"CC7(3,:) = 0.25_dp*CC7(3,:)/sqrt(Alpha_160*4*Pi)/MFd(3)",
"CC7p(2,:) = 0.25_dp*CC7p(2,:)/sqrt(Alpha_160*4*Pi)/MFd(2)",
"CC7p(3,:) = 0.25_dp*CC7p(3,:)/sqrt(Alpha_160*4*Pi)/MFd(3)",

"CC7SM(2,:) = 0.25_dp*CC7SM(2,:)/sqrt(Alpha_160*4*Pi)/MFd(2)",
"CC7SM(3,:) = 0.25_dp*CC7SM(3,:)/sqrt(Alpha_160*4*Pi)/MFd(3)",
"CC7pSM(2,:) = 0.25_dp*CC7pSM(2,:)/sqrt(Alpha_160*4*Pi)/MFd(2)",
"CC7pSM(3,:) = 0.25_dp*CC7pSM(3,:)/sqrt(Alpha_160*4*Pi)/MFd(3)"
};
\end{lstlisting}

\begin{lstlisting}[caption=Gluon\_wrapper.m]
ProcessWrapper = True; 
NameProcess = "Gluon2Q" 
ExternalFields = {bar[BottomQuark], BottomQuark, Gluon}; 

SumContributionsOperators["Gluon2Q"] = { 
{CC8, OG2qSL}, 
{CC8p, OG2qSR}}; 

NormalizationOperators["Gluon2Q"] ={
"CC8(2,:) = 0.25_dp*CC8(2,:)/sqrt(AlphaS_160*4*Pi)/MFd(2)",
"CC8(3,:) = 0.25_dp*CC8(3,:)/sqrt(AlphaS_160*4*Pi)/MFd(3)",
"CC8p(2,:) = 0.25_dp*CC8p(2,:)/sqrt(AlphaS_160*4*Pi)/MFd(2)",
"CC8p(3,:) = 0.25_dp*CC8p(3,:)/sqrt(AlphaS_160*4*Pi)/MFd(3)",

"CC8SM(2,:) = 0.25_dp*CC8SM(2,:)/sqrt(AlphaS_160*4*Pi)/MFd(2)",
"CC8SM(3,:) = 0.25_dp*CC8SM(3,:)/sqrt(AlphaS_160*4*Pi)/MFd(3)",
"CC8pSM(2,:) = 0.25_dp*CC8pSM(2,:)/sqrt(AlphaS_160*4*Pi)/MFd(2)",
"CC8pSM(3,:) = 0.25_dp*CC8pSM(3,:)/sqrt(AlphaS_160*4*Pi)/MFd(3)"

};
\end{lstlisting}


\begin{center}
$\bar \ell_\beta \left( q^2 \gamma^\mu  + i m_{\ell_\alpha} \sigma^{\mu \nu} q_\nu \right) \ell_\alpha A_\mu$
\end{center}
\begin{center}
\begin{tabular}{|ccc||ccc|}
\hline 
Variable & Operator & Name & Variable & Operator & Name\\
\hline 
K2L & $e m_{\ell_\alpha} (\bar \ell_\beta \sigma_{\mu\nu} P_L \ell_\alpha) q^{\nu} A^\mu$ & $K_2^L$  & K1L & $q^2 (\bar \ell_\beta \gamma_\mu P_L \ell_\alpha) A^\mu$ & $K_1^L$\\
K2R & $e m_{\ell_\alpha} (\bar \ell_\beta \sigma_{\mu\nu} P_R \ell_\alpha) q^{\nu} A^\mu$ & $K_2^L$ & K1R & $q^2 (\bar \ell_\beta \gamma_\nu P_R \ell_\alpha) A^\mu$ &  $K_1^R$\\
\hline 
\end{tabular}
\end{center}
These operators are derived by \NewPackage with the following input files
\begin{lstlisting}[caption=PhotonLLp.m]
NameProcess="Gamma2l";

ConsideredProcess = "2Fermion1Vector";
FermionOrderExternal={1,2};
NeglectMasses={3};  


ExternalFields= {bar[ChargedLepton], ChargedLepton,Photon};
CombinationGenerations = {{2,1},{3,1},{3,2}};


AllOperators={
   {OA2lSL,Op[6] Pair[ec[3],k[1]]}, 
   {OA2lSR,Op[7] Pair[ec[3],k[1]]},
   {OA1L,Op[6,ec[3]] Pair[k[3],k[3]]},
   {OA1R,Op[7,ec[3]] Pair[k[3],k[3]]}
};

OutputFile = "Gamma2l.m";

Filters = {}; 
\end{lstlisting}

The normalization is changed to match the standard definitions by 
\begin{lstlisting}[caption=Photon\_wrapper\_LFV.m]
ProcessWrapper = True; 
NameProcess = "Gamma2l" 
ExternalFields = {bar[ChargedLepton], ChargedLepton, Photon}; 

SumContributionsOperators["Gamma2l"] = { 
{K1L, OA1L}, 
{K1R, OA1R}, 
{K2L, OA2lSL}, 
{K2R, OA2lSR}}; 

NormalizationOperators["Gamma2l"] ={
"K1L = K1L/sqrt(Alpha_MZ*4*Pi)",
"K1R = K1R/sqrt(Alpha_MZ*4*Pi)",
"K2L(2,:) = -0.5_dp*K2L(2,:)/sqrt(Alpha_MZ*4*Pi)/MFe(2)",
"K2L(3,:) = -0.5_dp*K2L(3,:)/sqrt(Alpha_MZ*4*Pi)/MFe(3)",
"K2R(2,:) = -0.5_dp*K2R(2,:)/sqrt(Alpha_MZ*4*Pi)/MFe(2)",
"K2R(3,:) = -0.5_dp*K2R(3,:)/sqrt(Alpha_MZ*4*Pi)/MFe(3)"
};
\end{lstlisting}


\begin{center}
$(\bar \ell \Gamma \ell) Z$
\end{center}
\begin{center}
\begin{tabular}{|ccc||ccc|}
\hline 
Variable & Operator & Name & Variable & Operator & Name \\
\hline 
OZ2lVL & $(\bar{\ell} \, \gamma^\mu P_L \ell) Z_\mu$ & $R_1^L$ & OZ2lSL & $(\bar{\ell} p^\mu P_L \ell) Z_\mu$ & $R_2^L$  \\
OZ2lVR & $(\bar{\ell} \, \gamma^\mu P_R \ell) Z_\mu$ & $R_1^R$ & OZ2lSR & $(\bar{\ell} p^\mu P_R \ell) Z_\mu$ & $R_2^R$ \\
\hline 
\end{tabular}
\end{center}
In the following we omit flavor indices for the sake of
simplicity. These operators are derived by \NewPackage with the
following input files
\begin{lstlisting}[caption=Z2l.m]
NameProcess="Z2l";

ConsideredProcess = "2Fermion1Vector";
FermionOrderExternal={1,2};
NeglectMasses={1,2};  


ExternalFields= {ChargedLepton,bar[ChargedLepton],Zboson};
CombinationGenerations = {{1,2},{1,3},{2,3}};


AllOperators={
   {OZ2lSL,Op[7] Pair[ec[3],k[1]]}, {OZ2lSR,Op[6] Pair[ec[3],k[1]]},
   {OZ2lVL,Op[7,ec[3]]}, {OZ2lVR,Op[6,ec[3]]}
};

OutputFile = "Z2l.m";

Filters = {}; 
\end{lstlisting}

\begin{center}
$(\bar{\ell} \Gamma \ell) h$
\end{center}
\begin{center}
\begin{tabular}{|ccc||ccc|}
\hline 
Variable & Operator & Name & Variable & Operator & Name\\
\hline 
OH2lSL & $\bar{\ell} P_L \ell \, h$ &$S_L$ & OH2lSR & $\bar{\ell} P_R \ell \, h$ & $S_R$ \\
\hline 
\end{tabular}
\end{center}
These operators are derived by \NewPackage with the following input files
\begin{lstlisting}[caption=H2l.m]
NameProcess="H2l";

ConsideredProcess = "2Fermion1Scalar";
FermionOrderExternal={1,2};
NeglectMasses={1,2};  


ExternalFields= {ChargedLepton,bar[ChargedLepton],HiggsBoson};
CombinationGenerations = {{1,2,ALL},{1,3,ALL},{2,3,ALL}};


AllOperators={{OH2lSL,Op[7]},
              {OH2lSR,Op[6]}
};

OutputFile = "H2l.m";

Filters = {}; 
\end{lstlisting}

\begin{center}
$(\bar{d} \Gamma d) S$ and $(\bar{d} \Gamma d) P$
\end{center}
\begin{center}
\begin{tabular}{|ccc||ccc|}
\hline 
Variable & Operator & Name & Variable & Operator & Name\\
\hline 
OH2qSL & $\bar{d} P_L d \, S$ &$H_L^S$ & OH2qSR & $\bar{d} P_R d \, S$ & $H_R^S$ \\
OAh2qSL & $\bar{d} P_L d \, P$ &$H_L^P$ & OAh2qSR & $\bar{d} P_R d \, P$ & $H_R^P$ \\
\hline 
\end{tabular}
\end{center}
These auxiliary~\footnote{The $(\bar{d} \Gamma d) S$ and $(\bar{d}
  \Gamma d) P$ operators have been introduced to compute double
  penguin corrections to $\Delta M_{B_q}$, where $S$ and $P$ appear as
  intermediate (virtual) particles. They should not be used in
  processes where the scalar or pseudoscalar states are real
  particles because the loop functions are calculated with 
  vanishing external momenta.} operators are derived by \NewPackage with the following
input files
\begin{lstlisting}[caption=H2q.m]
NameProcess="H2q";

(* operators needed for double penguins with internal scalars *)
(* we neglect therefore the mass of the scalar in the loop functions *)
(* and treat it as massless *)

ConsideredProcess = "2Fermion1Scalar";
FermionOrderExternal={2,1};
NeglectMasses={3};   


ExternalFields= {DownQuark,bar[DownQuark],HiggsBoson};
CombinationGenerations = {{2,1,ALL},{3,1,ALL},{3,2,ALL}};


AllOperators={{OH2qSL,Op[7]},
              {OH2qSR,Op[6]}
};

OutputFile = "H2q.m";

Filters = {}; 
\end{lstlisting}

\begin{lstlisting}[caption=A2q.m]
NameProcess="A2q";

(* operators needed for double penguins with internal scalars *)
(* we neglect therefore the mass of the scalar in the loop functions *)
(* and treat it as massless *)

ConsideredProcess = "2Fermion1Scalar";
FermionOrderExternal={2,1};
NeglectMasses={3};  


ExternalFields= {DownQuark,bar[DownQuark],PseudoScalar};
CombinationGenerations = {{2,1,ALL},{3,1,ALL},{3,2,ALL}};


AllOperators={{OAh2qSL,Op[7]},
              {OAh2qSR,Op[6]}
};

OutputFile = "A2q.m";

Filters = {}; 
\end{lstlisting}


\subsection{4-Fermion operators}
All operators listed below carry four indices and have dimension
$3\times3\times3\times3$. In addition, the user can access the
different contributions of all operators from tree-level diagrams, as
well as penguin and box diagrams. The name conventions are as follows:
for each operator {\tt op} the additional parameter exist
\begin{itemize}
 \item {\tt TSop}: tree-level contributions with scalar propagator
 \item {\tt TVop}: tree-level contributions with scalar propagator
 \item {\tt PSop}: sum of penguin and self-energy contributions with scalar propagator
 \item {\tt PVop}: sum of penguin and self-energy contributions with scalar propagator
 \item {\tt Bop}: box contributions. 
\end{itemize}
We will denote the 4-fermion operators involving two leptons and two
down-type quarks depending on whether they lead to LFV or to QFV
processes: $\ell \ell d d$ for LFV and $d d \ell \ell$ for QFV.

\begin{center}
$(\bar{d} \Gamma d) (\bar{\ell} \Gamma^\prime \ell)$ 
and $(\bar{d} \Gamma d) (\bar{\nu} \Gamma^\prime \nu)$
\end{center}

\begin{center}
\begin{tabular}{|ccc||ccc|}
\hline 
Variable & Operator & Name & Variable & Operator & Name\\
\hline 
OddllSLL & $(\bar{d} P_L d) (\bar{\ell} P_L \ell)$ & $E_{LL}^S$ &&&\\
OddllSRR & $(\bar{d} P_R d) (\bar{\ell} P_R \ell)$ & $E_{RR}^S$ &&&\\
OddllSLR & $(\bar{d} P_L d) (\bar{\ell} P_R \ell)$ & $E_{LR}^S$ &&&\\
OddllSRL & $(\bar{d} P_R d) (\bar{\ell} P_L \ell)$ & $E_{RL}^S$ &&&\\
\hline 
OddllVLL & $(\bar{d} \gamma_\mu P_L d) (\bar{\ell} \gamma^\mu P_L \ell)$ & $E_{LL}^V$ & OddvvVLL & $(\bar{d} \gamma_\mu P_L d) (\bar{\nu} \gamma^\mu P_R \nu)$ & $F_{LL}^V$ \\
OddllVRR & $(\bar{d} \gamma_\mu P_R d) (\bar{\ell} \gamma^\mu P_R \ell)$ & $E_{RR}^V$ & OddvvVRR & $(\bar{d} \gamma_\mu P_R d) (\bar{\nu} \gamma^\mu P_R \nu)$ & $F_{RR}^V$ \\
OddllVLR & $(\bar{d} \gamma_\mu P_L d) (\bar{\ell} \gamma^\mu P_R \ell)$ & $E_{LR}^V$ & OddvvVLR & $(\bar{d} \gamma_\mu P_L d) (\bar{\nu} \gamma^\mu P_R \nu)$ & $F_{LR}^V$ \\
OddllVRL & $(\bar{d} \gamma_\mu P_R d) (\bar{\ell} \gamma^\mu P_L \ell)$ & $E_{RL}^V$ & OddvvVRL & $(\bar{d} \gamma_\mu P_R d) (\bar{\nu} \gamma^\mu P_L \nu)$ & $F_{RL}^V$ \\
\hline 
OddllTLL & $(\bar{d} \sigma_{\mu\nu} P_L d) (\bar{\ell} \sigma^{\mu\nu} P_L \ell)$ & $E_{LL}^T$ &&&\\
OddllTRR & $(\bar{d} \sigma_{\mu\nu} P_R d) (\bar{\ell} \sigma^{\mu\nu} P_R \ell)$ & $E_{RR}^T$ &&&\\
OddllTLR & $(\bar{d} \sigma_{\mu\nu} P_L d) (\bar{\ell} \sigma^{\mu\nu} P_R \ell)$ & $E_{LR}^T$ &&&\\
OddllTRL & $(\bar{d} \sigma_{\mu\nu} P_R d) (\bar{\ell} \sigma^{\mu\nu} P_L \ell)$ & $E_{RL}^T$ &&&\\
\hline 
\end{tabular}
\end{center}
These operators are derived by \NewPackage with the following input files
\begin{lstlisting}[caption=2d2L.m]
NameProcess="2d2L";

ConsideredProcess = "4Fermion";
FermionOrderExternal={2,1,4,3};
NeglectMasses={1,2,3,4};  


ExternalFields= {DownQuark,bar[DownQuark],ChargedLepton,bar[ChargedLepton]};

CombinationGenerations = {{3,1,1,1}, {3,1,2,2}, {3,1,3,3},  
                          {3,2,1,1}, {3,2,2,2}, {3,2,3,3}};


AllOperators={{OddllSLL,Op[7].Op[7]},
              {OddllSRR,Op[6].Op[6]},
              {OddllSRL,Op[6].Op[7]},
              {OddllSLR,Op[7].Op[6]},

              {OddllVRR,Op[7,Lor[1]].Op[7,Lor[1]]},
              {OddllVLL,Op[6,Lor[1]].Op[6,Lor[1]]},
              {OddllVRL,Op[7,Lor[1]].Op[6,Lor[1]]},
              {OddllVLR,Op[6,Lor[1]].Op[7,Lor[1]]},

              {OddllTLL,Op[-7,Lor[1],Lor[2]].Op[-7,Lor[1],Lor[2]]},
              {OddllTLR,Op[-7,Lor[1],Lor[2]].Op[-6,Lor[1],Lor[2]]},
              {OddllTRL,Op[-6,Lor[1],Lor[2]].Op[-7,Lor[1],Lor[2]]},
              {OddllTRR,Op[-6,Lor[1],Lor[2]].Op[-6,Lor[1],Lor[2]]}
}; 
\end{lstlisting}

\begin{lstlisting}[caption=2d2nu.m]
NameProcess="2d2nu";

ConsideredProcess = "4Fermion";
FermionOrderExternal={2,1,4,3};
NeglectMasses={1,2,3,4};  

ExternalFields= {DownQuark,bar[DownQuark],Neutrino,bar[Neutrino]};

CombinationGenerations = Flatten[Table[{{2,1, neutrino1, neutrino2}, 
      {3,1, neutrino1, neutrino2},{3,2, neutrino1, neutrino2}},
      {neutrino1,1,3},{neutrino2,1,3}],2];


AllOperators={{OddvvVRR,Op[7,Lor[1]].Op[7,Lor[1]]},
              {OddvvVLL,Op[6,Lor[1]].Op[6,Lor[1]]},
              {OddvvVRL,Op[7,Lor[1]].Op[6,Lor[1]]},
              {OddvvVLR,Op[6,Lor[1]].Op[7,Lor[1]]}
}; 
\end{lstlisting}


\begin{center}
$(\bar{\ell} \Gamma \ell) (\bar{d} \Gamma^\prime d)$  and $(\bar{\ell} \Gamma \ell) (\bar{u} \Gamma^\prime u)$
\end{center}

\begin{center}
\begin{tabular}{|ccc||ccc|}
\hline 
Variable & Operator & Name & Variable & Operator & Name \\
\hline 
OllddSLL & $(\bar{\ell} P_L \ell) (\bar{d} P_L d)$ & $B_{LL}^S$                 &         OlluuSLL & $(\bar{\ell} P_L \ell) (\bar{u} P_L u)$ & $C_{LL}^S$ \\
OllddSRR & $(\bar{\ell} P_R \ell) (\bar{d} P_R d)$ & $B_{RR}^S$                 &         OlluuSRR & $(\bar{\ell} P_R \ell) (\bar{u} P_R u)$ & $C_{RR}^S$ \\
OllddSRL & $(\bar{\ell} P_R \ell) (\bar{d} P_L d)$ & $B_{RL}^S$                 &         OlluuSRL & $(\bar{\ell} P_R \ell) (\bar{u} P_L u)$ & $C_{RL}^S$ \\
OllddSLR & $(\bar{\ell} P_L \ell) (\bar{d} P_R d)$ & $B_{LR}^S$                 &         OlluuSLR & $(\bar{\ell} P_L \ell) (\bar{u} P_R u)$ & $C_{LR}^S$ \\
\hline                                                                           
OllddVLL & $(\bar{\ell} \gamma_\mu P_L \ell) (\bar{d} \gamma^\mu P_L d)$ & $B_{LL}^V$    &      OlluuVLL & $(\bar{\ell} \gamma_\mu P_L \ell) (\bar{u} \gamma^\mu P_L u)$ & $C_{LL}^V$ \\
OllddVRR & $(\bar{\ell} \gamma_\mu P_R \ell) (\bar{d} \gamma^\mu P_R d)$ & $B_{RR}^V$    &      OlluuVRR & $(\bar{\ell} \gamma_\mu P_R \ell) (\bar{u} \gamma^\mu P_R u)$ & $C_{RR}^V$ \\
OllddVLR & $(\bar{\ell} \gamma_\mu P_L \ell) (\bar{d} \gamma^\mu P_R d)$ & $B_{LR}^V$    &      OlluuVLR & $(\bar{\ell} \gamma_\mu P_L \ell) (\bar{u} \gamma^\mu P_R u)$ & $C_{LR}^V$ \\
OllddVRL & $(\bar{\ell} \gamma_\mu P_R \ell) (\bar{d} \gamma^\mu P_L d)$ & $B_{RL}^V$    &      OlluuVRL & $(\bar{\ell} \gamma_\mu P_R \ell) (\bar{u} \gamma^\mu P_L u)$ & $C_{RL}^V$ \\
\hline                                                                            
OllddTLL & $(\bar{\ell} \sigma_{\mu\nu} P_L \ell) (\bar{d} \sigma^{\mu\nu} P_L d)$ & $B_{LL}^T$    &     OlluuTLL & $(\bar{\ell} \sigma_{\mu\nu} P_L \ell) (\bar{u} \sigma^{\mu\nu} P_L u)$ & $C_{LL}^T$ \\ 
OllddTRR & $(\bar{\ell} \sigma_{\mu\nu} P_R \ell) (\bar{d} \sigma^{\mu\nu} P_R d)$ & $B_{RR}^T$    &     OlluuTRR & $(\bar{\ell} \sigma_{\mu\nu} P_R \ell) (\bar{u} \sigma^{\mu\nu} P_R u)$ & $C_{RR}^T$ \\
OllddTLR & $(\bar{\ell} \sigma_{\mu\nu} P_L \ell) (\bar{d} \sigma^{\mu\nu} P_R d)$ & $B_{LR}^T$    &     OlluuTLR & $(\bar{\ell} \sigma_{\mu\nu} P_L \ell) (\bar{u} \sigma^{\mu\nu} P_R u)$ & $C_{LR}^T$ \\ 
OllddTRL & $(\bar{\ell} \sigma_{\mu\nu} P_R \ell) (\bar{d} \sigma^{\mu\nu} P_L d)$ & $B_{RL}^T$    &     OlluuTRL & $(\bar{\ell} \sigma_{\mu\nu} P_R \ell) (\bar{u} \sigma^{\mu\nu} P_L u)$ & $C_{RL}^T$ \\
\hline 
\end{tabular}
\end{center}
\begin{lstlisting}[caption=2L2d.m]
NameProcess="2L2d";

ConsideredProcess = "4Fermion";
FermionOrderExternal={2,1,4,3};
NeglectMasses={1,2,3,4};  


ExternalFields= {ChargedLepton,bar[ChargedLepton],DownQuark,bar[DownQuark]};
CombinationGenerations = {{2,1,1,1}, {3,1,1,1}, {3,2,1,1},
                          {2,1,2,2}, {3,1,2,2}, {3,2,2,2}};


AllOperators={{OllddSLL,Op[7].Op[7]},
              {OllddSRR,Op[6].Op[6]},
              {OllddSRL,Op[6].Op[7]},
              {OllddSLR,Op[7].Op[6]},

              {OllddVRR,Op[7,Lor[1]].Op[7,Lor[1]]},
              {OllddVLL,Op[6,Lor[1]].Op[6,Lor[1]]},
              {OllddVRL,Op[7,Lor[1]].Op[6,Lor[1]]},
              {OllddVLR,Op[6,Lor[1]].Op[7,Lor[1]]},

              {OllddTLL,Op[-7,Lor[1],Lor[2]].Op[-7,Lor[1],Lor[2]]},
              {OllddTLR,Op[-7,Lor[1],Lor[2]].Op[-6,Lor[1],Lor[2]]},
              {OllddTRL,Op[-6,Lor[1],Lor[2]].Op[-7,Lor[1],Lor[2]]},
              {OllddTRR,Op[-6,Lor[1],Lor[2]].Op[-6,Lor[1],Lor[2]]}
}; 
\end{lstlisting}

\begin{lstlisting}[caption=2L2u.m]
NameProcess="2L2u";

ConsideredProcess = "4Fermion";
FermionOrderExternal={2,1,4,3};
NeglectMasses={1,2,3,4};  
 

ExternalFields= {ChargedLepton,bar[ChargedLepton],UpQuark,bar[UpQuark]};
CombinationGenerations = {{2,1,1,1},{3,1,1,1},{3,2,1,1}};



AllOperators={{OlluuSLL,Op[7].Op[7]},
              {OlluuSRR,Op[6].Op[6]},
              {OlluuSRL,Op[6].Op[7]},
              {OlluuSLR,Op[7].Op[6]},

              {OlluuVRR,Op[7,Lor[1]].Op[7,Lor[1]]},
              {OlluuVLL,Op[6,Lor[1]].Op[6,Lor[1]]},
              {OlluuVRL,Op[7,Lor[1]].Op[6,Lor[1]]},
              {OlluuVLR,Op[6,Lor[1]].Op[7,Lor[1]]},

              {OlluuTLL,Op[-7,Lor[1],Lor[2]].Op[-7,Lor[1],Lor[2]]},
              {OlluuTLR,Op[-7,Lor[1],Lor[2]].Op[-6,Lor[1],Lor[2]]},
              {OlluuTRL,Op[-6,Lor[1],Lor[2]].Op[-7,Lor[1],Lor[2]]},
              {OlluuTRR,Op[-6,Lor[1],Lor[2]].Op[-6,Lor[1],Lor[2]]}
}; 
\end{lstlisting}


\begin{center}
$(\bar{d} \Gamma d) (\bar{d} \Gamma^\prime d)$ and $(\bar{\ell} \Gamma \ell) (\bar{\ell} \Gamma^\prime \ell)$
\end{center}

\begin{center}
\begin{tabular}{|ccc||ccc|}
\hline 
Variable & Operator &  Name & Variable & Operator & Name \\
\hline 
O4dSLL & $(\bar{d} P_L d) (\bar{d} P_L d)$ & $D_{LL}^S$                    &     O4lSLL & $(\bar{\ell} P_L \ell) (\bar{\ell} P_L \ell)$ & $A_{LL}^S$           \\
O4dSRR & $(\bar{d} P_R d) (\bar{d} P_R d)$ & $D_{RR}^S$                    &     O4lSRR & $(\bar{\ell} P_R \ell) (\bar{\ell} P_R \ell)$ & $A_{RR}^S$           \\
O4dSLR & $(\bar{d} P_L d) (\bar{d} P_R d)$ & $D_{LR}^S$                    &     O4lSLR & $(\bar{\ell} P_L \ell) (\bar{\ell} P_R \ell)$ & $A_{LR}^S$           \\
O4dSRL & $(\bar{d} P_R d) (\bar{d} P_L d)$ & $D_{RL}^S$                    &     O4lSRL & $(\bar{\ell} P_R \ell) (\bar{\ell} P_L \ell)$ & $A_{RL}^S$           \\
\hline                                                                          
O4dVLL & $(\bar{d} \gamma_\mu P_L d) (\bar{d} \gamma^\mu P_L d)$ & $D_{LL}^V$     &     O4lVLL & $(\bar{\ell} \gamma_\mu P_L \ell) (\bar{\ell} \gamma^\mu P_L \ell)$ & $A_{LL}^V$    \\
O4dVRR & $(\bar{d} \gamma_\mu P_R d) (\bar{d} \gamma^\mu P_R d)$ & $D_{RR}^V$     &     O4lVRR & $(\bar{\ell} \gamma_\mu P_R \ell) (\bar{\ell} \gamma^\mu P_R \ell)$ & $A_{RR}^V$    \\
O4dVLR & $(\bar{d} \gamma_\mu P_L d) (\bar{d} \gamma^\mu P_R d)$ & $D_{LR}^V$     &     O4lVLR & $(\bar{\ell} \gamma_\mu P_L \ell) (\bar{\ell} \gamma^\mu P_R \ell)$ & $A_{LR}^V$    \\
O4dVRL & $(\bar{d} \gamma_\mu P_R d) (\bar{d} \gamma^\mu P_L d)$ & $D_{RL}^V$     &     O4lVRL & $(\bar{\ell} \gamma_\mu P_R \ell) (\bar{\ell} \gamma^\mu P_L \ell)$ & $A_{RL}^V$    \\
\hline                                                                          
O4dTLL & $(\bar{d} \sigma_{\mu\nu} P_L d) (\bar{d} \sigma^{\mu\nu} P_L d)$ & $D_{LL}^T$    &    O4lTLL & $(\bar{\ell} \sigma_{\mu\nu} P_L \ell) (\bar{\ell} \sigma^{\mu\nu} P_L \ell)$ & $A_{LL}^T$ \\
O4dTRR & $(\bar{d} \sigma_{\mu\nu} P_R d) (\bar{d} \sigma^{\mu\nu} P_R d)$ & $D_{RR}^T$    &    O4lTRR & $(\bar{\ell} \sigma_{\mu\nu} P_R \ell) (\bar{\ell} \sigma^{\mu\nu} P_R \ell)$ & $A_{RR}^T$ \\
O4dTLR & $(\bar{d} \sigma_{\mu\nu} P_L d) (\bar{d} \sigma^{\mu\nu} P_R d)$ & $D_{LR}^T$    &    O4lTLR & $(\bar{\ell} \sigma_{\mu\nu} P_L \ell) (\bar{\ell} \sigma^{\mu\nu} P_R \ell)$ & $A_{LR}^T$ \\
O4dTRL & $(\bar{d} \sigma_{\mu\nu} P_R d) (\bar{d} \sigma^{\mu\nu} P_L d)$ & $D_{RL}^T$    &    O4lTRL & $(\bar{\ell} \sigma_{\mu\nu} P_R \ell) (\bar{\ell} \sigma^{\mu\nu} P_L \ell)$ & $A_{RL}^T$ \\
\hline 
\end{tabular}
\end{center}

\begin{lstlisting}[caption=4d.m]
NameProcess="4d";

ConsideredProcess = "4Fermion";
FermionOrderExternal={2,1,4,3};
NeglectMasses={1,2,3,4};  


ExternalFields= {DownQuark,bar[DownQuark],DownQuark,bar[DownQuark]};

ColorFlow = ColorDelta[1,2] ColorDelta[3,4];

CombinationGenerations = {{3,1,3,1},{3,2,3,2},{2,1,2,1}};


AllOperators={{O4dSLL,Op[7].Op[7]},
              {O4dSRR,Op[6].Op[6]},
              {O4dSRL,Op[6].Op[7]},
              {O4dSLR,Op[7].Op[6]},

              {O4dVRR,Op[7,Lor[1]].Op[7,Lor[1]]},
              {O4dVLL,Op[6,Lor[1]].Op[6,Lor[1]]},
              {O4dVRL,Op[7,Lor[1]].Op[6,Lor[1]]},
              {O4dVLR,Op[6,Lor[1]].Op[7,Lor[1]]},

              {O4dTLL,Op[-7,Lor[1],Lor[2]].Op[-7,Lor[1],Lor[2]]},
              {O4dTLR,Op[-7,Lor[1],Lor[2]].Op[-6,Lor[1],Lor[2]]},
              {O4dTRL,Op[-6,Lor[1],Lor[2]].Op[-7,Lor[1],Lor[2]]},
              {O4dTRR,Op[-6,Lor[1],Lor[2]].Op[-6,Lor[1],Lor[2]]}
};

Filters = {NoPenguins}; 
\end{lstlisting}

\begin{lstlisting}[caption=4L.m]
NameProcess="4L";

ConsideredProcess = "4Fermion";
FermionOrderExternal={2,1,4,3};
NeglectMasses={1,2,3,4};  

ExternalFields= {ChargedLepton,bar[ChargedLepton],ChargedLepton,bar[ChargedLepton]};
CombinationGenerations = {{2,1,1,1},{3,1,1,1},{3,2,2,2}};


AllOperators={{O4lSLL,Op[7].Op[7]},
              {O4lSRR,Op[6].Op[6]},
              {O4lSRL,Op[6].Op[7]},
              {O4lSLR,Op[7].Op[6]},

              {O4lVRR,Op[7,Lor[1]].Op[7,Lor[1]]},
              {O4lVLL,Op[6,Lor[1]].Op[6,Lor[1]]},
              {O4lVRL,Op[7,Lor[1]].Op[6,Lor[1]]},
              {O4lVLR,Op[6,Lor[1]].Op[7,Lor[1]]},

              {O4lTLL,Op[-7,Lor[1],Lor[2]].Op[-7,Lor[1],Lor[2]]},
              {O4lTLR,Op[-7,Lor[1],Lor[2]].Op[-6,Lor[1],Lor[2]]},
              {O4lTRL,Op[-6,Lor[1],Lor[2]].Op[-7,Lor[1],Lor[2]]},
              {O4lTRR,Op[-6,Lor[1],Lor[2]].Op[-6,Lor[1],Lor[2]]}
};

Filters = {NoCrossedDiagrams}; 
\end{lstlisting}


\begin{center}
$(\bar{d} \Gamma u) (\bar{\ell} \Gamma^\prime \nu)$
\end{center}

\begin{center}
\begin{tabular}{|ccc||ccc|}
\hline 
Variable & Operator & Name & Variable & Operator & Name \\
\hline 
OdulvVLL & $(\bar{d} \gamma_\mu P_L u) (\bar{\ell} \gamma^\mu P_L \nu)$ & $G_{LL}^V$    &  OdulvSLL & $(\bar{d} P_L u) (\bar{\ell} P_L \nu)$ & $G_{LL}^S$ \\
OdulvVRR & $(\bar{d} \gamma_\mu P_R u) (\bar{\ell} \gamma^\mu P_R \nu)$ & $G_{RR}^V$    &  OdulvSRR & $(\bar{d} P_R u) (\bar{\ell} P_R \nu)$ & $G_{RR}^S$ \\
OdulvVLR & $(\bar{d} \gamma_\mu P_L u) (\bar{\ell} \gamma^\mu P_R \nu)$ & $G_{LR}^V$    &  OdulvSLR & $(\bar{d} P_L u) (\bar{\ell} P_R \nu)$ & $G_{LR}^S$ \\
OdulvVRL & $(\bar{d} \gamma_\mu P_R u) (\bar{\ell} \gamma^\mu P_L \nu)$ & $G_{RL}^V$    &  OdulvSRL & $(\bar{d} P_R u) (\bar{\ell} P_L \nu)$ & $G_{RL}^S$ \\
\hline
\end{tabular}
\end{center}

\begin{lstlisting}[caption=du\_lv.m]
NameProcess="dulv";

ConsideredProcess = "4Fermion";
FermionOrderExternal={2,1,3,4};
NeglectMasses={1,2,3,4};  


ExternalFields= {DownQuark,bar[UpQuark], Neutrino, bar[ChargedLepton]};

CombinationGenerations = 
  Flatten[Table[{{3,1,i,j},{3,2,i,j},{2,2,i,j},{2,1,i,j}},{i,1,3},{j,1,3}],2];

Clear[i,j];


AllOperators={{OdulvSLL,Op[7].Op[7]},
              {OdulvSRR,Op[6].Op[6]},
              {OdulvSRL,Op[6].Op[7]},
              {OdulvSLR,Op[7].Op[6]},

              {OdulvVRR,Op[7,Lor[1]].Op[7,Lor[1]]},
              {OdulvVLL,Op[6,Lor[1]].Op[6,Lor[1]]},
              {OdulvVRL,Op[7,Lor[1]].Op[6,Lor[1]]},
              {OdulvVLR,Op[6,Lor[1]].Op[7,Lor[1]]}
};

Filters = {NoBoxes, NoPenguins}; 
\end{lstlisting}


\section{Application: Flavor observables implemented in \SARAH}
\subsection{Lepton flavor observables}
\label{sec:LFV}
Lepton flavor violation in the SM or MSSM without neutrino masses vanishes
exactly. Even adding Dirac neutrino masses to the SM predicts LFV
rates which are far beyond the experimental reach. However, many
extensions of the SM can introduce new sources for LFV of a size which
is testable nowadays. The best-known examples are SUSY and non-SUSY
models with high- or low-scale seesaw mechanism, models with
vector-like leptons and SUSY models with $R$-parity violation, see for
instance
Refs.~\cite{Hisano:1995cp,Deppisch:2002vz,Arganda:2005ji,Petcov:2005yh,Antusch:2006vw,Paradisi:2006jp,Hirsch:2008dy,Hirsch:2008gh,Borzumati:2009hu,Gross:2010ce,Biggio:2010me,Esteves:2010ff,Abada:2010kj,Abada:2011mg,Dreiner:2012mx,Cely:2012bz,Hirsch:2012yv,Davidson:2012ds,Hirsch:2012kv,Dinh:2012bp,Cannoni:2013gq,Krauss:2013gya,Arana-Catania:2013xma,Altmannshofer:2013zba,Crivellin:2013hpa,Celis:2013xja,Moroi:2013vya,Dinh:2013vya,Falkowski:2013jya,Toma:2013zsa,Benakli:2014cia,Teixeira:2014jza,Celis:2014asa,Crivellin:2014cta}.

We discuss in the following the implementation of the most important
LFV observables in \SARAH and \SPheno using the previously defined
operators which are calculated by \SPheno.

\subsubsection{$\boldsymbol{\ell_\alpha \to \ell_\beta \gamma}$}
The decay width is given by~\cite{Hisano:1995cp}
\begin{equation}
\Gamma \left( \ell_\alpha \to \ell_\beta \gamma \right) = \frac{\alpha m_{\ell_\alpha}^5}{4} \left( |K_2^L|^2 + |K_2^R|^2 \right) \, ,
\end{equation}
where $\alpha$ is the fine structure constant and the dipole Wilson
coefficients $K_2^{L,R}$ are defined in Eq.\eqref{eq:L-llg}.

\begin{lstlisting}[caption=LLgGamma.m]
NameProcess = "LLpGamma";
NameObservables = {{muEgamma, 701, "BR(mu->e gamma)"}, 
                   {tauEgamma, 702, "BR(tau->e gamma)"}, 
                   {tauMuGamma, 703, "BR(tau->mu gamma)"}};

NeededOperators = {K2L, K2R};

Body = "LLpGamma.f90"; 
\end{lstlisting}

\begin{lstlisting}[caption=LLgGamma.f90]
Real(dp) :: width
Integer :: i1, gt1, gt2

! ---------------------------------------------------------------- 
! l -> l' gamma
! Observable implemented by W. Porod, F. Staub and A. Vicente
! Based on J. Hisano et al, PRD 53 (1996) 2442 [hep-ph/9510309]
! ---------------------------------------------------------------- 

Do i1=1,3 

If (i1.eq.1) Then         ! mu -> e gamma
 gt1 = 2
 gt2 = 1
Elseif (i1.eq.2) Then     !tau -> e gamma
 gt1 = 3
 gt2 = 1
Else                      ! tau -> mu gamma
 gt1 = 3
 gt2 = 2
End if

width=0.25_dp*mf_l(gt1)**5*(Abs(K2L(gt1,gt2))**2 &
           & +Abs(K2R(gt1,gt2))**2)*Alpha

If (i1.eq.1) Then
muEgamma = width/(width+GammaMu)
Elseif (i1.eq.2) Then 
tauEgamma = width/(width+GammaTau)
Else
tauMuGamma = width/(width+GammaTau)
End if

End do
\end{lstlisting}


\subsubsection{$\boldsymbol{\ell_\alpha \to 3 \ell_\beta}$}
The decay width is given by
\begin{eqnarray}
\Gamma \left( \ell_\alpha \to 3 \ell_\beta \right) &=& \frac{m_{\ell_\alpha}^5}{512 \pi^3} \left[ e^4 \, \left( \left| K_2^L \right|^2 + \left| K_2^R \right|^2 \right) \left( \frac{16}{3} \log{\frac{m_{\ell_\alpha}}{m_{\ell_\beta}}} - \frac{22}{3} \right) \right. \label{L3Lwidth} \\
&+& \frac{1}{24} \left( \left| A_{LL}^S \right|^2 + \left| A_{RR}^S \right|^2 \right) + \frac{1}{12} \left( \left| A_{LR}^S \right|^2 + \left| A_{RL}^S \right|^2 \right) \nonumber \\
&+& \frac{2}{3} \left( \left| \hat A_{LL}^V \right|^2 + \left| \hat A_{RR}^V \right|^2 \right) + \frac{1}{3} \left( \left| \hat A_{LR}^V \right|^2 + \left| \hat A_{RL}^V \right|^2 \right) + 6 \left( \left| A_{LL}^T \right|^2 + \left| A_{RT}^T \right|^2 \right) \nonumber \\
&+& \frac{e^2}{3} \left( K_2^L A_{RL}^{S \ast} + K_2^R A_{LR}^{S \ast} + c.c. \right) - \frac{2 e^2}{3} \left( K_2^L \hat A_{RL}^{V \ast} + K_2^R \hat A_{LR}^{V \ast} + c.c. \right) \nonumber \\
&-& \frac{4 e^2}{3} \left( K_2^L \hat A_{RR}^{V \ast} + K_2^R \hat A_{LL}^{V \ast} + c.c. \right) \nonumber \\
&-& \left. \frac{1}{2} \left( A_{LL}^S A_{LL}^{T \ast} + A_{RR}^S A_{RR}^{T \ast} + c.c. \right) - \frac{1}{6} \left( A_{LR}^S \hat A_{LR}^{V \ast} + A_{RL}^S \hat A_{RL}^{V \ast} + c.c. \right) \right]  \nonumber \, .
\end{eqnarray}
Here we have defined
\begin{equation}
\hat A_{XY}^V = A_{XY}^V + e^2 K_1^X \qquad \left( X,Y = L,R \right) \, .
\end{equation}
The mass of the leptons in the final state has been neglected in this
formula, with the exception of the dipole terms $K_2^{L,R}$, where an
infrared divergence would otherwise occur due to the massless photon
propagator. Eq.\eqref{L3Lwidth} is in agreement with
\cite{Arganda:2005ji}, but also includes the coefficients $A_{LR}^S$
and $A_{RL}^S$.

\begin{lstlisting}[caption=Lto3Lp.m]
NameProcess = "Lto3Lp";
NameObservables = {{BRmuTo3e, 901, "BR(mu->3e)"}, 
                   {BRtauTo3e, 902, "BR(tau->3e)"}, 
                   {BRtauTo3mu, 903, "BR(tau->3mu)"}
                  };

ExternalStates =  {Electron}; 
NeededOperators = {K1L, K1R, K2L, K2R, 
 O4lSLL, O4lSRR, O4lSRL, O4lSLR , 
 O4lVRR, O4lVLL, O4lVRL, O4lVLR ,
 O4lTLL, O4lTRR };

Body = "Lto3Lp.f90"; 
\end{lstlisting}

\begin{lstlisting}[caption=Lto3Lp.f90]
Complex(dp) :: cK1L, cK1R, cK2L, cK2R
Complex(dp) :: CSLL, CSRR, CSLR, CSRL, CVLL, &
                    & CVRR, CVLR, CVRL, CTLL, CTRR
Real(dp) :: BRdipole, BRscalar, BRvector, BRtensor
Real(dp) :: BRmix1, BRmix2, BRmix3, BRmix4, GammaLFV
Real(dp) :: e2, e4
Integer :: i1, gt1, gt2, gt3, gt4

! ---------------------------------------------------------------- 
! l -> 3 l'
! Observable implemented by W. Porod, F. Staub and A. Vicente
! ---------------------------------------------------------------- 

e2 = (4._dp*Pi*Alpha_MZ)
e4 = e2**2

Do i1=1,3 

If (i1.eq.1) Then
 gt1 = 2
 gt2 = 1
Elseif (i1.eq.2) Then
 gt1 = 3
 gt2 = 1
Else 
 gt1 = 3
 gt2 = 2
End if
gt3 = gt2
gt4 = gt2

cK1L = K1L(gt1,gt2)
cK1R = K1R(gt1,gt2)

cK2L = K2L(gt1,gt2)
cK2R = K2R(gt1,gt2)

CSLL = O4lSLL(gt1,gt2,gt3,gt4)
CSRR = O4lSRR(gt1,gt2,gt3,gt4)
CSLR = O4lSLR(gt1,gt2,gt3,gt4)
CSRL = O4lSRL(gt1,gt2,gt3,gt4)

CVLL = O4lVLL(gt1,gt2,gt3,gt4)
CVRR = O4lVRR(gt1,gt2,gt3,gt4)
CVLR = O4lVLR(gt1,gt2,gt3,gt4)
CVRL = O4lVRL(gt1,gt2,gt3,gt4)

CVLL = CVLL + e2*cK1L
CVRR = CVRR + e2*cK1R
CVLR = CVLR + e2*cK1L
CVRL = CVRL + e2*cK1R

CTLL = O4lTLL(gt1,gt2,gt3,gt4)
CTRR = O4lTRR(gt1,gt2,gt3,gt4)

! Photonic dipole contributions
BRdipole = (Abs(cK2L)**2+Abs(cK2R)**2)&
&*(16._dp*Log(mf_l(gt1)/mf_l(gt2))-22._dp)/3._dp

! Scalar contributions
BRscalar = (Abs(CSLL)**2+Abs(CSRR)**2)/24._dp&
&+(Abs(CSLR)**2+Abs(CSRL)**2)/12._dp

! Vector contributions
BRvector = 2._dp*(Abs(CVLL)**2+Abs(CVRR)**2)/3._dp&
&+(Abs(CVLR)**2+Abs(CVRL)**2)/3._dp

! Tensor contributions
BRtensor = 6._dp*(Abs(CTLL)**2+Abs(CTRR)**2)

! Mix: dipole x scalar
BRmix1 = 2._dp/3._dp*Real(cK2L*Conjg(CSRL) + cK2R*Conjg(CSLR),dp)

! Mix: dipole x vector
BRmix2 = -4._dp/3._dp*Real(cK2L*Conjg(CVRL) + cK2R*Conjg(CVLR),dp) &
     & -8._dp/3._dp*Real(cK2L*Conjg(CVRR) + cK2R*Conjg(CVLL),dp)

! Mix: scalar x vector
BRmix3 = -1._dp/3._dp*Real(CSLR*Conjg(CVLR) + CSRL*Conjg(CVRL),dp)

! Mix: scalar x tensor
BRmix4 = -1._dp*Real(CSLL*Conjg(CTLL) + CSRR*Conjg(CTRR),dp)

GammaLFV = oo512pi3*mf_l(gt1)**5* &
     & (e4*BRdipole + BRscalar + BRvector + BRtensor &
     & + e2*BRmix1 + e2*BRmix2 + BRmix3 + BRmix4)

!----------------------------------------------------------------------
!taking alpha(Q=0) instead of alpha(m_Z) as this contains most of the
!running of the Wilson coefficients
!----------------------------------------------------------------------

If (i1.Eq.1) Then
 BRmuTo3e=GammaLFV/GammaMu
Else If (i1.Eq.2) Then
 BRtauTo3e=GammaLFV/GammaTau
Else 
 BRtauTo3mu=GammaLFV/GammaTau
End If
End do
\end{lstlisting}


\subsubsection{Coherent $\boldsymbol{\mu-e}$ conversion in nuclei}
The conversion rate, relative to the the muon capture rate, can be
expressed as \cite{Kuno:1999jp,Arganda:2007jw}
\begin{align}
{\rm CR} (\mu- e, {\rm Nucleus}) &= 
\frac{p_e \, E_e \, m_\mu^3 \, G_F^2 \, \alpha^3 
\, Z_{\rm eff}^4 \, F_p^2}{8 \, \pi^2 \, Z}  \nonumber \\
&\times \left\{ \left| (Z + N) \left( g_{LV}^{(0)} + g_{LS}^{(0)} \right) + 
(Z - N) \left( g_{LV}^{(1)} + g_{LS}^{(1)} \right) \right|^2 + 
\right. \nonumber \\
& \ \ \ 
 \ \left. \,\, \left| (Z + N) \left( g_{RV}^{(0)} + g_{RS}^{(0)} \right) + 
(Z - N) \left( g_{RV}^{(1)} + g_{RS}^{(1)} \right) \right|^2 \right\} 
\frac{1}{\Gamma_{\rm capt}}\,.
\end{align}   
$Z$ and $N$ are the number of protons and neutrons in the nucleus and
$Z_{\rm eff}$ is the effective atomic charge~\cite{Chiang:1993xz}.
Similarly, $G_F$ is the Fermi constant, $F_p$ is the nuclear matrix
element and $\Gamma_{\rm capt}$ represents the total muon capture
rate. $\alpha$ is the fine structure constant, $p_e$ and $E_e$ (
$\simeq m_\mu$ in the numerical evaluation) are the momentum and
energy of the electron and $m_\mu$ is the muon mass.  In the above,
$g_{XK}^{(0)}$ and $g_{XK}^{(1)}$ (with $X = L, R$ and $K = S, V$) can
be written in terms of effective couplings at the quark level as
\begin{align}
g_{XK}^{(0)} &= \frac{1}{2} \sum_{q = u,d,s} \left( g_{XK(q)} G_K^{(q,p)} +
g_{XK(q)} G_K^{(q,n)} \right)\,, \nonumber \\
g_{XK}^{(1)} &= \frac{1}{2} \sum_{q = u,d,s} \left( g_{XK(q)} G_K^{(q,p)} - 
g_{XK(q)} G_K^{(q,n)} \right)\,.
\end{align}
For coherent $\mu-e$ conversion in nuclei, only scalar ($S$) and
vector ($V$) couplings contribute. Furthermore, sizable contributions
are expected only from the $u,d,s$ quark flavors. The numerical values
of the relevant $G_K$ factors are~\cite{Kuno:1999jp,Kosmas:2001mv}
\begin{align}
&
G_V^{(u, p)}\, =\, G_V^{(d, n)\,} =\, 2 \,;\, \ \ \ \ 
G_V^{(d, p)}\, =\, G_V^{(u, n)}\, = 1\,; \nonumber \\
&
G_S^{(u, p)}\, =\, G_S^{(d, n)}\, =\, 5.1\,;\, \ \ 
G_S^{(d, p)}\, =\, G_S^{(u, n)}\, = \,4.3 \,;\, \nonumber \\
&
G_S^{(s, p)}\,=\, G_S^{(s, n)}\, = \,2.5\,.
\end{align}
Finally, the $g_{XK(q)}$ coefficients can be written in terms of the
Wilson coefficients in Eqs.\eqref{eq:L-llg}, \eqref{eq:L-2L2D} and
\eqref{eq:L-2L2U} as
\begin{eqnarray}
g_{LV(q)} &=& \frac{\sqrt{2}}{G_F} \left[ e^2 Q_q \left( K_1^L - K_2^R \right)- \frac{1}{2} \left( C_{\ell\ell qq}^{VLL} + C_{\ell\ell qq}^{VLR} \right) \right] \\
g_{RV(q)} &=& \left. g_{LV(q)} \right|_{L \to R} \\ 
g_{LS(q)} &=& - \frac{\sqrt{2}}{G_F} \frac{1}{2} \left( C_{\ell\ell qq}^{SLL} + C_{\ell\ell qq}^{SLR} \right) \\
g_{RS(q)} &=& \left. g_{LS(q)} \right|_{L \to R} \, .
\end{eqnarray}
Here $Q_q$ is the quark electric charge ($Q_d = -1/3$, $Q_u = 2/3$)
and $C_{\ell\ell qq}^{IXK} = B_{XY}^K \, \left( C_{XY}^K \right)$ for
d-quarks (u-quarks), with $X = L, R$ and $K = S, V$.

\begin{lstlisting}[caption=MuEconversion.m]
NameProcess = "MuEconversion";
NameObservables = {{CRmuEAl, 800, "CR(mu-e, Al)"}, 
                   {CRmuETi, 801, "CR(mu-e, Ti)"}, 
                   {CRmuESr, 802, "CR(mu-e, Sr)"}, 
                   {CRmuESb, 803, "CR(mu-e, Sb)"}, 
                   {CRmuEAu, 804, "CR(mu-e, Au)"}, 
                   {CRmuEPb, 805, "CR(mu-e, Pb)"}
                  };

NeededOperators = {K1L, K1R, K2L, K2R, 
  OllddSLL, OllddSRR, OllddSRL, OllddSLR, OllddVRR, OllddVLL, 
  OllddVRL, OllddVLR, OllddTLL, OllddTLR, OllddTRL, OllddTRR,
  OlluuSLL, OlluuSRR, OlluuSRL, OlluuSLR, OlluuVRR, OlluuVLL,
  OlluuVRL, OlluuVLR, OlluuTLL, OlluuTLR, OlluuTRL, OlluuTRR
};

Body = "MuEconversion.f90"; 
\end{lstlisting}

\begin{lstlisting}[caption=MuEconversion.f90]
Complex(dp) :: gPLV(3), gPRV(3)
Complex(dp),Parameter :: mat0(3,3)=0._dp 
Real(dp) :: Znuc,Nnuc, Zeff, Fp, GammaCapt, GSp(3), GSn(3), &
     & GVp(3), GVn(3), e2
Complex(dp) :: Lcont,Rcont,gLS(3),gRS(3),gLV(3),gRV(3),g0LS,g0RS, & 
     & g0LV,g0RV,g1LS,g1RS,g1LV,g1RV 
Integer :: i1, i2

! ---------------------------------------------------------------- 
! Coherent mu-e conversion in nuclei
! Observable implemented by W. Porod, F. Staub and A. Vicente
! Based on Y. Kuno, Y. Okada, Rev. Mod. Phys. 73 (2001) 151 [hep-ph/9909265]
! and E. Arganda et al, JHEP 0710 (2007) 104 [arXiv:0707.2955]
! ---------------------------------------------------------------- 

e2 = 4._dp*Pi*Alpha_MZ

! 1: uu
! 2: dd
! 3: ss

! vector couplings

gLV(1) = 0.5_dp*(OlluuVLL(2,1,1,1) + OlluuVLR(2,1,1,1))
gRV(1) = 0.5_dp*(OlluuVRL(2,1,1,1) + OlluuVRR(2,1,1,1))
gLV(2) = 0.5_dp*(OllddVLL(2,1,1,1) + OllddVLR(2,1,1,1))
gRV(2) = 0.5_dp*(OllddVRL(2,1,1,1) + OllddVRR(2,1,1,1))
gLV(3) = 0.5_dp*(OllddVLL(2,1,2,2) + OllddVLR(2,1,2,2))
gRV(3) = 0.5_dp*(OllddVRL(2,1,2,2) + OllddVRR(2,1,2,2))

gLV = -gLV*Sqrt(2._dp)/G_F
gRV = -gRV*Sqrt(2._dp)/G_F

gPLV(1) = (K1L(2,1)-K2R(2,1))*(2._dp/3._dp) 
gPRV(1) = (K1R(2,1)-K2L(2,1))*(2._dp/3._dp) 
gPLV(2) = (K1L(2,1)-K2R(2,1))*(-1._dp/3._dp)  
gPRV(2) = (K1R(2,1)-K2L(2,1))*(-1._dp/3._dp)  
gPLV(3) = (K1L(2,1)-K2R(2,1))*(-1._dp/3._dp)  
gPRV(3) = (K1R(2,1)-K2L(2,1))*(-1._dp/3._dp) 
gPLV = gPLV*Sqrt(2._dp)/G_F*e2
gPRV = gPRV*Sqrt(2._dp)/G_F*e2

gLV=gPLV+gLV
gRV=gPRV+gRV


! scalar couplings

gLS(1) = 0.5_dp*(OlluuSLL(2,1,1,1)+OlluuSLR(2,1,1,1))
gRS(1) = 0.5_dp*(OlluuSRL(2,1,1,1)+OlluuSRR(2,1,1,1))
gLS(2) = 0.5_dp*(OllddSLL(2,1,1,1)+OllddSLR(2,1,1,1))
gRS(2) = 0.5_dp*(OllddSRL(2,1,1,1)+OllddSRR(2,1,1,1))
gLS(3) = 0.5_dp*(OllddSLL(2,1,2,2)+OllddSLR(2,1,2,2))
gRS(3) = 0.5_dp*(OllddSRL(2,1,2,2)+OllddSRR(2,1,2,2))

gLS = -gLS*Sqrt(2._dp)/G_F
gRS = -gRS*Sqrt(2._dp)/G_F


Do i1=1,6 
 If(i1.eq.1) Then 
Znuc=13._dp 
Nnuc=14._dp 
Zeff=11.5_dp 
Fp=0.64_dp 
GammaCapt=4.64079e-19_dp 
Else If(i1.eq.2) Then 
Znuc=22._dp 
Nnuc=26._dp 
Zeff=17.6_dp 
Fp=0.54_dp 
GammaCapt=1.70422e-18_dp 
Else If(i1.eq.3) Then 
Znuc=38._dp 
Nnuc=42._dp 
Zeff=25.0_dp 
Fp=0.39_dp 
GammaCapt=4.61842e-18_dp 
Else If(i1.eq.4) Then 
Znuc=51._dp 
Nnuc=70._dp 
Zeff=29.0_dp 
Fp=0.32_dp 
GammaCapt=6.71711e-18_dp 
Else If(i1.eq.5) Then 
Znuc=79._dp 
Nnuc=118._dp 
Zeff=33.5_dp 
Fp=0.16_dp 
GammaCapt=8.59868e-18_dp 
Else If(i1.eq.6) Then 
Znuc=82._dp 
Nnuc=125._dp 
Zeff=34.0_dp 
Fp=0.15_dp 
GammaCapt=8.84868e-18_dp 
End If 

! numerical values 
! based on Y. Kuno, Y. Okada, Rev. Mod. Phys. 73 (2001) 151 [hep-ph/9909265]
! and T. S. Kosmas et al, PLB 511 (2001) 203 [hep-ph/0102101]
GSp=(/5.1,4.3,2.5/) 
GSn=(/4.3,5.1,2.5/) 
GVp=(/2.0,1.0,0.0/) 
GVn=(/1.0,2.0,0.0/) 

g0LS=0._dp 
g0RS=0._dp 
g0LV=0._dp 
g0RV=0._dp 
g1LS=0._dp 
g1RS=0._dp 
g1LV=0._dp 
g1RV=0._dp 
Do i2=1,3
g0LS=g0LS+0.5_dp*gLS(i2)*(GSp(i2)+GSn(i2))
g0RS=g0RS+0.5_dp*gRS(i2)*(GSp(i2)+GSn(i2))
g0LV=g0LV+0.5_dp*gLV(i2)*(GVp(i2)+GVn(i2))
g0RV=g0RV+0.5_dp*gRV(i2)*(GVp(i2)+GVn(i2))
g1LS=g1LS+0.5_dp*gLS(i2)*(GSp(i2)-GSn(i2))
g1RS=g1RS+0.5_dp*gRS(i2)*(GSp(i2)-GSn(i2))
g1LV=g1LV+0.5_dp*gLV(i2)*(GVp(i2)-GVn(i2))
g1RV=g1RV+0.5_dp*gRV(i2)*(GVp(i2)-GVn(i2))
End Do
Lcont=(Znuc+Nnuc)*(g0LV+g0LS)+(Znuc-Nnuc)*(g1LV-g1LS) 
Rcont=(Znuc+Nnuc)*(g0RV+g0RS)+(Znuc-Nnuc)*(g1RV-g1RS)

! Conversion rate
If (i1.eq.1) Then
 CRMuEAl =oo8pi2*mf_l(2)**5*G_F**2*Alpha**3*Zeff**4*Fp**2/Znuc*& 
   & (Abs(Lcont)**2+Abs(Rcont)**2)/GammaCapt 
Else if (i1.eq.2) Then 
 CRMuETi =oo8pi2*mf_l(2)**5*G_F**2*Alpha**3*Zeff**4*Fp**2/Znuc*& 
   & (Abs(Lcont)**2+Abs(Rcont)**2)/GammaCapt 
Else if (i1.eq.3) Then 
 CRMuESr =oo8pi2*mf_l(2)**5*G_F**2*Alpha**3*Zeff**4*Fp**2/Znuc*& 
   & (Abs(Lcont)**2+Abs(Rcont)**2)/GammaCapt 
Else if (i1.eq.4) Then 
 CRMuESb =oo8pi2*mf_l(2)**5*G_F**2*Alpha**3*Zeff**4*Fp**2/Znuc*& 
   & (Abs(Lcont)**2+Abs(Rcont)**2)/GammaCapt 
Else if (i1.eq.5) Then 
 CRMuEAu =oo8pi2*mf_l(2)**5*G_F**2*Alpha**3*Zeff**4*Fp**2/Znuc*& 
   & (Abs(Lcont)**2+Abs(Rcont)**2)/GammaCapt 
Else if (i1.eq.6) Then 
 CRMuEPb =oo8pi2*mf_l(2)**5*G_F**2*Alpha**3*Zeff**4*Fp**2/Znuc*& 
   & (Abs(Lcont)**2+Abs(Rcont)**2)/GammaCapt 
End if
End do 
\end{lstlisting}


\subsubsection{$\boldsymbol{\tau \to P \ell}$}

Our analytical expressions for $\tau \to P \ell$, where $\ell = e,
\mu$ and $P$ is a pseudoscalar meson, generalize the results in
\cite{Arganda:2008jj}. The decay width is given by
\begin{equation}
\Gamma \left( \tau \to \ell P \right) = \frac{1}{4 \pi} \frac{\lambda^{1/2}(m_\tau^2,m_\ell^2,m_P^2)}{m_\tau^2} \frac{1}{2} \sum_{i,f} |\mathcal{M}_{\tau \ell P}|^2 \, ,
\end{equation}
where the averaged squared amplitude can be written as
\begin{equation}
\frac{1}{2} \sum_{i,f} |\mathcal{M}_{\tau \ell P}|^2 = \frac{1}{4 m_\tau} \sum_{I,J = S,V} \left[ 2 m_\tau m_\ell \left( a_P^I a_P^{J \, \ast} - b_P^I b_P^{J \, \ast} \right) + (m_\tau^2 + m_\ell^2 - m_P^2) \left( a_P^I a_P^{J \, \ast} + b_P^I b_P^{J \, \ast} \right) \right] \, .
\end{equation}
The coefficients $a_P^{S,V}$ and $b_P^{S,V}$ can be expressed in terms
of the Wilson coefficients in Eqs.\eqref{eq:L-2L2D} and
\eqref{eq:L-2L2U} as
\begin{eqnarray}
a_P^S &=& \frac{1}{2} f_\pi \, \sum_{X = L,R} \left[ \frac{D_X^d(P)}{m_d} \left( B^S_{LX} + B^S_{RX} \right) + \frac{D_X^u(P)}{m_u} \left( C^S_{LX} + C^S_{RX} \right) \right] \\
b_P^S &=& \frac{1}{2} f_\pi \, \sum_{X = L,R} \left[ \frac{D_X^d(P)}{m_d} \left( B^S_{RX} - B^S_{LX} \right) + \frac{D_X^u(P)}{m_u} \left( C^S_{RX} - C^S_{LX} \right) \right] \\
a_P^V &=& \frac{1}{4} f_\pi \, C(P) (m_\tau - m_\ell) \left[ - B_{LL}^V + B_{LR}^V - B_{RL}^V + B_{RR}^V \right. \nonumber \\
&&  \left. + C_{LL}^V - C_{LR}^V + C_{RL}^V - C_{RR}^V \right] \\
b_P^V &=& \frac{1}{4} f_\pi \, C(P) (m_\tau + m_\ell) \left[ - B_{LL}^V + B_{LR}^V + B_{RL}^V - B_{RR}^V \right. \nonumber \\
&&  \left. + C_{LL}^V - C_{LR}^V - C_{RL}^V + C_{RR}^V \right] \, . \label{coeffsTauMesonLepton}
\end{eqnarray}
In these expressions $m_d$ and $m_u$ are the down- and up-quark
masses, respectively, $f_\pi$ is the pion decay constant and the
coefficients $C(P), D_{L,R}^{d,u}(P)$ take different forms for each
pseudoscalar meson $P$ \cite{Arganda:2008jj}. For $P = \pi$ one has
\begin{eqnarray}
C(\pi) &=& 1 \\
D_L^d(\pi) &=& - \frac{m_\pi^2}{4} \\
D_L^u(\pi) &=& \frac{m_\pi^2}{4} \, ,
\end{eqnarray}
for $P = \eta$
\begin{eqnarray}
C(\eta) &=& \frac{1}{\sqrt{6}} \left( \sin \theta_\eta + \sqrt{2} \cos \theta_\eta \right) \\
D_L^d(\eta) &=& \frac{1}{4 \sqrt{3}} \left[ (3 m_\pi^2 - 4 m_K^2) \cos \theta_\eta - 2 \sqrt{2} m_K^2 \sin \theta_\eta \right] \\
D_L^u(\eta) &=& \frac{1}{4 \sqrt{3}} m_\pi^2 \left( \cos \theta_\eta - \sqrt{2} \sin \theta_\eta \right) \, ,
\end{eqnarray}
and for $P = \eta^\prime$
\begin{eqnarray}
C(\eta^\prime) &=& \frac{1}{\sqrt{6}} \left( \sqrt{2} \sin \theta_\eta - \cos \theta_\eta \right) \\
D_L^d(\eta^\prime) &=& \frac{1}{4 \sqrt{3}} \left[ (3 m_\pi^2 - 4 m_K^2) \sin \theta_\eta + 2 \sqrt{2} m_K^2 \cos \theta_\eta \right] \\
D_L^u(\eta^\prime) &=& \frac{1}{4 \sqrt{3}} m_\pi^2 \left( \sin \theta_\eta + \sqrt{2} \cos \theta_\eta \right) \, .
\end{eqnarray}
Here $m_\pi$ and $m_K$ are the masses of the neutral pion and Kaon,
respectively, and $\theta_\eta$ is the $\eta-\eta^\prime$ mixing
angle. In addition, $D_R^{d,u}(P) = - \left(D_L^{d,u}(P)\right)^\ast$.

Notice that the Wilson coefficients in Eq.\eqref{coeffsTauMesonLepton}
include all pseudoscalar and axial contributions to $\tau \to \ell
P$. Therefore, this goes beyond some well-known results in the
literature, see for example \cite{Paradisi:2005tk,Arganda:2008jj},
where box contributions were neglected.

\begin{lstlisting}[caption=TauLMeson.m]
NameProcess = "TauLMeson";
NameObservables = {{BrTautoEPi, 2001, "BR(tau->e pi)"}, 
                   {BrTautoEEta, 2002, "BR(tau->e eta)"}, 
                   {BrTautoEEtap, 2003, "BR(tau->e eta')"},
		   {BrTautoMuPi, 2004, "BR(tau->mu pi)"}, 
                   {BrTautoMuEta, 2005, "BR(tau->mu eta)"}, 
                   {BrTautoMuEtap, 2006, "BR(tau->mu eta')"}};

NeededOperators = {OllddSLL, OllddSRR, OllddSRL, OllddSLR, 
  OllddVRR, OllddVLL, OllddVRL, OllddVLR,
  OlluuSLL, OlluuSRR, OlluuSRL, OlluuSLR, 
  OlluuVRR, OlluuVLL, OlluuVRL, OlluuVLR
};

Body = "TauLMeson.f90"; 
\end{lstlisting}

\begin{lstlisting}[caption=TauLMeson.f90]
Real(dp) :: Fpi, thetaEta, mPi, mK, mEta, mEtap, meson_abs_T2, cont, &
     & mP, CP, factor, BR
Complex(dp) :: BSLL, BSLR, BSRL, BSRR, BVLL, BVLR, BVRL, BVRR, &
     & CSLL, CSLR, CSRL, CSRR, CVLL, CVLR, CVRL, CVRR, aP(2), bP(2), &
     & DLdP, DRdP, DLuP, DRuP
Integer :: i1, i2, out, k1, k2

! ---------------------------------------------------------------- 
! tau -> l meson
! Observable implemented by W. Porod, F. Staub and A. Vicente
! Generalizes the analytical expressions in
! E. Arganda et al, JHEP 0806 (2008) 079 [arXiv:0803.2039]
! ---------------------------------------------------------------- 

Fpi=0.0924_dp! Pion decay constant in GeV 
thetaEta=-Pi/10._dp! eta-eta' mixing angle 
mPi=0.13497_dp! Pion mass in GeV 
mK=0.49761_dp! Kaon mass in GeV 
mEta=0.548_dp! Eta mass in GeV 
mEtap=0.958_dp! Eta' mass in GeV 

!Mesons:
!1:Pi0 
!2:Eta 
!3:Eta' 
Do i1=1,3 
   If(i1.eq.1) Then !1:Pi0 
      mP = mPi 
      CP = 1._dp 
      DLdP = - mPi**2/4._dp
      DRdP = - Conjg(DLdP)
      DLuP = mPi**2/4._dp
      DRuP = - Conjg(DLuP)
   Else If(i1.eq.2) Then !2:Eta 
      mP = mEta 
      CP = (Sin(thetaEta)+Sqrt(2._dp)*Cos(thetaEta))/Sqrt(6._dp) 
      DLdP = 1._dp/(4._dp*Sqrt(3._dp))*((3._dp*mPi**2-4._dp*mK**2) &
        & *Cos(thetaEta)-2._dp*Sqrt(2._dp)*mK**2*Sin(thetaEta))
      DRdP = - Conjg(DLdP)
      DLuP = 1._dp/(4._dp*Sqrt(3._dp))*mPi**2*(Cos(thetaEta)       &
        & -Sqrt(2._dp)*Sin(thetaEta))
      DRuP = - Conjg(DLuP)
   Else If(i1.eq.3) Then !3:Eta' 
      mP = mEtap 
      CP = (Sqrt(2._dp)*Sin(thetaEta)-Cos(thetaEta))/Sqrt(6._dp) 
      DLdP = 1._dp/(4._dp*Sqrt(3._dp))*((3._dp*mPi**2-4._dp*mK**2) & 
        & *Sin(thetaEta)+2._dp*Sqrt(2._dp)*mK**2*Cos(thetaEta))
      DRdP = - Conjg(DLdP)
      DLuP = 1._dp/(4._dp*Sqrt(3._dp))*mPi**2*(Sin(thetaEta)+      &
        & Sqrt(2._dp)*Cos(thetaEta))
      DRuP = - Conjg(DLuP)
   End If

!Leptons:
!1:e
!2:mu 
Do i2=1,2 
If (i2.eq.1) Then         ! tau -> e P
 out = 1
Elseif (i2.eq.2) Then     ! tau -> mu P
 out = 2
End if

! d-quark coefficients

BSLL = OllddSLL(3,out,1,1)
BSLR = OllddSLR(3,out,1,1)
BSRL = OllddSRL(3,out,1,1)
BSRR = OllddSRR(3,out,1,1)
BVLL = OllddVLL(3,out,1,1)
BVLR = OllddVLR(3,out,1,1)
BVRL = OllddVRL(3,out,1,1)
BVRR = OllddVRR(3,out,1,1)

! u-quark coefficients

CSLL = OlluuSLL(3,out,1,1)
CSLR = OlluuSLR(3,out,1,1)
CSRL = OlluuSRL(3,out,1,1)
CSRR = OlluuSRR(3,out,1,1)
CVLL = OlluuVLL(3,out,1,1)
CVLR = OlluuVLR(3,out,1,1)
CVRL = OlluuVRL(3,out,1,1)
CVRR = OlluuVRR(3,out,1,1)

! aP, bP scalar
aP(1) = Fpi/2._dp*(DLdP/mf_d(1)*(BSLL+BSRL) + DRdP/mf_d(1)*(BSLR+BSRR) &
         & + DLuP/mf_u(1)*(CSLL+CSRL) + DRuP/mf_u(1)*(CSLR+CSRR))
bP(1) = Fpi/2._dp*(DLdP/mf_d(1)*(BSRL-BSLL) + DRdP/mf_d(1)*(BSRR-BSLR) &
         & + DLuP/mf_u(1)*(CSRL-CSLL) + DRuP/mf_u(1)*(CSRR-CSLR))

! aP, bP vector
aP(2) = Fpi/4._dp*CP*(mf_l(3)-mf_l(out))*(-BVLL+BVLR-BVRL+BVRR+        &
         & CVLL-CVLR+CVRL-CVRR)
bP(2) = Fpi/4._dp*CP*(mf_l(3)+mf_l(out))*(-BVLL+BVLR+BVRL-BVRR+        &
         & CVLL-CVLR-CVRL+CVRR)

! averaged squared amplitude 
meson_abs_T2=0._dp 
Do k1=1,2 
   Do k2=1,2 
      cont=2._dp*mf_l(out)*mf_l(3)*(aP(k1)*conjg(aP(k2))               &
           & -bP(k1)*conjg(bP(k2)))+                                   & 
           & (mf_l(3)**2+mf_l(out)**2-mP**2)*(aP(k1)*conjg(aP(k2))+    &
           & bP(k1)*conjg(bP(k2))) 
      meson_abs_T2=meson_abs_T2+cont 
   End Do
End Do
meson_abs_T2=meson_abs_T2/(2._dp*mf_l(3)) 

! branching ratio 
factor=oo4pi*Sqrt(lamb(mf_l(3)**2,mf_l(out)**2,mP**2))                 &
            & /(mf_l(3)**2*GammaTau)*0.5_dp
BR=factor*meson_abs_T2
If (i1.eq.1) Then !pi
   If(i2.eq.1) Then 
      BrTautoEPi = BR
   Else
      BrTautoMuPi = BR
   End If
Elseif (i1.eq.2) Then !eta
   If(i2.eq.1) Then 
      BrTautoEEta = BR
   Else
      BrTautoMuEta = BR
   End If
Else !eta'
   If(i2.eq.1) Then 
      BrTautoEEtap = BR
   Else
      BrTautoMuEtap = BR
   End If
End if

End Do
End Do

Contains 

Real(dp) Function lamb(x,y,z) 
Real(dp),Intent(in)::x,y,z 
 lamb=(x+y-z)**2-4._dp*x*y 
End Function lamb 
\end{lstlisting}


\subsubsection{$\boldsymbol{h \to \ell_\alpha \ell_\beta}$}

The decay width is given by~\cite{Arganda:2004bz}
\begin{eqnarray}
\Gamma \left( h \to \ell_\alpha \ell_\beta \right) &\equiv& \Gamma \left( h \to \ell_\alpha \bar \ell_\beta \right) + \Gamma \left( h \to \bar \ell_\alpha \ell_\beta \right) = \\
&& \frac{1}{16 \pi m_h} \left[ \left(1-\left(\frac{m_{\ell_\alpha} + m_{\ell_\beta}}{m_h}\right)^2\right)\left(1-\left(\frac{m_{\ell_\alpha} - m_{\ell_\beta}}{m_h}\right)^2\right)\right]^{1/2} \nonumber \\
&& \times \left[ \left( m_h^2 - m_{\ell_\alpha}^2 - m_{\ell_\beta}^2 \right) \left( |S_L|^2 + |S_R|^2 \right)_{\alpha \beta} - 4 m_{\ell_\alpha} m_{\ell_\beta} \text{Re}(S_L S_R^\ast)_{\alpha \beta} \right] \nonumber \\
&& + (\alpha \leftrightarrow \beta) \nonumber
\end{eqnarray}

\begin{lstlisting}[caption=hLLp.m]
NameProcess = "hLLp";
NameObservables = {{BrhtoMuE, 1101, "BR(h->e mu)"}, 
                   {BrhtoTauE, 1102, "BR(h->e tau)"}, 
                   {BrhtoTauMu, 1103, "BR(h->mu tau)"}};

NeededOperators = {OH2lSL, OH2lSR};

Body = "hLLp.f90"; 

\end{lstlisting}

\begin{lstlisting}[caption=hLLp.f90]
Real(dp) :: width1, width2, width, mh, gamh, kinfactor
Complex(dp) :: SL1, SR1, SL2, SR2
Integer :: i1, gt1, gt2, hLoc

! ---------------------------------------------------------------- 
! h -> l l'
! Observable implemented by W. Porod, F. Staub and A. Vicente
! Based on E. Arganda et al, PRD 71 (2005) 035011 [hep-ph/0407302]
! ---------------------------------------------------------------- 

!! NEXT LINE HAVE TO BE PARSED BY SARAH
! Checking if there are several generations of Scalars and what is the SM-like doublet
@ If[getGen[HiggsBoson]>1, "hLoc = MaxLoc(Abs("<>ToString[HiggsMixingMatrix]<>"(2,:)),1)", "hLoc = 1"]

@ "mh = "<>ToString[SPhenoMass[HiggsBoson]]<>If[getGen[HiggsBoson]>1, "(hLoc)", ""]

@ "gamh ="<>ToString[SPhenoWidth[HiggsBoson]]<>If[getGen[HiggsBoson]>1, "(hLoc)", ""]

If (.not.L_BR) gamh = 4.5E-3_dp  ! Decays not calculated; using SM value

Do i1=1,3 

If (i1.eq.1) Then         ! h -> e mu
 gt1 = 1
 gt2 = 2
Elseif (i1.eq.2) Then     ! h -> e tau
 gt1 = 1
 gt2 = 3
Else                      ! h -> mu tau
 gt1 = 2
 gt2 = 3
End if

! width = Gamma(h -> \bar{l1} l2) + Gamma(h -> l1 \bar{l2})

SL1 = OH2lSL(gt1,gt2,hLoc)
SR1 = OH2lSR(gt1,gt2,hLoc)
SL2 = OH2lSL(gt2,gt1,hLoc)
SR2 = OH2lSR(gt2,gt1,hLoc)

kinfactor = (1-(mf_l(gt1)+mf_l(gt2)/mh)**2)*&
       & (1-(mf_l(gt1)-mf_l(gt2)/mh)**2)

width1 = (mh**2-mf_l(gt1)**2-mf_l(gt2)**2)*(Abs(SL1)**2+Abs(SR1)**2) & 
     & - 4._dp*mf_l(gt1)*mf_l(gt2)*Real(SL1*Conjg(SR1),dp)
width2 = (mh**2-mf_l(gt1)**2-mf_l(gt2)**2)*(Abs(SL2)**2+Abs(SR2)**2) & 
     & - 4._dp*mf_l(gt1)*mf_l(gt2)*Real(SL2*Conjg(SR2),dp)

! decay width
width = oo16pi/mh * sqrt(kinfactor) * (width1+width2)

If (i1.eq.1) Then
BrhtoMuE = width/(width+gamh)
Elseif (i1.eq.2) Then 
BrhtoTauE = width/(width+gamh)
Else
BrhtoTauMu = width/(width+gamh)
End if

End do
\end{lstlisting}


\subsubsection{$\boldsymbol{Z \to \ell_\alpha \ell_\beta}$}
The decay width is given by~\cite{Bi:2000xp}
\begin{eqnarray}
\Gamma \left( Z \to \ell_\alpha \ell_\beta \right) &\equiv& \Gamma \left( Z \to \ell_\alpha \bar \ell_\beta \right) + \Gamma \left( Z \to \bar \ell_\alpha \ell_\beta \right) = \\
&& \frac{m_Z}{48 \pi} \left[ 2 \left( |R_1^L|^2 + |R_1^R|^2 \right) + \frac{m_Z^2}{4} \left( |R_2^L|^2 + |R_2^R|^2 \right) \right] \, , \nonumber 
\end{eqnarray}
where the charged lepton masses have been neglected.

\begin{lstlisting}[caption=ZLLp.m]
NameProcess = "ZLLp";
NameObservables = {{BrZtoMuE, 1001, "BR(Z->e mu)"}, 
                   {BrZtoTauE, 1002, "BR(Z->e tau)"}, 
                   {BrZtoTauMu, 1003, "BR(Z->mu tau)"}};

NeededOperators = {OZ2lSL, OZ2lSR,OZ2lVL,OZ2lVR};

Body = "ZLLp.f90"; 
\end{lstlisting}

\begin{lstlisting}[caption=ZLLp.f90]
Real(dp) :: width
Integer :: i1, gt1, gt2

! ---------------------------------------------------------------- 
! Z -> l l'
! Observable implemented by W. Porod, F. Staub and A. Vicente
! Based on X. -J. Bi et al, PRD 63 (2001) 096008 [hep-ph/0010270]
! ---------------------------------------------------------------- 

Do i1=1,3 

If (i1.eq.1) Then         ! Z -> e mu
 gt1 = 1
 gt2 = 2
Elseif (i1.eq.2) Then     !Z -> e tau
 gt1 = 1
 gt2 = 3
Else                      ! Z -> mu tau
 gt1 = 2
 gt2 = 3
End if

! decay width
width = oo48pi*(2*(Abs(OZ2lVL(gt1,gt2))**2 +            &
  &                    Abs(OZ2lVR(gt1,gt2))**2)*mZ      & 
  & + (Abs(OZ2lSL(gt1,gt2))**2+Abs(OZ2lSR(gt1,gt2))**2) &
  & * mZ * mZ2 * 0.25_dp) 

If (i1.eq.1) Then
BrZtoMuE = width/(width+gamZ)
Elseif (i1.eq.2) Then 
BrZtoTauE = width/(width+gamZ)
Else
BrZtoTauMu = width/(width+gamZ)
End if

End do
\end{lstlisting}


\subsection{Quark flavor observables}
\label{sec:QFV}

QFV has been observed and its description in the SM due to the CKM
matrix is well established.  However, the large majority of BSM models
causes additional contributions which have to be studied carefully,
see for instance
Refs.~\cite{Ciuchini:1998ix,Buchalla:1993bv,Misiak:1999yg,Dedes:2001fv,Altmannshofer:2013foa,Dedes:2008iw,Lunghi:2006hc,Bobeth:2001jm,Huber:2005ig,Logan:2000iv,Buras:2002vd,Hou:1992sy,Ibrahim:1999hh,Barbieri:2012tu,Altmannshofer:2012az,Becirevic:2012fy,Becirevic:2012jf,Buras:2013ooa,Dorsner:2013tla,Descotes-Genon:2013wba,Buras:2013dea,Barbieri:2014tja,Buras:2014sba,Konig:2014iqa,Greljo:2014dka}.

We give also here a description of the implementation of the different
observables using the operators present in the \SPheno output of
\SARAH.

\subsubsection{$\boldsymbol{B_{s,d}^0 \to \ell^+ \ell^-}$} 
Our analytical results for $B_{s,d}^0 \to \ell^+ \ell^-$ follow
\cite{Dedes:2008iw}. The $B^0 \equiv B_{s,d}^0$ decay width to a pair
of charged leptons can be written as
\begin{equation}
\Gamma \left(B^0 \to \ell_\alpha^+ \ell_\beta^- \right) = \frac{|\mathcal{M_{B\ell\ell}}|^2}{16 \pi M_{B}} \left[ \left(1-\left(\frac{m_{\ell_\alpha} + m_{\ell_\beta}}{m_B}\right)^2\right)\left(1-\left(\frac{m_{\ell_\alpha} - m_{\ell_\beta}}{m_B}\right)^2\right)\right]^{1/2} \, .
\end{equation}
Here
\begin{eqnarray}
|\mathcal{M_{B\ell\ell}}|^2 &=& 2 |F_S|^2 \left[ m_B^2 - \left(m_{\ell_\alpha} + m_{\ell_\beta}\right)^2 \right] + 2 |F_P|^2 \left[ m_B^2 - \left(m_{\ell_\alpha} - m_{\ell_\beta}\right)^2 \right] \nonumber \\
&& + 2 |F_V|^2 \left[ m_B^2 \left(m_{\ell_\alpha} - m_{\ell_\beta}\right)^2 - \left(m_{\ell_\alpha}^2 - m_{\ell_\beta}^2\right)^2 \right] \nonumber \\
&& + 2 |F_A|^2 \left[ m_B^2 \left(m_{\ell_\alpha} + m_{\ell_\beta}\right)^2 - \left(m_{\ell_\alpha}^2 - m_{\ell_\beta}^2\right)^2 \right] \nonumber \\
&& + 4 \, \text{Re}(F_S F_V^\ast) \left(m_{\ell_\alpha} - m_{\ell_\beta}\right) \left[ m_B^2 + \left(m_{\ell_\alpha} + m_{\ell_\beta}\right)^2 \right] \nonumber \\
&& + 4 \, \text{Re}(F_P F_A^\ast) \left(m_{\ell_\alpha} + m_{\ell_\beta}\right) \left[ m_B^2 - \left(m_{\ell_\alpha} - m_{\ell_\beta}\right)^2 \right] \, , \label{AmpBll}
\end{eqnarray}
and the $F_X$ coefficients are defined in terms of our Wilson
coefficients as\footnote{Notice that our effective Lagrangian differs
  from the one in \cite{Dedes:2008iw} by a $1/(4 \pi)^2$ factor. This
  relative factor has been absorbed in the expression for
  $\mathcal{M_{B\ell\ell}}$, see Eq.\eqref{AmpBll}.}
\begin{eqnarray}
F_S &=& \frac{i}{4} \frac{m_B^2 f_B}{m_d + m_{d^\prime}} \left( E_{LL}^S + E_{LR}^S - E_{RR}^S - E_{RL}^S \right) \\
F_P &=& \frac{i}{4} \frac{m_B^2 f_B}{m_d + m_{d^\prime}} \left( - E_{LL}^S + E_{LR}^S - E_{RR}^S + E_{RL}^S \right) \\
F_V &=& - \frac{i}{4} f_B \left( E_{LL}^V + E_{LR}^V - E_{RR}^V - E_{RL}^V \right) \\
F_A &=& - \frac{i}{4} f_B \left( - E_{LL}^V + E_{LR}^V - E_{RR}^V + E_{RL}^V \right) \, ,
\end{eqnarray}
where $f_B \equiv f_{B_{d,s}^0}$ is the $B_{d,s}^0$ decay constant and
$m_{d,d^\prime}$ are the masses of the quarks contained in the $B$
meson, $B_d^0 \equiv \bar b d$ and $B_s^0 \equiv \bar b s$. In the
lepton flavor conserving case, $\alpha = \beta$, the $F_V$
contribution vanishes. In this case, the results in
\cite{Dedes:2008iw} are in agreement with previous computations
\cite{Bobeth:2002ch,Isidori:2002qe}.

\begin{lstlisting}[caption=B0ll.m]
NameProcess = "B0toLL";
NameObservables = {{BrB0dEE, 4000, "BR(B^0_d->e e)"}, 
                   {ratioB0dEE, 4001, "BR(B^0_d->e e)/BR(B^0_d->e e)_SM"},  
                   {BrB0sEE, 4002, "BR(B^0_s->e e)"}, 
                   {ratioB0sEE, 4003, "BR(B^0_s->e e)/BR(B^0_s->e e)_SM"}, 
                   {BrB0dMuMu, 4004, "BR(B^0_d->mu mu)"}, 
                   {ratioB0dMuMu, 4005, "BR(B^0_d->mu mu)/BR(B^0_d->mu mu)_SM"}, 
                   {BrB0sMuMu, 4006, "BR(B^0_s->mu mu)"}, 
                   {ratioB0sMuMu, 4007, "BR(B^0_s->mu mu)/BR(B^0_s->mu mu)_SM"}, 
                   {BrB0dTauTau, 4008, "BR(B^0_d->tau tau)"}, 
                   {ratioB0dTauTau, 4009, "BR(B^0_d->tau tau)/BR(B^0_d->tau tau)_SM"}, 
                   {BrB0sTauTau, 4010, "BR(B^0_s->tau tau)"}, 
                   {ratioB0sTauTau, 4011, "BR(B^0_s->tau tau)/BR(B^0_s->tau tau)_SM"} };


NeededOperators = {OddllSLL, OddllSRR, OddllSRL, OddllSLR,
                   OddllVRR, OddllVLL, OddllVRL, OddllVLR,
                   OddllSLLSM, OddllSRRSM, OddllSRLSM, OddllSLRSM,
                   OddllVRRSM, OddllVLLSM, OddllVRLSM, OddllVLRSM};

Body = "B0ll.f90"; 
\end{lstlisting}

\begin{lstlisting}[caption=B0ll.f90]
Real(dp) :: AmpSquared,AmpSquared2, AmpSquared_SM, AmpSquared2_SM, &
              & width_SM, width
Real(dp) :: MassB0s, MassB0d, fBs, fBd, TauB0s, TauB0d 
Real(dp) :: hbar=6.58211899E-25_dp 
Real(dp) :: MassB0,MassB02,fB0,GammaB0 
Complex(dp) :: CS(4), CV(4), CT(4)
Complex(dp) :: FS=0._dp, FP=0._dp, FV=0._dp, FA=0._dp 
Integer :: i1, gt1, gt2, gt3, gt4

! ---------------------------------------------------------------- 
! B0 -> l l
! Observable implemented by W. Porod, F. Staub and A. Vicente
! Based on A. Dedes et al, PRD 79 (2009) 055006 [arXiv:0812.4320]
! ---------------------------------------------------------------- 

! Using global hadronic data
fBd = f_B0d_CONST  
fBs = f_B0s_CONST  
TauB0d = tau_B0d 
TauB0s = tau_B0s
MassB0d = mass_B0d 
MassB0s = mass_B0s 

Do i1=1,6
gt1 = 3
If (i1.eq.1) Then ! B0d -> e+ e-
  MassB0 = MassB0d
  MassB02 = MassB0d**2
  fB0 = fBd
  GammaB0 = (hbar)/(TauB0d)
  gt2 = 1
  gt3 = 1
  gt4 = 1
Else if (i1.eq.2) Then ! B0s -> e+ e-
  MassB0 = MassB0s
  MassB02 = MassB0s**2
  fB0 = fBs
  GammaB0 = (hbar)/(TauB0s)
  gt2 = 2
  gt3 = 1
  gt4 = 1
Else if (i1.eq.3) Then ! B0d -> mu+ mu-
  MassB0 = MassB0d
  MassB02 = MassB0d**2
  fB0 = fBd
  GammaB0 = (hbar)/(TauB0d)
  gt2 = 1
  gt3 = 2
  gt4 = 2
Else if (i1.eq.4) Then ! B0s -> mu+ mu-
  MassB0 = MassB0s
  MassB02 = MassB0s**2 
  fB0 = fBs
  GammaB0 = (hbar)/(TauB0s)
  gt2 = 2
  gt3 = 2
  gt4 = 2
Else if (i1.eq.5) Then ! B0d -> tau+ tau-
  MassB0 = MassB0d
  MassB02 = MassB0d**2
  fB0 = fBd
  GammaB0 = (hbar)/(TauB0d)
  gt2 = 1
  gt3 = 3
  gt4 = 3
Else if (i1.eq.6) Then ! B0s -> tau+ tau-
  MassB0 = MassB0s
  MassB02 = MassB0s**2 
  fB0 = fBs
  GammaB0 = (hbar)/(TauB0s)
  gt2 = 2
  gt3 = 3
  gt4 = 3
End if

! BSM contributions

CS(1) = OddllSRR(gt1,gt2,gt3,gt4)
CS(2) = OddllSRL(gt1,gt2,gt3,gt4)
CS(3) = OddllSLL(gt1,gt2,gt3,gt4)
CS(4) = OddllSLR(gt1,gt2,gt3,gt4)

CV(1) = OddllVLL(gt1,gt2,gt3,gt4)
CV(2) = OddllVLR(gt1,gt2,gt3,gt4)
CV(3) = OddllVRR(gt1,gt2,gt3,gt4)
CV(4) = OddllVRL(gt1,gt2,gt3,gt4)

FS= 0.25_dp*MassB02*fB0/(MFd(gt1)+MFd(gt2))*( CS(1)+CS(2)-CS(3)-CS(4)) 
FP= 0.25_dp*MassB02*fB0/(MFd(gt1)+MFd(gt2))*(-CS(1)+CS(2)-CS(3)+CS(4)) 
FV= -0.25_dp*fB0*( CV(1)+CV(2)-CV(3)-CV(4)) 
FA= -0.25_dp*fB0*(-CV(1)+CV(2)-CV(3)+CV(4)) 

AmpSquared = 2 * abs(FS)**2 * (MassB02 - (mf_l(gt3)+mf_l(gt4))**2) &
     & + 2 *abs(FP)**2 * (MassB02 - (mf_l(gt3)-mf_l(gt4))**2) &
     & + 2 *abs(FV)**2 * (MassB02*(mf_l(gt4)-mf_l(gt3))**2    &
             & - (mf_l2(gt4)-mf_l2(gt3))**2) &
     & + 2 *abs(FA)**2 * (MassB02*(mf_l(gt4)+mf_l(gt3))**2 -  &
             & (mf_l2(gt4)-mf_l2(gt3))**2) &
     & + 4 *REAL(FS*conjg(FV)) *(mf_l(gt3)-mf_l(gt4)) *(MassB02 &
             & + (mf_l(gt3)+mf_l(gt4))**2) &
     & + 4 *REAL(FP*conjg(FA)) *(mf_l(gt3)+mf_l(gt4)) *(MassB02 &
             & - (mf_l(gt3)-mf_l(gt4))**2) 

width = oo16pi * AmpSquared / MassB0 * &
    & sqrt(1-((mf_l(gt4)+mf_l(gt3))/MassB0)**2) &
    & * sqrt(1-((mf_l(gt4)-mf_l(gt3))/MassB0)**2)*(Alpha/Alpha_160)**4


! SM contributions

CS(1) = OddllSRRSM(gt1,gt2,gt3,gt4)
CS(2) = OddllSRLSM(gt1,gt2,gt3,gt4)
CS(3) = OddllSLLSM(gt1,gt2,gt3,gt4)
CS(4) = OddllSLRSM(gt1,gt2,gt3,gt4)

CV(1) = OddllVLLSM(gt1,gt2,gt3,gt4)
CV(2) = OddllVLRSM(gt1,gt2,gt3,gt4)
CV(3) = OddllVRRSM(gt1,gt2,gt3,gt4)
CV(4) = OddllVRLSM(gt1,gt2,gt3,gt4)

FS= 0.25_dp*MassB02*fB0/(MFd(gt1)+MFd(gt2))*( CS(1)+CS(2)-CS(3)-CS(4)) 
FP= 0.25_dp*MassB02*fB0/(MFd(gt1)+MFd(gt2))*(-CS(1)+CS(2)-CS(3)+CS(4)) 
FV= -0.25_dp*fB0*( CV(1)+CV(2)-CV(3)-CV(4)) 
FA= -0.25_dp*fB0*(-CV(1)+CV(2)-CV(3)+CV(4)) 

AmpSquared = 2 * abs(FS)**2 * (MassB02 - (mf_l(gt3)+mf_l(gt4))**2) &
     & + 2 *abs(FP)**2 * (MassB02 - (mf_l(gt3)-mf_l(gt4))**2) &
     & + 2 *abs(FV)**2 * (MassB02*(mf_l(gt4)-mf_l(gt3))**2 -  &
        &  (mf_l2(gt4)-mf_l2(gt3))**2) &
     & + 2 *abs(FA)**2 * (MassB02*(mf_l(gt4)+mf_l(gt3))**2 -  &
        &  (mf_l2(gt4)-mf_l2(gt3))**2) &
     & + 4 *REAL(FS*conjg(FV)) *(mf_l(gt3)-mf_l(gt4)) *(MassB02 &
        & + (mf_l(gt3)+mf_l(gt4))**2) &
     & + 4 *REAL(FP*conjg(FA)) *(mf_l(gt3)+mf_l(gt4)) *(MassB02 &
        & - (mf_l(gt3)-mf_l(gt4))**2)

width_SM = oo16pi * AmpSquared / MassB0 * sqrt(1-((mf_l(gt4)+   &
     & mf_l(gt3))/MassB0)**2) &
     & * sqrt(1-((mf_l(gt4)-mf_l(gt3))/MassB0)**2)*(Alpha/Alpha_160)**4 


If (i1.Eq.1) Then
  BrB0dEE= width / GammaB0
  ratioB0dEE= width / width_SM
Else If (i1.Eq.2) Then
  BrB0sEE= width / GammaB0
  ratioB0sEE= width / width_SM
Else If (i1.Eq.3) Then
  BrB0dMuMu= width / GammaB0
  ratioB0dMuMu= width / width_SM
Else If (i1.Eq.4) Then
  BrB0sMuMu= width / GammaB0
  ratioB0sMuMu= width / width_SM
Else If (i1.Eq.5) Then
  BrB0dTauTau= width / GammaB0
  ratioB0dTauTau= width / width_SM
Else If (i1.Eq.6) Then
  BrB0sTauTau= width / GammaB0
  ratioB0sTauTau= width / width_SM
End If

End do 
\end{lstlisting}


\subsubsection{$\boldsymbol{\bar B \to X_s \gamma}$}
The branching ratio for $\bar B \to X_s \gamma$, with a cut $E_\gamma
> 1.6$ GeV in the $\bar B$ rest frame, can be obtained
as~\cite{Kagan:1998ym,Lunghi:2006hc}
\begin{eqnarray}
\BR & \hspace*{-0.1cm} \left( \bar B \to X_s \gamma \right)_{E_\gamma > 1.6 \text{GeV}} = 10^{-4} \bigg[ a_{SM} + a_{77} \left( |\delta C_7^{(0)}|^2 + |\delta C_7^{\prime (0)}|^2 \right) + a_{88} \left( |\delta C_8^{(0)}|^2 + |\delta C_8^{\prime (0)}|^2 \right) \nonumber \\
& + \text{Re} \left( a_7 \, \delta C_7^{(0)} + a_8 \, \delta C_8^{(0)} + a_{78} \left( \delta C_7^{(0)} \, \delta C_8^{(0) \ast} + \delta C_7^{\prime (0)} \, \delta C_8^{\prime (0) \, \ast} \right) \right) \bigg] \, , \label{BRBXsGamma}
\end{eqnarray}
where $a_{SM} = 3.15$ is the NNLO SM prediction
\cite{Misiak:2006zs,Misiak:2006ab}, the other $a$ coefficients in
Eq.\eqref{BRBXsGamma} are found to be
\begin{eqnarray}
a_{77} &=& 4.743 \nonumber \\
a_{88} &=& 0.789 \nonumber \\
a_7 &=& -7.184 + 0.612 \, i \nonumber \\
a_8 &=& -2.225 - 0.557 \, i \nonumber \\
a_{78} &=& 2.454 - 0.884 \, i 
\end{eqnarray}
and we have defined $\delta C_i^{(0)} = C_i^{(0)} - C_i^{(0) \,
  \text{SM}}$. Finally, the $C_i^{(0)}$ coefficients can be written in
terms of $Q_{1,2}^{L,R}$ in Eqs.\eqref{eq:L-qqgamma} and
\eqref{eq:L-qqgluon} as
\begin{eqnarray}
C_7^{(0)} &=& n_{CQ} \, Q_1^R \\
C_7^{\prime (0)} &=& n_{CQ} \, Q_1^L \\
C_8^{(0)} &=& n_{CQ} \, Q_2^R \\
C_8^{\prime (0)} &=& n_{CQ} \, Q_2^L \label{eq:nCQprev}
\end{eqnarray}
where $n_{CQ}^{-1} = - \frac{G_F}{4 \sqrt{2} \pi^2} V_{tb}
V_{ts}^\ast$ and $V$ is the Cabibbo-Kobayashi-Maskawa (CKM) matrix.

\begin{lstlisting}[caption=bsGamma.m]
NameProcess = "bsGamma";
NameObservables = {{BrBsGamma, 200, "BR(B->X_s gamma)"},
                   {ratioBsGamma, 201, "BR(B->X_s gamma)/BR(B->X_s gamma)_SM"}};

NeededOperators = {CC7, CC7p, CC8, CC8p,
                   CC7SM, CC7pSM, CC8SM, CC8pSM};

Body = "bsGamma.f90";  
\end{lstlisting}

\begin{lstlisting}[caption=bsGamma.f90]
Integer :: gt1, gt2
Complex(dp) :: norm, delta_C7_0, delta_C7p_0, delta_C8_0, delta_C8p_0
Real(dp) :: NNLO_SM 

! ---------------------------------------------------------------- 
! \bar{B} -> X_s gamma (Egamma > 1.6 GeV)
! Observable implemented by W. Porod, F. Staub and A. Vicente
! Based on E. Lunghi, J. Matias, JHEP 0704 (2007) 058 [hep-ph/0612166]
! ---------------------------------------------------------------- 

gt1=3 !b
gt2=2 !s

! normalization of our Wilson coefficients
! relative to the ones used in hep-ph/0612166
norm = -CKM_160(3,3)*Conjg(CKM_160(gt1,gt2))*Alpha_160/ &
         & (8._dp*Pi*sinW2_160*mW2)

! Wilson coefficients
delta_C7_0 =(CC7(gt1,gt2)-CC7SM(gt1,gt2))/norm
delta_C7p_0=(CC7p(gt1,gt2)-CC7pSM(gt1,gt2))/norm
delta_C8_0 =(CC8(gt1,gt2)-CC8SM(gt1,gt2))/norm
delta_C8p_0=(CC8p(gt1,gt2)-CC8pSM(gt1,gt2))/norm

! NNLO SM prediction
! as obtained in M. Misiak et al, PRL 98 (2007) 022002
! and M. Misiak and M. Steinhauser, NPB 764 (2007) 62
NNLO_SM=3.15_dp

BrBsGamma=NNLO_SM+4.743_dp*(Abs(delta_C7_0)**2+Abs(delta_C7p_0)**2)&
&+0.789_dp*(Abs(delta_C8_0)**2+Abs(delta_C8p_0)**2)&
&+Real((-7.184_dp,0.612_dp)*delta_C7_0&
&+(-2.225_dp,-0.557_dp)*delta_C8_0+(2.454_dp,-0.884_dp)*&
&(delta_C7_0*conjg(delta_C8_0)+delta_C7p_0*conjg(delta_C8p_0)),dp)

! ratio BSM/SM
ratioBsGamma = BrBsGamma/NNLO_SM

! branching ratio
BrBsGamma=1E-4_dp*BrBsGamma 
\end{lstlisting}


\subsubsection{$\boldsymbol{\bar B \to X_s \ell^+ \ell^-}$}
Our results for $\bar B \to X_s \ell^+ \ell^-$ are based on
\cite{Huber:2005ig}, expanded with the addition of prime operators
contributions \cite{Lunghi:private}. The branching ratios for the
$\ell = e$ case can be written as
\begin{eqnarray}
10^7 \, \BR && \hspace*{-0.2cm} \left( \bar B \to X_s e^+ e^- \right) = 2.3148 - 0.001658 \IM (R_{10}) + 0.0005 \IM(R_{10} R_8^\ast + R_{10}^\prime R_8^{\prime \, \ast}) \nonumber \\
&& + 0.0523 \IM(R_7) + 0.02266 \IM(R_7 R_8^\ast + R_7^\prime R_8^{\prime \, \ast}) + 0.00496 \IM(R_7 R_9^\ast + R_7^\prime R_9^{\prime \, \ast}) \nonumber \\
&& + 0.00518 \IM(R_8) + 0.0261 \IM(R_8 R_9^\ast + R_8^\prime R_9^{\prime \, \ast}) - 0.00621 \IM(R_9) - 0.5420 \RE(R_{10}) \nonumber \\
&& - 0.03340 \RE(R_7) + 0.0153 \RE(R_7 R_{10}^\ast + R_7^\prime R_{10}^{\prime \, \ast}) + 0.0673 \RE(R_7 R_8^\ast + R_7^\prime R_8^{\prime \, \ast}) \nonumber \\
&& - 0.86916 \RE(R_7 R_9^\ast + R_7^\prime R_9^{\prime \, \ast}) - 0.0135 \RE(R_8) + 0.00185 \RE(R_8 R_{10} + R_8^\prime R_{10}^{\prime \, \ast}) \nonumber \\
&& - 0.09921 \RE(R_8 R_9^\ast + R_8^\prime R_9^{\prime \, \ast}) + 2.833 \RE(R_9) - 0.10698 \RE(R_9 R_{10}^\ast + R_9^\prime R_{10}^{\prime \, \ast}) \nonumber \\
&& + 11.0348 \left( |R_{10}|^2 + |R_{10}^\prime|^2 \right) + 0.2804 \left( |R_7|^2 + |R_7^\prime|^2 \right) \nonumber \\
&& + 0.003763 \left( |R_8|^2 + |R_8^\prime|^2 \right) + 1.527 \left( |R_9|^2 + |R_9^\prime|^2 \right) \, ,
\end{eqnarray}
whereas for the $\ell = \mu$ case one gets
\begin{eqnarray}
10^7 \, \BR && \hspace*{-0.2cm} \left( \bar B \to X_s \mu^+ \mu^- \right) = 2.1774 - 0.001658 \IM (R_{10}) + 0.0005 \IM(R_{10} R_8^\ast + R_{10}^\prime R_8^{\prime \, \ast}) \nonumber \\
&& + 0.0534 \IM(R_7) + 0.02266 \IM(R_7 R_8^\ast + R_7^\prime R_8^{\prime \, \ast}) + 0.00496 \IM(R_7 R_9^\ast + R_7^\prime R_9^{\prime \, \ast}) \nonumber \\
&& + 0.00527 \IM(R_8) + 0.0261 \IM(R_8 R_9^\ast + R_8^\prime R_9^{\prime \, \ast}) - 0.0115 \IM(R_9) - 0.5420 \RE(R_{10}) \nonumber \\
&& + 0.0208 \RE(R_7) + 0.0153 \RE(R_7 R_{10}^\ast + R_7^\prime R_{10}^{\prime \, \ast}) + 0.0648 \RE(R_7 R_8^\ast + R_7^\prime R_8^{\prime \, \ast}) \nonumber \\
&& - 0.8545 \RE(R_7 R_9^\ast + R_7^\prime R_9^{\prime \, \ast}) - 0.00938 \RE(R_8) + 0.00185 \RE(R_8 R_{10} + R_8^\prime R_{10}^{\prime \, \ast}) \nonumber \\
&& - 0.0981 \RE(R_8 R_9^\ast + R_8^\prime R_9^{\prime \, \ast}) + 2.6917 \RE(R_9) - 0.10698 \RE(R_9 R_{10}^\ast + R_9^\prime R_{10}^{\prime \, \ast}) \nonumber \\
&& + 10.7652 \left( |R_{10}|^2 + |R_{10}^\prime|^2 \right) + 0.2880 \left( |R_7|^2 + |R_7^\prime|^2 \right) \nonumber \\
&& + 0.003763 \left( |R_8|^2 + |R_8^\prime|^2 \right) + 1.527 \left( |R_9|^2 + |R_9^\prime|^2 \right) \, .
\end{eqnarray}
Here we have defined the ratios of Wilson coefficients
\begin{equation}
R_{7,8} = \frac{Q_{1,2}^R}{Q_{1,2}^{R, \text{SM}}} \qquad , \qquad R_{7,8}^\prime = \frac{Q_{1,2}^L}{Q_{1,2}^{L, \text{SM}}}
\end{equation}
as well as
\begin{equation}
R_{9,10} = \frac{E_{LL}^V \pm E_{LR}^V}{E_{LL}^{V, \text{SM}} \pm E_{LR}^{V, \text{SM}}} \qquad , \qquad R_{9,10}^\prime = \frac{E_{RR}^V \pm E_{RL}^V}{E_{RR}^{V, \text{SM}} \pm E_{RL}^{V, \text{SM}}} \, .
\end{equation}

\begin{lstlisting}[caption=BtoSLL.m]
NameProcess = "BtoSLL";
NameObservables = {{BrBtoSEE, 5000, "BR(B-> s e e)"}, 
                   {ratioBtoSEE, 5001, "BR(B-> s e e)/BR(B-> s e e)_SM"}, 
                   {BrBtoSMuMu, 5002, "BR(B-> s mu mu)"} , 
                   {ratioBtoSMuMu, 5003, "BR(B-> s mu mu)/BR(B-> s mu mu)_SM"}};

NeededOperators = {OddllVRR, OddllVLL, OddllVRL, OddllVLR, 
                   CC7, CC7p, CC8, CC8p,
                   OddllVRRSM, OddllVLLSM, OddllVRLSM, OddllVLRSM, 
                   CC7SM, CC7pSM, CC8SM, CC8pSM
};

Body = "BtoSLL.f90"; 
\end{lstlisting}
\begin{lstlisting}[caption=BtoSLL.f90]
Complex(dp) :: c7(2), c7p(2), c8(2), c8p(2), r7, r7p, r8, r8p, norm, &
     & r9(2), r9p(2), r10(2), r10p(2),                               & 
     & c9ee(2), c9pee(2), c10ee(2), c10pee(2),                       &
     & c9_cee(2), c9p_cee(2), c10_cee(2), c10p_cee(2),               &
     & c9mm(2), c9pmm(2), c10mm(2), c10pmm(2), c9_cmm(2),            &
     & c9p_cmm(2), c10_cmm(2), c10p_cmm(2) 

! ---------------------------------------------------------------- 
! \bar{B} -> X_s l+ l-
! Observable implemented by W. Porod, F. Staub and A. Vicente
! Based on T. Huber et al, NPB 740 (2006) 105, [hep-ph/0512066]
! Prime operators added after private communication with E. Lunghi
! ---------------------------------------------------------------- 

! Wilson coefficients

c7(1) = CC7(3,2)
c7(2) =  CC7SM(3,2)
c7p(1) = CC7p(3,2)
c7p(2) = CC7pSM(3,2)

c8(1) = CC8(3,2)
c8(2) = CC8SM(3,2)
c8p(1) = CC8p(3,2)
c8p(2) = CC8pSM(3,2)

c9ee(1) = OddllVLL(3,2,1,1)+OddllVLR(3,2,1,1)
c9ee(2) = (OddllVLLSM(3,2,1,1)+OddllVLRSM(3,2,1,1))
c9mm(1) = OddllVLL(3,2,2,2)+OddllVLR(3,2,2,2)
c9mm(2) =  (OddllVLLSM(3,2,2,2)+OddllVLRSM(3,2,2,2))
c9pee(1) = OddllVRR(3,2,1,1)+OddllVRL(3,2,1,1)
c9pee(2) =  (OddllVRRSM(3,2,1,1)+OddllVRLSM(3,2,1,1))
c9pmm(1) = OddllVRR(3,2,2,2)+OddllVRL(3,2,2,2)
c9pmm(2) =  (OddllVRRSM(3,2,2,2)+OddllVRLSM(3,2,2,2))

c10ee(1) = OddllVLL(3,2,1,1)-OddllVLR(3,2,1,1)
c10ee(2) =  (OddllVLLSM(3,2,1,1)-OddllVLRSM(3,2,1,1))
c10mm(1) = OddllVLL(3,2,2,2)-OddllVLR(3,2,2,2)
c10mm(2) =  (OddllVLLSM(3,2,2,2)-OddllVLRSM(3,2,2,2))
c10pee(1) = OddllVRR(3,2,1,1)-OddllVRL(3,2,1,1)
c10pee(2) =  (OddllVRRSM(3,2,1,1)-OddllVRLSM(3,2,1,1))
c10pmm(1) = OddllVRR(3,2,2,2)-OddllVRL(3,2,2,2)
c10pmm(2) = (OddllVRRSM(3,2,2,2)-OddllVRLSM(3,2,2,2))

! ratios

r7 = c7(1) / c7(2)
r7p = c7p(1) / c7(2)
r8 = c8(1) / c8(2)
r8p = c8p(1) / c8(2)

r9(1) = c9ee(1)/c9ee(2)
r9(2) = c9mm(1)/c9mm(2)
r9p(1) = c9pee(1)/c9ee(2)
r9p(2) = c9pmm(1)/c9mm(2)

r10(1) = c10ee(1)/c10ee(2)
r10(2) = c10mm(1)/c10mm(2)
r10p(1) = c10pee(1)/c10ee(2)
r10p(2) = c10pmm(1)/c10mm(2)

BrBtoSEE = (2.3148_dp - 1.658e-3_dp * Aimag(R10(1))                   &
 & + 5.e-4_dp * Aimag(r10(1)*Conjg(r8) + r10p(1)*Conjg(r8p) )         &
 & + 5.23e-2_dp * Aimag(r7) + 5.18e-3_dp * Aimag(r8)                  &
 & + 2.266e-2_dp * Aimag(r7 * Conjg(r8) + r7p * Conjg(r8p) )          &
 & + 4.96e-3_dp * Aimag(r7 * Conjg(r9(1)) + r7p * Conjg(r9p(1)) )     &
 & + 2.61e-2_dp * Aimag(r8 * Conjg(r9(1)) + r8p * Conjg(r9p(1)) )     &
 & - 6.21e-3_dp * Aimag(r9(1)) - 0.5420_dp * Real( r10(1), dp)        &
 & - 3.340e-2_dp * Real(r7,dp) - 1.35e-2_dp * Real(r8,dp)             &
 & + 1.53e-2_dp * Real(r7*Conjg(r10(1)) + r7p*Conjg(r10p(1)), dp )    &
 & + 6.73e-2_dp * Real(r7 * Conjg(r8) + r7p * Conjg(r8p), dp )        &
 & - 0.86916_dp * Real(r7*Conjg(r9(1)) + r7p*Conjg(r9p(1)), dp )      &
 & + 1.85e-3_dp * Real(r8*Conjg(r10(1)) + r8p*Conjg(r10p(1)), dp )    &
 & - 9.921e-2_dp * Real(r8*Conjg(r9(1)) + r8p*Conjg(r9p(1)), dp )     &
 & + 2.833_dp* Real(r9(1),dp) + 0.2804_dp * (Abs(r7)**2 + Abs(r7p)**2)&
 & - 0.10698_dp * Real( r9(1) * Conjg(r10(1))                         &
 &                    + r9p(1) * Conjg(r10p(1)), dp)                  &
 & + 11.0348_dp * (Abs(r10(1))**2 + Abs(r10p(1))**2 )                 &
 & + 1.527_dp * (Abs(r9(1))**2 + Abs(r9p(1))**2 )                     &
 & + 3.763e-3_dp * (Abs(r8)**2 + Abs(r8p)**2 ) ) 

  ! ratio BR(B -> Xs mu+ mu-)/BR(B -> Xs e+ e-)_SM
  ratioBtoSee = BrBtoSEE/16.5529_dp

  ! branching ratio B -> Xs e+ e-
  BrBtoSEE = BrBtoSEE* 1.e-7_dp

BrBtoSMuMu = (2.1774_dp - 1.658e-3_dp * Aimag(R10(2))                 &
  & + 5.e-4_dp * Aimag(r10(2)*Conjg(r8) + r10p(2)*Conjg(r8p) )        &
  & + 5.34e-2_dp * Aimag(r7) + 5.27e-3_dp * Aimag(r8)                 &
  & + 2.266e-2_dp * Aimag(r7 * Conjg(r8) + r7p * Conjg(r8p) )         &
  & + 4.96e-3_dp * Aimag(r7 * Conjg(r9(2)) + r7p * Conjg(r9p(2)) )    &
  & + 2.61e-2_dp * Aimag(r8 * Conjg(r9(2)) + r8p * Conjg(r9p(2)) )    &
  & - 1.15e-2_dp * Aimag(r9(2)) - 0.5420_dp * Real( r10(2), dp)       &
  & + 2.08e-2_dp * Real(r7,dp) - 9.38e-3_dp * Real(r8,dp)             &
  & + 1.53e-2_dp * Real(r7*Conjg(r10(2)) + r7p*Conjg(r10p(2)), dp )   &
  & + 6.848e-2_dp * Real(r7 * Conjg(r8) + r7p * Conjg(r8p), dp )      &
  & - 0.8545_dp * Real(r7*Conjg(r9(2)) + r7p*Conjg(r9p(2)), dp )      &
  & + 1.85e-3_dp * Real(r8*Conjg(r10(2)) + r8p*Conjg(r10p(2)), dp )   &
  & - 9.81e-2_dp * Real(r8*Conjg(r9(2)) + r8p*Conjg(r9p(2)), dp )     &
  & + 2.6917_dp * Real(r9(2),dp) + 0.2880_dp*(Abs(r7)**2+Abs(r7p)**2) &
  & - 0.10698_dp * Real( r9(2) * Conjg(r10(2))                        &
  &                    + r9p(2) * Conjg(r10p(2)), dp)                 &
  & + 10.7652_dp * (Abs(r10(2))**2 + Abs(r10p(2))**2 )                &
  & + 1.4884_dp * (Abs(r9(2))**2 + Abs(r9p(2))**2 )                   &
  & + 3.81e-3_dp * (Abs(r8)**2 + Abs(r8p)**2 ) )

  ! ratio BR(B -> Xs mu+ mu-)/BR(B -> Xs mu+ mu-)_SM
  ratioBtoSMuMu = BrBtoSMuMu/16.0479_dp

  ! branching ratio B -> Xs mu+ mu-
  BrBtoSMuMu = BrBtoSMuMu* 1.e-7_dp
\end{lstlisting}


\subsubsection{$\boldsymbol{B^+ \to K^+ \ell^+ \ell^-}$}
Our results for $B^+ \to K^+ \ell^+ \ell^-$ are based on the
expressions given in \cite{Altmannshofer:2013foa}. The branching ratio
for $B^+ \to K^+ \mu^+ \mu^-$ in the high-$q^2$ region, $q^2$ being
the dilepton invariant mass squared, can be written as
\begin{equation} \label{eq:BKmumu}
\BR \left( B^+ \to K^+ \mu^+ \mu^- \right)_{q^2 \in [14.18,22] \text{GeV}^2} \simeq 1.11 + 0.22 \left( C_7^{\text{NP}} + C_7^\prime \right) + 0.27 \left( C_9^{\text{NP}} + C_9^\prime \right) - 0.27 \left( C_{10}^{\text{NP}} + C_{10}^\prime \right) \, .
\end{equation}
The coefficients in Eq. \eqref{eq:BKmumu} can be related to the ones
in our generic Lagrangian as
\begin{eqnarray}
C_7^{\text{NP}} &=& n_{CQ} \, \left( Q_1^R - Q_1^{R,\text{SM}} \right) \\
C_7^\prime &=& n_{CQ} \, Q_1^L \\
C_9^{\text{NP}} &=& n_{CQ} \, \left[ \left( E_{LL}^V + E_{LR}^V \right) - \left( E_{LL}^{V,\text{SM}} + E_{LR}^{V,\text{SM}} \right) \right] \\
C_9^\prime &=& n_{CQ} \, \left( E_{RR}^V + E_{RL}^V \right) \\
C_{10}^{\text{NP}} &=& n_{CQ} \, \left[ \left( E_{LL}^V - E_{LR}^V \right) - \left( E_{LL}^{V,\text{SM}} - E_{LR}^{V,\text{SM}} \right) \right] \\
C_{10}^\prime &=& n_{CQ} \, \left( E_{RR}^V - E_{RL}^V \right)
\end{eqnarray}
where the normalization factor $n_{CQ}$ was already defined after
Eq. \eqref{eq:nCQprev}.

\begin{lstlisting}[caption=BtoKLL.m]
NameProcess = "BtoKLL";
NameObservables = {{BrBtoKmumu, 6000, "BR(B -> K mu mu)"}, 
                   {ratioBtoKmumu, 6001, "BR(B -> K mu mu)/BR(B -> K mu mu)_SM"}};

NeededOperators = {OddllVRR, OddllVLL, OddllVRL, OddllVLR, CC7, CC7p,
                   OddllVRRSM, OddllVLLSM, OddllVRLSM, OddllVLRSM, CC7SM, CC7pSM
};

Body = "BtoKLL.f90";  
\end{lstlisting}

\begin{lstlisting}[caption=BtoKLL.f90]
Complex(dp) :: c7NP, c7p, c9NP, c9p, c10NP, c10p, norm
Real(dp) ::  GF

! ---------------------------------------------------------------- 
! B^+ -> K^+ l+ l- (14.18 GeV^2 < q^2 < 22 GeV^2)
! Observable implemented by W. Porod, F. Staub and A. Vicente
! Based on W. Altmannshofer, D. M. Straub, EPJ C 73 (2013) 2646
! [arXiv:1308.1501]
! ---------------------------------------------------------------- 

c7NP = (CC7(3,2) - CC7SM(3,2))
c7p = CC7p(3,2)
c9NP = (OddllVLL(3,2,1,1)+OddllVLR(3,2,1,1) - &
            & (OddllVLLSM(3,2,1,1)+OddllVLRSM(3,2,1,1)))
c9p =  (OddllVRR(3,2,1,1)+OddllVRL(3,2,1,1))
c10NP = (OddllVLL(3,2,1,1)-OddllVLR(3,2,1,1) - &
            &  (OddllVLLSM(3,2,1,1)-OddllVLRSM(3,2,1,1)))
c10p = (OddllVRR(3,2,1,1)-OddllVRL(3,2,1,1))


! running GF
GF = (Alpha_160*4._dp*Pi/sinW2_160)/mw**2*sqrt2/8._dp

! normalization of our Wilson coefficients
! relative to the ones used in arXiv:1308.1501
norm = - oo16pi2*4._dp*GF/sqrt2*CKM_160(3,3)*Conjg(CKM_160(3,2))

! Branching ratio in the high-q^2 region
! q^2 in [14.18,22] GeV^2
BrBtoKmumu = (1.11_dp + 0.22_dp*(c7Np+c7p)/norm + &
   & 0.27_dp*(c9NP+c9p)/norm - 0.27_dp*(c10NP+c10p)/norm)

! ratio relative to SM
ratioBtoKmumu = BrBtoKmumu/1.11_dp

! total BR
BrBtoKmumu = BrBtoKmumu*1.0E-7_dp
\end{lstlisting}


\subsubsection{$\boldsymbol{\bar B \to X_{d,s} \nu \bar \nu}$}
The branching ratio for $\bar B \to X_q \nu \bar \nu$, with $q = d,s$,
is given by~\cite{Bobeth:2001jm}
\begin{eqnarray}
\BR \left( \bar B \to X_q \nu \bar \nu \right) &=& \frac{\alpha^2}{4 \pi^2 \sin^4 \theta_W} \frac{|V_{tb} V_{tq}^\ast|^2}{|V_{cb}|^2} \frac{\BR \left( \bar B \to X_c e \bar \nu_e \right) \kappa(0)}{f(\hat m_c) \kappa(\hat m_c)} \label{BXnunu} \\
&& \times \sum_f \left[ \left( |c_L|^2 + |c_R|^2 \right) f(\hat m_q) - 4 \RE \left(c_L c_R^\ast \right) \hat m_q \tilde f(\hat m_q) \right] \nonumber \, .
\end{eqnarray}
The sum runs over the three neutrinos and $\hat m_i \equiv
m_i/m_b$. The functions $f(\hat m_c)$ and $\kappa(\hat m_c)$ represent
the phase-space and the 1-loop QCD corrections, respectively. In case
of $\kappa(\hat m_c)$, one needs the numerical values $\kappa(0) =
0.83$ and $\kappa(\hat m_c) = 0.88$. The functions $f(x)$ and $\tilde
f(x)$ take the form
\begin{eqnarray}
f(x) &=& 1 - 8x^2 + 8 x^6-x^8 -24 x^4 \log x \\
\tilde f(x) &=& 1 + 9x^2 - 9x^4-x^6+12x^2(1+x^2)\log x \, .
\end{eqnarray}
Finally, $\BR \left( \bar B \to X_c e \bar \nu_e \right)_{\text{exp}}
= 0.101$ \cite{Beringer:1900zz} and the coefficients $c_L$ and $c_R$
are given by
\begin{eqnarray}
c_L &=& n^q_{BX \nu\nu} \, F_{LL}^V \\
c_R &=& n^q_{BX \nu\nu} \, F_{RL}^V \, ,
\end{eqnarray}
where $\left(n^q_{BX \nu\nu}\right)^{-1} = \frac{4 G_F}{\sqrt{2}}
\frac{\alpha}{2 \pi \sin^2 \theta_W} V_{tb}^\ast V_{tq}$ is the
relative factor between our Wilson coefficients and the ones in
\cite{Bobeth:2001jm}.

\begin{lstlisting}[caption=BtoQnunu.m]
NameProcess = "BtoQnunu";
NameObservables = {{BrBtoSnunu, 7000, "BR(B->s nu nu)"},
                   {ratioBtoSnunu, 7001, "BR(B->s nu nu)/BR(B->s nu nu)_SM"}, 
                   {BrBtoDnunu, 7002, "BR(B->D nu nu)"}, 
                   {ratioBtoDnunu, 7003, "BR(B->D nu nu)/BR(B->D nu nu)_SM"}};

NeededOperators = {OddvvVRR, OddvvVLL, OddvvVRL, OddvvVLR,
                   OddvvVRRSM, OddvvVLLSM, OddvvVRLSM, OddvvVLRSM};

Body = "BtoQnunu.f90";  
\end{lstlisting}

\begin{lstlisting}[caption=BtoQnunu.f90]
Complex(dp) :: cL, cR, br, br_SM, cL_SM, cR_SM, norm
Real(dp) :: f_mq, tf_mq, kappa_0, kappa_c, f_mc, BrBXeNu, sw2, mq
Real(dp) :: prefactor, factor1, factor2, GF
Integer :: out, i1, i2 

! ---------------------------------------------------------------- 
! \bar{B} -> X_{d,s} nu nu
! Observable implemented by W. Porod, F. Staub and A. Vicente
! Based on C. Bobeth et al, NPB 630 (2002) 87 [hep-ph/0112305] 
! ---------------------------------------------------------------- 

kappa_0 = 0.830_dp
kappa_c = 0.88_dp
f_mc = 0.53_dp
BrBXeNu = 0.101_dp ! PDG central value

sw2 = sinw2_160
GF = (Alpha_160*4._dp*Pi/sinW2_160)/mw**2*sqrt2/8._dp

Do out = 1,2
If (out.eq.1) Then ! B -> X_d nu nu
 mq = mf_d(1)/mf_d(3)
 norm = Alpha_160*4._dp*GF/sqrt2/(2._dp*pi*sinw2_160)* &
             & Conjg(CKM_160(3,3)*Conjg( CKM_160(3,1) ))
Else ! B -> X_s nu nu
 mq = mf_d(2)/mf_d(3)
 norm = Alpha_160*4._dp*GF/sqrt2/(2._dp*pi*sinw2_160)* &
             & Conjg(CKM_160(3,3)*Conjg( CKM_160(3,2) ))
End if 

! f and tilde f functions
f_mq = 1._dp - 8._dp*mq**2 + 8._dp*mq**6 - &
            & mq**8 -24._dp*mq**4*Log(mq)
tf_mq = 1._dp + 9._dp*mq**2 - 9._dp*mq**4 - mq**6 + &
            & 12._dp*mq**2*(1._dp + mq**2)*Log(mq)

prefactor =  Alpha_mz**2/(4._dp*pi**2*sw2**2)*Abs(CKM_160(3,3)/ &
               & CKM_160(2,3))**2*BrBXeNu/(f_mc*kappa_c)*kappa_0
factor1 = f_mq
factor2 = - 4._dp*mq*tf_mq

br = 0._dp
br_SM = 0._dp

 Do i1= 1,3
  Do i2 = 1,3

   ! BSM
   cL = OddvvVLL(3,out,i1,i2)/norm
   cR = OddvvVRL(3,out,i1,i2)/norm
   br = br + factor1*(Abs(cL)**2 + Abs(cR)**2) +    &
               &  factor2*Real(cL*Conjg(cR),dp)

   ! SM
   cL = OddvvVLLSM(3,out,i1,i2)/norm
   cR = OddvvVRLSM(3,out,i1,i2)/norm
   br_SM = br_SM + factor1*(Abs(cL)**2 + Abs(cR)**2) + &
            & factor2*Real(cL*Conjg(cR),dp)

  End Do
 End do
If (out.eq.1) Then ! B -> X_d nu nu
  BrBtoDnunu = prefactor*br*Abs(CKM_160(3,1))**2 
  ratioBtoDnunu = br/br_SM
Else ! B -> X_s nu nu
  BrBtoSnunu = prefactor*br*Abs(CKM_160(3,2))**2  
  ratioBtoSnunu = br/br_SM
End if 
End Do 
\end{lstlisting}


\subsubsection{$\boldsymbol{K \to \pi \nu \bar \nu}$}
Following \cite{Bobeth:2001jm}, the branching ratios for rare Kaon
decays involving neutrinos in the final state can be written as
\begin{eqnarray}
\BR \left( K^+ \to \pi^+ \nu \bar \nu \right) &=& 2 r_1 \, r_2 \, r_{K^+} \sum_f \left[ \left( \text{Im} \lambda_t X_f \right)^2 + \left( \text{Re} \lambda_c X_{NL} + \text{Re} \lambda_t X_f \right)^2 \right] \\
\BR \left( K_L \to \pi^0 \nu \bar \nu \right) &=& 2 r_1 \, r_{K_L} \sum_f \left( \text{Im} \lambda_t X_f \right)^2 \, ,
\end{eqnarray}
where the sums are over the three neutrino species, $X_{NL} = 9.78
\cdot 10^{-4}$ is the SM NLO charm
correction~\cite{Buchalla:1993wq,Buchalla:1998ba}, $\lambda_t =
V_{ts}^\ast V_{td}$ and $\lambda_c = V_{cs}^\ast V_{cd}$, the
coefficients $r_1$, $r_2$, $r_{K^+}$ and $r_{K_L}$ take the numerical
values
\begin{eqnarray}
r_1 &=& 1.17 \cdot 10^{-4} \nonumber \\
r_2 &=& 0.24 \nonumber \\
r_{K^+} &=& 0.901 \nonumber \\
r_{K_L} &=& 0.944
\end{eqnarray}
and $X_f$ contains the Wilson coefficients contributing to the
processes, $F^V_{LL}$ and $F^V_{RL}$, as
\begin{equation}
X_f = n_{K \pi \nu\nu} \, \left( F^V_{LL} + F^V_{RL} \right) \, .
\end{equation}
Here $n_{K \pi \nu\nu}^{-1} = \frac{4 G_F}{\sqrt{2}} \frac{\alpha}{2
  \pi \sin^2 \theta_W} V_{ts}^\ast V_{td}$.
  
\begin{lstlisting}[caption=KtoPInunu.m]
NameProcess = "KtoPInunu";
NameObservables = {{BrKptoPipnunu, 8000, "BR(K^+ -> pi^+ nu nu)"}, 
                   {ratioKptoPipnunu, 8001, "BR(K^+ -> pi^+ nu nu)/BR(K^+ -> pi^+ nu nu)_SM"}, 
                   {BrKltoPinunu, 8002, "BR(K_L -> pi^0 nu nu)"}, 
                   {ratioKltoPinunu, 8003, "BR(K_L -> pi^0 nu nu)/BR(K_L -> pi^0 nu nu)_SM"}};

NeededOperators = {OddvvVRR, OddvvVLL, OddvvVRL, OddvvVLR,
                   OddvvVRRSM, OddvvVLLSM, OddvvVRLSM, OddvvVLRSM};

Body = "KtoPInunu.f90";  
\end{lstlisting}

\begin{lstlisting}[caption=KtoPInunu.f90]
Complex(dp) :: br, r1, r2, rKp, rKl, Xx, XNL, Lt, Lc
Complex(dp) :: Xx_SM, br_SM, norm
Real(dp) :: GF
Integer :: out, i1, i2 

! ---------------------------------------------------------------- 
! K -> pi nu nu
! Observable implemented by W. Porod, F. Staub and A. Vicente
! Based on C. Bobeth et al, NPB 630 (2002) 87 [hep-ph/0112305] 
! ---------------------------------------------------------------- 

GF = (Alpha_160*4._dp*Pi/sinW2_160)/mw**2*sqrt2/8._dp
norm = Alpha_160*4._dp*GF/sqrt2/(2._dp*pi*sinw2_160) &
          & *Conjg(CKM_160(3,2))*CKM_160(3,1)

r1 = 1.17E-4_dp
r2 = 0.24_dp
rKp =  0.901
rKl = 0.944

! SM NLO charm correction
! See G. Buchalla and A. Buras, NPB 412 (1994) 106 and NPB 548 (1999) 309
XNL = 9.78E-4_dp

! out = 1 : K^+ -> pi^+ nu nu
! out = 2 : K_L -> pi^0 nu nu

Do out = 1,2
br = 0._dp
br_SM = 0._dp
 Do i1= 1,3
  Do i2 = 1,3
   Xx = ((OddvvVLL(2,1,i1,i2)+OddvvVRL(2,1,i1,i2))/norm) 
   Xx_SM = ((OddvvVLLSM(2,1,i1,i2)+OddvvVRLSM(2,1,i1,i2))/norm) 
   Lt = Conjg(CKM_160(3,2))*CKM_160(3,1)
   Lc = Conjg(CKM_160(2,2))*CKM_160(2,1)
   If (out.eq.1) Then 
     br = br + Aimag(Xx*Lt)**2 + (Real(Lc*XNL,dp) + Real(Xx*Lt,dp))**2
     br_SM = br_SM + Aimag(Xx_SM*Lt)**2 + &
                 & (Real(Lc*XNL,dp) + Real(Xx_SM*Lt,dp))**2
   Else
    br = br + Abs(Aimag(Xx*Lt))**2
    br_SM = br_SM + Abs(Aimag(Xx_SM*Lt))**2
   End if   
  End Do
 End do
If (out.eq.1) Then ! K^+ -> pi^+ nu nu
 BrKptoPipnunu = 2._dp*r1*r2*rKp*br
 RatioKptoPipnunu = br/br_SM
 ! SM expectation: (7.2 +/- 2.1)*10^-11 (hep-ph/0112135)
Else ! K_L -> pi^0 nu nu
 BrKltoPinunu = 2._dp*r1*rKl*br
 RatioKltoPinunu = br/br_SM
 ! SM expectation: (3.1 +/- 1.0)*10^-11 (hep-ph/0408142)
End if 
End Do 
\end{lstlisting}


\subsubsection{$\boldsymbol{\Delta M_{B_{s,d}}}$}
The $B_q^0 - \bar B_q^0$ mass difference can be written as
\cite{Buras:2001mb,Buras:2002vd}
\begin{equation} \label{eq:MBq}
\Delta M_{B_q} = \frac{G_F^2 m_W^2}{6 \pi^2} m_{B_q} \eta_B f_{B_q}^2 \hat B_{B_q} |V_{tq}^{\text{eff}}|^2 |F_{tt}^q| \, ,
\end{equation}
where $q = s, d$, $m_{B_q}$ and $f_{B_q}$ are the $B_q^0$ mass and
decay constant, respectively, $\eta_B = 0.55$ is a QCD factor
\cite{Buras:1990fn,Urban:1997gw}, $\hat B_{B_q}$ is a non-perturbative
parameter (with values $\hat B_{B_d} = 1.26$ and $\hat B_{B_s} =
1.33$, obtained from recent lattice computations \cite{Laiho:2009eu})
and $|V_{tq}^{\text{eff}}|^2 = \left( V_{tb}^\ast V_{tq}
\right)^2$. $F_{tt}^q$ is given by
\begin{eqnarray}
F_{tt}^q &=& S_0(x_t) + \frac{1}{4 r} C_{\text{new}}^{VLL} \label{Ftt} \\
&& + \frac{1}{4 r} C_1^{VRR} + \bar{P}_1^{LR} C_1^{LR} + \bar{P}_2^{LR} C_2^{LR} \nonumber \\
&& + \bar{P}_1^{SLL} \left( C_1^{SLL} + C_1^{SRR} \right) + \bar{P}_2^{SLL} \left( C_2^{SLL} + C_2^{SRR} \right) \nonumber
\end{eqnarray}
where $r = 0.985$ \cite{Buras:1990fn}, $x_t = \frac{m_t^2}{m_W^2}$,
with $m_t$ the top quark mass, the $\bar P$ coefficients take the
numerical values
\begin{eqnarray}
\bar{P}_1^{LR} &=& -0.71 \nonumber \\
\bar{P}_2^{LR} &=& 0.90  \nonumber \\
\bar{P}_1^{SLL} &=& -0.37  \nonumber \\
\bar{P}_2^{SLL} &=& -0.72
\end{eqnarray}
and the function
\begin{equation} \label{eq:S0-inamilim}
S_0(x_t) = \frac{4 x_t - 11 x_t^2 + x_t^3}{4 (1-x_t)^2} - \frac{3 x_t^3 \log x_t}{2(1-x_t)^3}
\end{equation}
was introduced by Inami and Lim in \cite{Inami:1980fz} and given, for
example, in \cite{Buras:1998raa}. Finally, the coefficients in
Eq. \eqref{Ftt} are related to the $D_{XY}^I$ coefficients in
Eq.\eqref{eq:L-4D} as
\begin{eqnarray}
C_{\text{new}}^{VLL} &=& n^q_\Delta \left( D_{LL}^V - D_{LL}^{V,\text{SM}} \right) \\
C_1^{VRR} &=& n^q_\Delta D_{RR}^V \\
C_1^{LR} &=& n^q_\Delta \left( D_{LR}^V + D_{RL}^V \right) \\
C_2^{LR} &=& n^q_\Delta \left( D_{LR}^S + D_{RL}^S + \delta_2^{LR} \right) \\
C_1^{SLL} &=& n^q_\Delta \left( D_{LL}^S + \delta_1^{SLL} \right) \\
C_1^{SRR} &=& n^q_\Delta \left( D_{RR}^S + \delta_1^{SRR} \right) \\
C_2^{SLL} &=& n^q_\Delta D_{LL}^T \\
C_2^{SRR} &=& n^q_\Delta D_{RR}^T
\end{eqnarray}
where the factor $\left(n^q_\Delta\right)^{-1} = \frac{G_F^2 m_W^2}{16
  \pi^2} |V_{tq}^{\text{eff}}|^2$ normalizes our Wilson coefficients
to the ones in \cite{Buras:2001mb,Buras:2002vd}. The corrections
$\delta_2^{LR}$, $\delta_1^{SLL}$ and $\delta_1^{SRR}$ are induced by
double penguin diagrams mediated by scalar and pseudoscalar
states~\cite{Buras:2001mb,Buras:2002vd}. These 2-loop contributions
may have a sizable impact in some models, and their inclusion is
necessary in order to achieve a precise result for $\Delta
M_{B_q}$. They can be written as
\begin{eqnarray}
\delta_2^{LR} &=& - \frac{H_L^{S,P} \left(H_R^{S,P}\right)^\ast}{m_{S,P}^2} \label{eq:dp1} \\
\delta_1^{SLL} &=& - \frac{\left(H_L^{S,P}\right)^2}{2 \, m_{S,P}^2} \label{eq:dp2} \\
\delta_1^{SRR} &=& - \frac{\left(H_L^{S,P}\right)^2}{2 \, m_{S,P}^2} \label{eq:dp3}
\end{eqnarray}
where $H_L^{S,P}$ and $H_R^{S,P}$ are defined in
Eq.\eqref{eq:L-ddSP}. The double penguin corrections in
Eqs.\eqref{eq:dp1}-\eqref{eq:dp3} are obtained by summing up over all
scalar and pseudoscalar states in the model.

\begin{lstlisting}[caption=DeltaMBq.m]
NameProcess = "DeltaMBq";
NameObservables = {{DeltaMBs, 1900, "Delta(M_Bs)"}, 
                   {ratioDeltaMBs, 1901, "Delta(M_Bs)/Delta(M_Bs)_SM"},  
                   {DeltaMBq, 1902, "Delta(M_Bd)"},  
                   {ratioDeltaMBq, 1903, "Delta(M_Bd)/Delta(M_Bd)_SM"}};

ExternalStates =  {Fd}; 
NeededOperators = {O4dSLL, O4dSRR, O4dSRL, O4dSLR, O4dVRR, O4dVLL, 
                  O4dVLLSM, O4dVRL, O4dVLR, O4dTLL, O4dTLR, O4dTRL, O4dTRR};

IncludeSMprediction["DeltaMBq"] = False;

Body = "DeltaMBq.f90"; 
\end{lstlisting}

\begin{lstlisting}[caption=DeltaMBq.f90]
Complex(dp) :: MBq, etaB, FBq2, BBq, Ftt, Veff2, r, &
     & P1bLR, P2bLR, P1bSLL, P2bSLL, norm, &
     & CVLLnew, C1VRR, C1LR, C2LR, C1SLL, C1SRR, C2SLL, C2SRR
Real(dp) ::  hbar, xt, GF
Real(dp) :: mS
Complex(dp) :: HL, HR, AL, AR
Integer :: i1, iS

! ---------------------------------------------------------------- 
! Delta M_{Bd,Bs}
! Observable implemented by W. Porod, F. Staub and A. Vicente
! Based on A. J. Buras et al, NPB 619 (2001) 434 [hep-ph/0107048]
! and NPB 659 (2003) 3 [hep-ph/0210145]
! ---------------------------------------------------------------- 

hbar =  6.58211889e-25_dp
xt = mf_u2_160(3)/mw2
r = 0.985_dp
P1bLR = -0.71_dp
P2bLR = 0.90_dp
P1bSLL = -0.37_dp
P2bSLL = -0.72_dp

! QCD factor, see A. J. Buras et al, NPB 47 (1990) 491
! and J. Urban et al, NPB 523 (1998) 40 
etaB = 0.55_dp

GF =  (Alpha_160*4._dp*Pi/sinW2_160)/mw**2*sqrt2/8._dp

Do i1 = 1,2

If (i1.eq.1) Then ! Delta M_Bd
 MBq = mass_B0d
 FBq2 = f_B0d_CONST**2
 BBq = 1.26_dp ! see arXiv:0910.2928
 Veff2 = Conjg(Conjg(CKM_160(3,3))*CKM_160(3,1))**2
Else ! Delta M_Bs
 MBq = mass_B0s
 FBq2 = f_B0s_CONST**2
 BBq = 1.33_dp ! see arXiv:0910.2928
 Veff2 = Conjg(Conjg(CKM_160(3,3))*CKM_160(3,2))**2
End if

! normalization factor
norm = GF**2*mw2/(16._dp*Pi**2)*Veff2

! Wilson coefficients
CVLLnew = (O4dVLL(3,i1,3,i1)-O4dVLLSM(3,i1,3,i1))/norm ! we remove the SM contribution
C1VRR = O4dVRR(3,i1,3,i1)/norm
C1LR = (O4dVLR(3,i1,3,i1)+O4dVRL(3,i1,3,i1))/norm
C2LR = (O4dSLR(3,i1,3,i1)+O4dSRL(3,i1,3,i1))/norm
C1SLL = O4dSLL(3,i1,3,i1)/norm
C1SRR = O4dSRR(3,i1,3,i1)/norm
C2SLL  = O4dTLL(3,i1,3,i1)/norm
C2SRR  = O4dTRR(3,i1,3,i1)/norm


! Double Higgs penguins
@ If[getGen[HiggsBoson] > 1, "Do iS = 1, "<>ToString[getGen[HiggsBoson]],""]
@ If[getGen[HiggsBoson] > 1, "HL = OH2qSL(3,i1,iS)",  "HL = OH2qSL(3,i1)"]
@ If[getGen[HiggsBoson] > 1, "HR = OH2qSR(3,i1,iS)",  "HR = OH2qSR(3,i1)"]
@ If[getGen[HiggsBoson] > 1, "mS = "<>SPhenoMassSq[HiggsBoson,iS],  "mS = "<>SPhenoMassSq[HiggsBoson]]
C2LR = C2LR - HL*Conjg(HR)/(mS*norm)
C1SLL = C1SLL - 0.5_dp*HL**2/(mS*norm)
C1SRR = C1SRR - 0.5_dp*HR**2/(mS*norm)
@ If[getGen[HiggsBoson] > 1,"End Do",""]


@ If[getGen[PseudoScalar] > 1, "Do iS = "<>ToString[getGenSPhenoStart[PseudoScalar]]<>", "<>ToString[getGen[PseudoScalar]],""]
@ If[getGen[PseudoScalar] > 1, "AL = OAh2qSL(3,i1,iS)",  "AL = OAh2qSL(3,i1)"]
@ If[getGen[PseudoScalar] > 1, "AR = OAh2qSR(3,i1,iS)",  "AR = OAh2qSR(3,i1)"]
@ If[getGen[PseudoScalar] > 1, "mS = "<>SPhenoMassSq[PseudoScalar,iS],  "mS = "<>SPhenoMassSq[PseudoScalar]]
C2LR = C2LR - AL*Conjg(AR)/(mS*norm)
C1SLL = C1SLL - 0.5_dp*AL**2/(mS*norm)
C1SRR = C1SRR - 0.5_dp*AR**2/(mS*norm)
@ If[getGen[PseudoScalar] > 1,"End Do",""]


Ftt = S0xt(xt) + CVLLnew/(4._dp*r) + &
     & C1VRR/(4._dp*r) + P1bLR*C1LR + P2bLR*C2LR  + &
     & P1bSLL*(C1SLL + C1SRR) + P2bSLL*(C2SLL + C2SRR)

If (i1.eq.1) Then ! Delta M_Bd
   ratioDeltaMBq = Abs(Ftt/S0xt(xt))
   DeltaMBq = G_F**2*mw2/(6._dp*Pi**2)*    &
      & MBq*etaB*BBq*FBq2*Veff2*Abs(Ftt)*1.e-12_dp/hbar
Else ! Delta M_Bs
   ratioDeltaMBs = Abs(Ftt/S0xt(xt))
   DeltaMBs =  G_F**2*mw2/(6._dp*Pi**2)*   & 
      & MBq*etaB*BBq*FBq2*Veff2*Abs(Ftt)*1.e-12_dp/hbar
End if
 
End Do

Contains 

  Real(dp) Function S0xt(x) ! See for example hep-ph/9806471
    Implicit None
    Real(dp), Intent(in) :: x
    S0xt = 1._dp - 2.75_dp * x + 0.25_dp * x**2 &
        & - 1.5_dp * x**2 * Log(x) / (1-x)
    S0xt = x*S0xt / (1 -x)**2
  End  Function S0xt 
\end{lstlisting}


\subsubsection{$\boldsymbol{\Delta M_{K}}$ and $\boldsymbol{\varepsilon_K}$}
$\Delta M_{K}$ and $\varepsilon_K$, the observables associated to $K^0
- \bar K^0$ mixing, can be written
as~\cite{Buras:1998raa,Crivellin:2012jv}
\begin{eqnarray}
\Delta M_{K} &=& 2 \, \text{Re} \, \langle \bar K^0 | H_\text{eff}^{\Delta S = 2} | K^0 \rangle \label{deltaMK} \\
\varepsilon_K &=& \frac{e^{i \pi/4}}{\sqrt{2} \Delta M_{K}} \, \text{Im} \, \langle \bar K^0 | H_\text{eff}^{\Delta S = 2} | K^0 \rangle \label{epsK} \, .
\end{eqnarray}
The matrix element in Eqs. \eqref{deltaMK} and \eqref{epsK} is given by
\begin{eqnarray}
\langle \bar K^0 | H_\text{eff}^{\Delta S = 2} | K^0 \rangle &=& f_V \left( D_{LL}^V + D_{RR}^V \right) + f_S \left( D_{LL}^S + D_{RR}^S \right) + f_T \left( D_{LL}^T + D_{RR}^T \right) \nonumber \\
&& + f_{LR}^1 \left( D_{LR}^S + D_{RL}^S \right) + f_{LR}^2 \left( D_{LR}^V + D_{RL}^V \right) \, .
\end{eqnarray}
The $f$ coefficients are
\begin{eqnarray}
f_V &=& \frac{1}{3} m_K f_K^2 B_1^{VLL}(\mu) \\
f_S &=& - \frac{5}{24} \left( \frac{m_K}{m_s(\mu) + m_d(\mu)} \right)^2 m_K f_K^2 B_1^{SLL}(\mu) \\
f_T &=& - \frac{1}{2} \left( \frac{m_K}{m_s(\mu) + m_d(\mu)} \right)^2 m_K f_K^2 B_2^{SLL}(\mu) \\
f_{LR}^1 &=& - \frac{1}{6} \left( \frac{m_K}{m_s(\mu) + m_d(\mu)} \right)^2 m_K f_K^2 B_1^{LR}(\mu) \\
f_{LR}^2 &=& \frac{1}{4} \left( \frac{m_K}{m_s(\mu) + m_d(\mu)} \right)^2 m_K f_K^2 B_2^{LR}(\mu)
\end{eqnarray}
where $\mu = 2$ GeV is the energy scale at which the matrix element is
computed and $f_K$ the Kaon decay constant. The values of the quark
masses at $\mu = 2$ GeV are given by $m_d(\mu) = 7$ MeV and $m_s(\mu)
= 125$ MeV (see table 1 in \cite{Ciuchini:1998ix}), whereas the
$B_i^X$ coefficients have the following values at $\mu = 2$ GeV
\cite{Buras:2001ra}: $B_1^{VLL}(\mu) = 0.61$, $B_1^{SLL}(\mu) = 0.76$,
$B_2^{SLL}(\mu) = 0.51$, $B_1^{LR}(\mu) = 0.96$ and $B_2^{LR}(\mu) =
1.3$.

As in \cite{Crivellin:2012jv}, we treat the SM contribution
separately. We define $D_{LL}^V = D_{LL}^{V,SM} + D_{LL}^{V,BSM}$. For
$D_{LL}^{V,BSM}$ one just subtracts the SM contributions to
$D_{LL}^V$, whereas for $D_{LL}^{V,SM}$ one can use the results in
\cite{Herrlich:1993yv,Herrlich:1995hh,Herrlich:1996vf}, where the
relevant QCD corrections are included,
\begin{equation}
D_{LL}^{V,SM} = \frac{G_F^2 m_W^2}{4 \pi^2} \left[ \lambda_c^{\ast \, 2} \eta_1 S_0(x_c) + 
\lambda_t^{\ast \, 2} \eta_2 S_0(x_t) + 2 \lambda_c^\ast \lambda_t^\ast \eta_3 S_0(x_c,x_t) \right] \, .
\end{equation}
Here $x_i = m_i^2/m_w^2$, $\lambda_i = V_{is}^\ast V_{id}$ and
$S_0(x)$ and $S_0(x,y)$ are the Inami-Lim
functions~\cite{Inami:1980fz}.  $S_0(x)$ was already defined in Eq.
\eqref{eq:S0-inamilim}, whereas $S_0(x_c,x_t)$ is given
by~\cite{Buras:1998raa}
\begin{equation} \label{eq:S0-inamilim2}
S_0(x_c,x_t) = x_c \left[ \log \frac{x_t}{x_c} - \frac{3 x_t}{4(1-x_t)} - \frac{3 x_t^2 \log x_t}{4 (1-x_t)^2} \right] \, .
\end{equation}
In the last expression we have kept only terms linear in $x_c \ll
1$. Finally, the $\eta_i$ coefficients comprise short distance QCD
corrections. Their numerical values are $\eta_{1,2,3} = \left( 1.44 ,
0.57 , 0.47 \right)$ \cite{Herrlich:1996vf}~\footnote{Note that we
have chosen a value for $\eta_1$ which results from our numerical
values for $\alpha_s(m_Z)$ and $m_c(m_c)$, see table 5
in \cite{Herrlich:1996vf}.}.

\begin{lstlisting}[caption=KKmix.m]
NameProcess = "KKmix";
NameObservables = {{DeltaMK, 9100, "Delta(M_K)"}, 
                   {ratioDeltaMK, 9102, "Delta(M_K)/Delta(M_K)_SM"}, 
                   {epsK, 9103, "epsilon_K"}, 
                   {ratioepsK, 9104, "epsilon_K/epsilon_K^SM"}};

NeededOperators = {O4dSLL, O4dSRR, O4dSRL, O4dSLR, O4dVRR, O4dVLL, O4dVRL, 
                   O4dVLR, O4dTLL, O4dTLR, O4dTRL, O4dTRR,
                   O4dSLLSM, O4dSRRSM, O4dSRLSM, O4dSLRSM, O4dVRRSM, O4dVLLSM, O4dVRLSM, O4dVLRSM, 
                   O4dTLLSM, O4dTLRSM, O4dTRLSM, O4dTRRSM};

Body = "KKmix.f90";  
\end{lstlisting}

\begin{lstlisting}[caption=KKmix.f90]
Real(dp) :: b_VLL, b_SLL1, b_SLL2, b_LR1, b_LR2
Real(dp) :: ms_mu, md_mu
Complex(dp) :: CVLL, CVRR, CSLL, CSRR, CTLL, CTRR, CLR1, CLR2
Complex(dp) :: fV, fS, fT, fLR1, fLR2, cVLLSM
Complex(dp) :: f_k, M_K, H2eff, DeltaMK_SM, epsK_SM
Real(dp) :: norm, hbar, xt, xc, GF
Integer :: i1
Real(dp), Parameter :: eta_tt = 0.57_dp, eta_ct = 0.47_dp, &
                         & eta_cc = 1.44_dp
! Parameters from S. Herrlich and U. Nierste NPB 476 (1996) 27

! ---------------------------------------------------------------- 
! Delta M_K and epsilon_K 
! Observables implemented by W. Porod, F. Staub and A. Vicente
! Based on A. Crivellin et al, Comput. Phys. Commun. 184 (2013) 1004 [arXiv:1203.5023]
! ---------------------------------------------------------------- 

! using globally defined hadronic parameters
M_K = mass_K0
f_K = f_k_CONST

xt = mf_u(3)**2 / mW**2
xc = mf_u(2)**2 / mW**2

GF = (Alpha_160*4._dp*Pi/sinW2_160)/mw**2*sqrt2/8._dp

! Coefficients at mu = 2 GeV
! See A. J. Buras et al, NPB 605 (2001) 600 [hep-ph/0102316] 
b_VLL = 0.61_dp 
b_SLL1 = 0.76_dp 
b_SLL2 = 0.51_dp 
b_LR1 = 0.96_dp
b_LR2 = 1.3_dp

! Quark mass values at mu = 2 GeV
! See M. Ciuchini et al, JHEP 9810 (1998) 008 [hep-ph/9808328] - Table 1
md_mu = 0.007_dp
ms_mu = 0.125_dp

fV = 1._dp/3._dp*M_K*f_k**2*b_VLL
fS = -5._dp/24._dp*M_K*f_K**2*(M_K/(ms_mu+md_mu))**2*b_SLL1
fT = -1._dp/2._dp*M_K*f_K**2*(M_K/(ms_mu+md_mu))**2*b_SLL2
fLR1 = -1._dp/6._dp*M_K*f_K**2*(M_K/(ms_mu+md_mu))**2*b_LR1
fLR2 = 1._dp/4._dp*M_K*f_K**2*(M_K/(ms_mu+md_mu))**2*b_LR2

! SM contribution
! Based on the results by S. Herrlich and U. Nierste
! NPB 419 (1994) 292, PRD 52 (1995) 6505 and NPB 476 (1996) 27
cVLLSM = eta_cc * (Conjg(CKM_160(2,2))*CKM_160(2,1))**2 * S0xt(xc)    &
     & + eta_tt * (Conjg(CKM_160(3,2))*CKM_160(3,1))**2 * S0xt(xt)    &
     & + Conjg(CKM_160(2,2)*CKM_160(3,2))*(CKM_160(2,1)*CKM_160(3,1)) &
     &   * 2._dp * eta_ct * S0_2(xc,xt)

cVLLSM = Conjg(cVLLSM)  ! we compute (d\bar{s})(d\bar{s}) and not (\bar{d}s)(\bar{d}s)
cVLLSM = oo4pi2*(GF*mW)**2*cVLLSM ! normalization

! BSM contributions (+SM in CVLL)
CVLL = O4dVLL(2,1,2,1)-O4dVLLSM(2,1,2,1)+cVLLSM
CVRR = O4dVRR(2,1,2,1)
CSLL = O4dSLL(2,1,2,1)
CSRR = O4dSRR(2,1,2,1)
CTLL = O4dTLL(2,1,2,1)
CTRR = O4dTRR(2,1,2,1)
CLR1 = O4dSLR(2,1,2,1)+O4dSRL(2,1,2,1)  
CLR2 = O4dVLR(2,1,2,1)+O4dVRL(2,1,2,1)

! BSM
H2eff = fV*(CVLL+CVRR) + fS*(CSLL+CSRR) +fT*(CTLL+CTRR) &
              & + fLR1*CLR1 + fLR2*CLR2

DeltaMK = Abs(2._dp*Real(H2eff,dp))
epsK = 1._dp/(sqrt2*DeltaMK)*Abs(Aimag(H2eff))

! SM
H2eff = fV*cVLLSM

DeltaMK_SM = Abs(2._dp*Real(H2eff,dp))
epsK_SM = 1._dp/(sqrt2*DeltaMK_SM)*Abs(Aimag(H2eff))

ratioDeltaMK = DeltaMK/DeltaMK_SM
ratioepsK = epsK/epsK_SM

Contains 

! Inami - Lim functions

 Real(dp) Function S0xt(x)
   Implicit None
   Real(dp), Intent(in) :: x
   S0xt = 1._dp - 2.75_dp * x + 0.25_dp * x**2 - &
              &  1.5_dp * x**2 * Log(x) / (1-x)
   S0xt = x*S0xt / (1 -x)**2
 End  Function S0xt
 
 Real(dp) Function S0_2(xc, xt)
   Implicit None
   Real(dp), Intent(in) :: xc, xt
   S0_2 = Log(xt/xc) - 0.75_dp *  xt /(1-xt) &
        & - 0.75_dp * xt**2 * Log(xt) / (1-xt)**2
   S0_2 = xc *  S0_2
 End  Function S0_2 
\end{lstlisting}


\subsubsection{$\boldsymbol{P \to \ell \nu}$}

Although $P \to \ell \nu$, where $P = qq'$ is a pseudoscalar meson,
does not violate quark flavor, we have included it in the list of
observables for practical reasons, as it can be computed with the same
ingredients as the QFV observables. The decay width for the process $P
\to \ell_\alpha \nu$ is given by \cite{Barranco:2013tba}
\begin{eqnarray}
\Gamma \left( P \to \ell_\alpha \nu \right) &=& \frac{|G_F f_P (m_P^2 - m_{\ell_\alpha}^2)|^2}{8 \pi m_P^3} \label{Plnu} \\
&& \times \sum_\nu \left| V_{qq'} m_{\ell_\alpha} + \frac{m_{\ell_\alpha}}{2 \sqrt{2}} \left( G_{LL}^V - G_{RL}^V \right) + \frac{m_P^2}{2 \sqrt{2} (m_q + m_{q'})} \left( G_{RR}^S - G_{LR}^S \right) \right|^2 \, . \nonumber
\end{eqnarray}
Here $f_P$ is the meson decay constant, $m_q$ and $m_{q'}$ are the
masses of the quarks in the meson and the Wilson coefficients
$G_{XY}^I$ are defined in Eq.\eqref{eq:L-DULV}. The sum in
Eq.\eqref{Plnu} is over the three neutrinos (whose masses are
neglected).

Each $P \to \ell_\alpha \nu$ decay width is plagued by hadronic
uncertainties. However, by taking the ratios
\begin{equation}
R_P = \frac{\Gamma \left( P \to e \nu \right)}{\Gamma \left( P \to \mu \nu \right)}
\end{equation}
the hadronic uncertainties cancel out to a good approximation,
allowing for a precise theoretical determination. In case of $R_K$,
the SM prediction includes small electromagnetic corrections that
account for internal bremsstrahlung and structure-dependent effects
\cite{Cirigliano:2007xi}. This leads to an impressive theoretical
uncertainty of $\delta R_K / R_K \sim 0.1 \%$, making $R_P$ the
perfect observable to search for lepton flavor universality violation
\cite{Abada:2012mc}.

\begin{lstlisting}[caption=Plnu.m]
NameProcess = "Plnu";
NameObservables = {{BrDmunu, 300, "BR(D->mu nu)"}, 
                   {ratioDmunu, 301, "BR(D->mu nu)/BR(D->mu nu)_SM"},
                   {BrDsmunu, 400, "BR(Ds->mu nu)"}, 
                   {ratioDsmunu, 401, "BR(Ds->mu nu)/BR(Ds->mu nu)_SM"},
                   {BrDstaunu, 402, "BR(Ds->tau nu)"},
                   {ratioDstaunu, 403, "BR(Ds->tau nu)/BR(Ds->tau nu)_SM"},
                   {BrBmunu, 500, "BR(B->mu nu)"},
                   {ratioBmunu, 501, "BR(B->mu nu)/BR(B->mu nu)_SM"},
                   {BrBtaunu, 502, "BR(B->tau nu)"},
                   {ratioBtaunu, 503, "BR(B->tau nu)/BR(B->tau nu)_SM"},
                   {BrKmunu, 600, "BR(K->mu nu)"},
                   {ratioKmunu, 601, "BR(K->mu nu)/BR(K->mu nu)_SM"},
                   {RK, 602 ,"R_K = BR(K->e nu)/(K->mu nu)"},
                   {RKSM, 603 ,"R_K^SM = BR(K->e nu)_SM/(K->mu nu)_SM"}};

NeededOperators = {OdulvSLL, OdulvSRR, OdulvSRL, OdulvSLR,
                   OdulvVRR, OdulvVLL, OdulvVRL, OdulvVLR,
                   OdulvSLLSM, OdulvSRRSM, OdulvSRLSM, OdulvSLRSM,
                   OdulvVRRSM, OdulvVLLSM, OdulvVRLSM, OdulvVLRSM
};

Body = "Plnu.f90"; 
\end{lstlisting}

\begin{lstlisting}[caption=Plnu.f90]
Integer :: gt1, gt2, i1, i2, iP
Complex(dp) :: br, br_SM
Real(dp) :: m_M, f_M, tau_M, mlep, mq1, mq2, hbar, ratio, &
     & BrKenuSM, BRKenu, QED

! ---------------------------------------------------------------- 
! P -> l nu
! Observable implemented by W. Porod, F. Staub and A. Vicente
! Based on J. Barranco et al, arXiv:1303.3896
! ---------------------------------------------------------------- 

hbar = 6.58211889e-25_dp

! Electromagnetic correction to R_K
! See V. Cirigliano, I. Rosell, PRL 99 (2007) 231801 [arXiv:0707.3439]
QED = -3.6e-2_dp

! meson parameters

Do iP=1,4
If (iP.eq.1) Then ! Ds-meson
 gt1 = 2
 gt2 = 2
 m_M = mass_Dsp
 f_M = f_DSp_CONST
 tau_M = tau_DSp/hbar
Elseif (iP.eq.2) Then ! B-meson
 gt1 = 3
 gt2 = 1
 m_M = mass_Bp
 f_M = f_Bp_CONST
 tau_M = tau_Bp/hbar
Elseif (iP.eq.3) Then ! Kaon
 gt1 = 2
 gt2 = 1
 m_M = mass_Kp
 f_M =  f_Kp_CONST
 tau_M = tau_Kp/hbar
Elseif (iP.eq.4) Then ! D-meson
 gt1 = 1
 gt2 = 2
 m_M = mass_Dp
 f_M =  f_Dp_CONST
 tau_M = tau_Dp/hbar
End if

 mq1 =  mf_u_160(gt2)
 mq2 =  mf_d_160(gt1)

Do i1=1,3
br = 0._dp
br_SM = 0._dp
mlep = mf_l(i1)

Do i2=1,3
 br = br + ((OdulvVLL(gt1,gt2,i1,i2)-OdulvVLR(gt1,gt2,i1,i2))*mlep/   &
                   &           (2._dp*sqrt2)                          &
   & + m_M**2*(OdulvSRL(gt1,gt2,i1,i2)-OdulvSLL(gt1,gt2,i1,i2))/      &
                   & (2._dp*sqrt2*(mq1+mq2)))
 br_SM = br_SM+ (OdulvVLLSM(gt1,gt2,i1,i2)-OdulvVLRSM(gt1,gt2,i1,i2)) &
                   & *mlep/(2._dp*sqrt2)
End Do

ratio = Abs(br/br_SM)**2
br = oo8pi*tau_M*(f_M)**2*M_M*Abs(br)**2*(1._dp - mlep**2/M_M**2)**2 ! G_F already in coefficients included


If (iP.eq.1) Then  !! Ds-meson 
 If (i1.eq.2) Then  ! Ds->mu nu
  BrDsmunu =   br
  ratioDsmunu = ratio
  Elseif (i1.eq.3) Then ! Ds->tau nu
  BrDstaunu =  br
  ratioDstaunu = ratio
 End if
Elseif (iP.eq.2) Then !! B-meson
 If (i1.eq.2) Then  ! B->mu nu
  BrBmunu =   br
  ratioBmunu = ratio
  Else              ! B->tau nu
  BrBtaunu =  br
  ratioBtaunu = ratio
 End if
Else If (iP.eq.3) Then !! Kaon 
 If (i1.eq.1) Then  ! K->e nu
  BrKenu =   br
  BrKenuSM = BrKenu*ratio
  Elseif (i1.eq.2) Then  ! K->mu nu
  BrKmunu =  br
  ratioKmunu = ratio
  RK = BrKenu/BrKmunu*(1+QED)
  RKSM = BrKenuSM/BrKmunu*ratio*(1+QED)
 End if
Else If (iP.eq.4) Then  !! D-meson 
 If (i1.eq.2) Then  ! D->mu nu
  BrDmunu =   br
  ratioDmunu = ratio
 End if
End if
End Do
End Do 
\end{lstlisting}


\section{Models}
\label{app:models}
The following models are included in the public version of \SARAH and can now be used together with the \FlavorKit
to get predictions for the different observables. 

\subsection{Supersymmetric Models}
\begin{itemize}
 \item Minimal supersymmetric standard model (see Ref.~\cite{Martin:1997ns} and references therein)
    \begin{itemize} 
     \item With general flavor and CP structure ({\tt MSSM})
     \item Without flavor violation ({\tt MSSM/NoFV})
     \item With explicit CP violation in the Higgs sector ({\tt MSSM/CPV})
     \item In SCKM basis ({\tt MSSM/CKM})
    \end{itemize}
   \item Singlet extensions: 
   \begin{itemize}
    \item Next-to-minimal supersymmetric standard model ({\tt NMSSM}, {\tt NMSSM/NoFV}, {\tt NMSSM/CPV}, {\tt NMSSM/CKM}) (see Refs.~\cite{Maniatis:2009re,Ellwanger:2009dp} and references therein)
    \item near-to-minimal supersymmetric standard model  ({\tt near-MSSM})  \cite{Barger:2006dh}
    \item General singlet extended, supersymmetric standard model ({\tt SMSSM})   \cite{Barger:2006dh,Ross:2012nr}
    \item DiracNMSSM ({\tt DiracNMSSM})   \cite{Lu:2013cta,Kaminska:2014wia}
  \end{itemize}
  \item Triplet extensions 
  \begin{itemize} 
    \item Triplet extended MSSM ({\tt TMSSM})  \cite{DiChiara:2008rg}
    \item Triplet extended NMSSM ({\tt TNMSSM})  \cite{Agashe:2011ia}
  \end{itemize}
   \item Models with $R$-parity violation  \cite{Hall:1983id,Dreiner:1997uz,Allanach:2003eb,Bhattacharyya:1997vv,Barger:1989rk,Allanach:1999ic,Hirsch:2000ef,Barbier:2004ez}
  \begin{itemize}
    \item bilinear RpV ({\tt MSSM-RpV/Bi}) 
    \item Lepton number violation ({\tt MSSM-RpV/LnV})
    \item Only trilinear lepton number violation ({\tt MSSM-RpV/TriLnV})
    \item Baryon number violation ({\tt MSSM-RpV/BnV})  
    \item $\mu\nu$SSM ({\tt munuSSM}) \cite{LopezFogliani:2005yw,Bartl:2009an}
  \end{itemize}
   \item Additional $U(1)'s$ 
  \begin{itemize}
    \item $U(1)$-extended MSSM ({\tt UMSSM})   \cite{Barger:2006dh}
    \item secluded MSSM ({\tt secluded-MSSM})   \cite{Chiang:2009fs}
    \item minimal $B-L$ model ({\tt B-L-SSM})    \cite{Khalil:2007dr,FileviezPerez:2010ek,O'Leary:2011yq,Basso:2012gz}
    \item minimal singlet-extended $B-L$ model ({\tt N-B-L-SSM})
  \end{itemize}
   \item SUSY-scale seesaw extensions
    \begin{itemize}
      \item inverse seesaw ({\tt inverse-Seesaw}) \cite{Malinsky:2005bi,BhupalDev:2012ru}
      \item linear seesaw ({\tt LinSeesaw})  \cite{Malinsky:2005bi,DeRomeri:2012qd}
      \item singlet extended inverse seesaw ({\tt inverse-Seesaw-NMSSM}) \cite{Gogoladze:2012jp}
      \item inverse seesaw with $B-L$ gauge group ({\tt B-L-SSM-IS})  \cite{Basso:2012ew}
      \item minimal $U(1)_R \times U(1)_{B-L}$ model with inverse seesaw  ({\tt BLRinvSeesaw}) \cite{Hirsch:2011hg,Hirsch:2012kv}
\end{itemize}
 \item Models with Dirac Gauginos
   \begin{itemize}
    \item MSSM/NMSSM with Dirac Gauginos ({\tt DiracGauginos})  \cite{Belanger:2009wf,Benakli:2010gi,Benakli:2012cy}
    \item minimal $R$-Symmetric SSM ({\tt MRSSM}) \cite{Kribs:2007ac}
    \item Minimal Dirac Gaugino supersymmetric standard model ({\tt MDGSSM}) \cite{Benakli:2014cia}
   \end{itemize}
 \item High-scale extensions
\begin{itemize}
 \item Seesaw 1 - 3 ($SU(5)$ version) ,
 ({\tt Seesaw1},{\tt Seesaw2},{\tt Seesaw3})  \cite{Borzumati:2009hu,Rossi:2002zb,Hirsch:2008dy,Esteves:2009vg,Esteves:2010ff}
 \item Left/right model ($\Omega$LR) ({\tt Omega}) \cite{Esteves:2010si,Esteves:2011gk}
 \item Quiver model ({\tt QEW12}, {\tt  QEWmld2L3}) \cite{Bharucha:2013ela}
\end{itemize}
\end{itemize}

\subsection{Non-Supersymmetric Models}
\begin{itemize}
\item Standard Model (SM) ({\tt SM}), Standard model in CKM basis ({\tt SM/CKM})  (see for instance  Ref.~\cite{Hollik:2010id} and references therein)
\item inert Higgs doublet model ({\tt Inert})  \cite{LopezHonorez:2006gr}
\item B-L extended SM ({\tt B-L-SM}) \cite{Emam:2007dy,Basso:2008iv,Basso:2009hf}
\item B-L extended SM with inverse seesaw ({\tt B-L-SM-IS}) \cite{Khalil:2010iu}
\item SM extended by a scalar color octet ({\tt SM-8C}) \cite{Patel:2013zla}
\item Two Higgs doublet model ({\tt THDM})  (see for instance  Ref.~\cite{Branco:2011iw} and references therein)
\item Singlet extended SM ({\tt SSM}) \cite{O'Connell:2006wi}
\item Singlet Scalar DM ({\tt SSDM}) \cite{Goudelis:2009zz}
\end{itemize}

\bibliography{lit.bib}
\bibliographystyle{ArXiv}

\end{document}